\documentclass{pasa}%

\usepackage{natbib}
\usepackage{color}
\usepackage{fancybox}
\def\aap{{ A\&A}}

\def\aj{{AJ}}

\def\apj{{ApJ}}
\def\apjl{{ApJL}}

\def\mnras{{MNRAS}}

\def\pasp{{PASP}}
\def\aaps{{Astrophysics \& Space Science}}
\def\nat{Nature}

\def\apjs{{ApJS}}
\def\araa{{ARA\&A}}

\def\icarus{{ICARUS}}

\def\gca{{"Geochimica Cosmochimica Acta"}}
\def\physscr{{"Phys.~Scr."}}

\newcommand\mearth{{\,{\rm M}_{\oplus}}}
\newcommand\mj{{\,{\rm M}_{\rm J}}}
\newcommand\rsun{{\,{\rm R}_{\odot}}}
\newcommand\msun{{\,{\rm M}_{\odot}}}

\def\del#1{{}}

\usepackage{ulem}
\usepackage{enumerate}
\newcommand{\SNc}[1]{\textcolor{black}{ #1}}

\newcommand{\parg}[1]{\paragraph{#1}}

\title[Dawes Review: The Tidal Downsizing Theory of planet formation]{Dawes Review. The tidal downsizing hypothesis of planet formation.}
\author[S. Nayakshin]{
	Sergei Nayakshin
	\thanks{E-mail: sn85@le.ac.uk}
	\\
\affil{Department of Physics and Astronomy, University of Leicester, University Road, Leicester, LE1 7RH, UK}%
}%
\jid{PASA}
\doi{10.1017/pas.\the\year.xxx}
\jyear{\the\year}

\begin{document}%


\begin{abstract} 
Tidal Downsizing is the modern version of the Kuiper (1951) scenario of  planet formation. Detailed simulations of self-gravitating discs, gas fragments, dust grain dynamics, and planet evolutionary calculations are summarised here and used to build a predictive planet formation model and population synthesis. A new interpretation of exoplanetary and debris disc data,  the Solar System's origins, and the links between planets and brown dwarfs is offered. This interpretation is contrasted with the current observations and the predictions of the Core Accretion theory. Observations that can distinguish the two scenarios are pointed out. In particular, Tidal Downsizing predicts that presence of debris discs, sub-Neptune mass planets, planets more massive than $\sim 5$~Jupiter masses and brown dwarfs should not correlate strongly with the metallicity of the host. For gas giants of $\sim$ Saturn to a few Jupiter mass, a strong host star metallicity correlation is predicted only at separation less than a few AU from the host. Composition of massive cores is predicted to be dominated by rock rather than ices. \SNc{Debris discs made by Tidal Downsizing are distinct from those made by Core Accretion at birth: they have an innermost edge always larger than about 1 au, have smaller total masses and are usually in a dynamically excited state. It is argued that} planet formation in surprisingly young or very dynamic systems such as HL Tau and Kepler-444 \SNc{may be} a signature of Tidal Downsizing. 
Open questions and potential weaknesses of the hypothesis are pointed out.
 \end{abstract}
\maketitle%

The Dawes Reviews are substantial reviews of topical areas in astronomy, published by authors of international standing at the invitation of the PASA Editorial Board. The reviews recognise William Dawes (1762-1836), second lieutenant in the Royal Marines and the astronomer on the First Fleet. Dawes was not only an accomplished astronomer, but spoke five languages, had a keen interest in botany, mineralogy, engineering, cartography and music, compiled the first Aboriginal-English dictionary, and was an outspoken opponent of slavery.

\section{Introduction}

A planet is a celestial body moving in an elliptic orbit around a star. Although there does not appear to be a sharp boundary in terms of properties, objects more massive than $\approx 13\mj$ are called brown dwarfs (BDs) since they can fuse deuterium while planets are never sufficiently hot for that \citep{BurrowsEtal01}.  

Formation of a star begins when a large cloud dominated by molecular hydrogen collapses due to its self-gravity. The first hydrostatic object that forms in the centre of the collapsing cloud is a gaseous sphere of 1 to a few Jupiter masses; it grows rapidly by accretion of more gas from the cloud \citep{Larson69}. Due to an excess angular momentum, material accreting onto the protostar forms a disc of gas and dust. Planets form out of this (protoplanetary) disc, explaining  the flat architecture of both the Solar System and the extra-solar planetary systems \citep{FabryckyEtal14,WF14}.

The most widely accepted theory of planet formation is the Core Accretion (CA) scenario, pioneered by \cite{Safronov72}. In this scenario, microscopic grains in the protoplanetary disc combine to yield asteroid-sized bodies \citep[e.g.,][]{GoldreichWard73}, which then coalesce to form rocky and/or icy planetary cores \citep{Wetherill90,KL99}. These solid cores accrete gas from the disc when they become sufficiently massive \citep{Mizuno80,Stevenson82,IkomaEtal00,Rafikov06}, becoming gas giant planets \citep{PollackEtal96,AlibertEtal05,MordasiniEtal14}.

\cite{Kuiper51b} envisaged that a planet's life begins as that of stars, by gravitational instability, with formation of a few Jupiter mass gas clump in a massive protoplanetary disc. In difference to stars, young planets do not accrete more gas in this picture. They may actually loose most of their primordial gas if tidal forces from the host stars are stronger than self-gravity of the clumps. However, before the clumps are destroyed,  solid planetary cores are formed inside them when grains grow and sediment to the centre \citep{McCreaWilliams65}. In this scenario, the inner four planets in the Solar System are the remnant cores of such massive gas condesations. Jupiter, on the other hand, is an example of a gas clump that was not destroyed by the stellar tides because it was sufficiently far from the Sun. The other three giants in the Solar System are partially disrupted due to a strong negative feedback from their massive cores \citep[][and \S \ref{sec:SS_basic}]{HW75}.

It was later realised that gas clumps dense and yet cool enough for dust grain growth and sedimentation could not actually exist at the location of the Earth for more than a year, so Kuiper's suggestion lost popularity \citep{DW75}. However, recent simulations show that gas fragments migrate inward rapidly from their birth place at $\sim 100$~AU, potentially all the way into the star \citep[][more references in \S \ref{sec:rapid}]{BoleyEtal10}. Simulations also show that grain sedimentation and core formation  can occur inside the clumps while they are at separations of tens of AU, where the stellar tides are weaker. The clumps may eventually migrate to a few AU and could then be tidally disrupted. Kuiper's top-down scenario of planet formation is therefore made plausible by planet migration; it was recently re-invented \citep{BoleyEtal10} and re-branded "Tidal Downsizing" hypothesis \citep{Nayakshin10c}.

\SNc{This review presents the main ideas behind the Tidal Downsizing scenario, recent theoretical progress, detailed numerical simulations and a wide comparison to the current observational data. An attempt is made at finding a physically self-consistent set of assumptions within which Tidal Downsizing hypothesis could account for  all observational facts relevant to the process of planet formation.}

\SNc{Exploration of this extreme scenario is the quickest route to rejecting the Tidal Downsizing hypothesis or constraining its inner workings if it is successful. 
Further, it is possible that the final planet formation theory will combine elements of both Tidal Downsizing and Core Accretion, e.g., by having them operating at different epochs, scales, or systems. By pushing the Tidal Downsizing scenario to the limit we may locate the potential phase space divide between the two theories sooner.}

This review is structured as following. \S \ref{sec:TD_scenario} lists important physical processes underpinning the scenario and points out how they could combine to account for the Solar System's structure. \S\S 4-7 present detailed calculations that constrain these processes, whereas \S \ref{sec:dp_code} overviews a population synthesis approach for making statistical model predictions. \S\S \ref{sec:Z}-\ref{sec:kepler444} are devoted to the comparison of Tidal Downsizing's predictions with those of Core Accretion and the current observations. \S \ref{sec:SS} is a brief summary of the same for the Solar System. The Discussion (\S \ref{sec:discussion}) presents a summary of how Tidal Downsizing might relate to the exoplanetary data, observations that could distinguish between the Tidal Downsizing and the Core Accretion scenarios, open questions, and potential weaknesses of Tidal Downsizing.

\section{Observational characteristics of planetary systems}\label{sec:key_obs}

In terms of numbers, $\sim 90$\% of planets are those less massive than $\sim 20 \mearth$ \citep{MayorEtal11,HowardEtal12}. These smaller planets tend to be dominated by massive solid cores with gas envelopes accounting for a small fraction of their mass budget only, from tiny (like on Earth) to $\sim 10$\%.  There is a very sharp rollover in the planet mass function above the mass of $\sim 20\mearth$. On the other end of the mass scale, there are gas giant planets that are usually more massive than $\sim 100 \mearth$ and consist mainly of a H/He gas mixture enveloping a solid core. In terms of environment, planets should be able to form as close as $\lesssim 0.05$~AU from the host star \citep{MQ95} to as far away as tens and perhaps even hundreds of AU \citep{MaroisEtal08,BroganEtal15}.

Both small and large planets are not just smaller pieces of their host stars: their bulk compositions  are over-abundant in metals compared to their host stars \citep{Guillot05,MillerFortney11}. Planet formation process should also provide a route to forming smaller $\sim 1 - 1000$~km sized solid bodies, called planetesimals, such as those in the asteroid and the Kuiper belt in the Solar System and the debris discs around nearby stars \citep{Wyatt08}.

While gas giant planet detection frequency is a strongly increasing function of the host star's metallicity \citep{FischerValenti05}, the yield of observed smaller members of the planetary system -- massive solid cores \citep{BuchhaveEtal12,WangFischer14} and debris discs \citep{Moro-MartinEtal15} -- do not correlate with metallicity. 

One of the observational surprises of the last decade has been the robustness of the planet formation process. Planets must form in under 3 \citep{HaischEtal01} and perhaps even 1 Myr \citep[][ and \S \ref{sec:HLT}]{BroganEtal15}, and also in very dynamic environments around eccentric stellar binaries \citep[e.g.,][]{WelshEtal12} and also orbiting the primary in eccentric binary systems such as Kepler-444 \citep[][\S \ref{sec:kepler444}]{DupuyEtal16}. 

It was argued in the past that formation pathways of brown dwarfs (BDs) and of more massive stellar companions to stars should be distinct from those of planets \citep[e.g.,][]{WF14} because of their different metallicity correlations and other properties. However, observations now show a continuos transition from gas giant planets to brown dwarfs on small orbits in terms of their metal content, host star metallicity correlations, and the frequency of appearance (see \S \ref{sec:BD_vs_planets}). Also, observations show that planets and stellar companions are often members of same systems. There are stellar multiple systems whose orbital structure is very much like that of planetary systems \citep[e.g.,][]{TokovininEtal15}.  This suggests that we need a theory that can address formation of both planetary and stellar mass companions in one framework \citep[as believed by][]{Kuiper51b}.

%
%
%
%
%
%
%
%
%
%
%
%
%
%
%
%
%
%
%
%

\section{Tidal Downsizing hypothesis}\label{sec:TD_scenario}

\subsection{Basic steps}\label{sec:TD_basic}

\begin{figure}
\includegraphics[width=0.9\columnwidth]{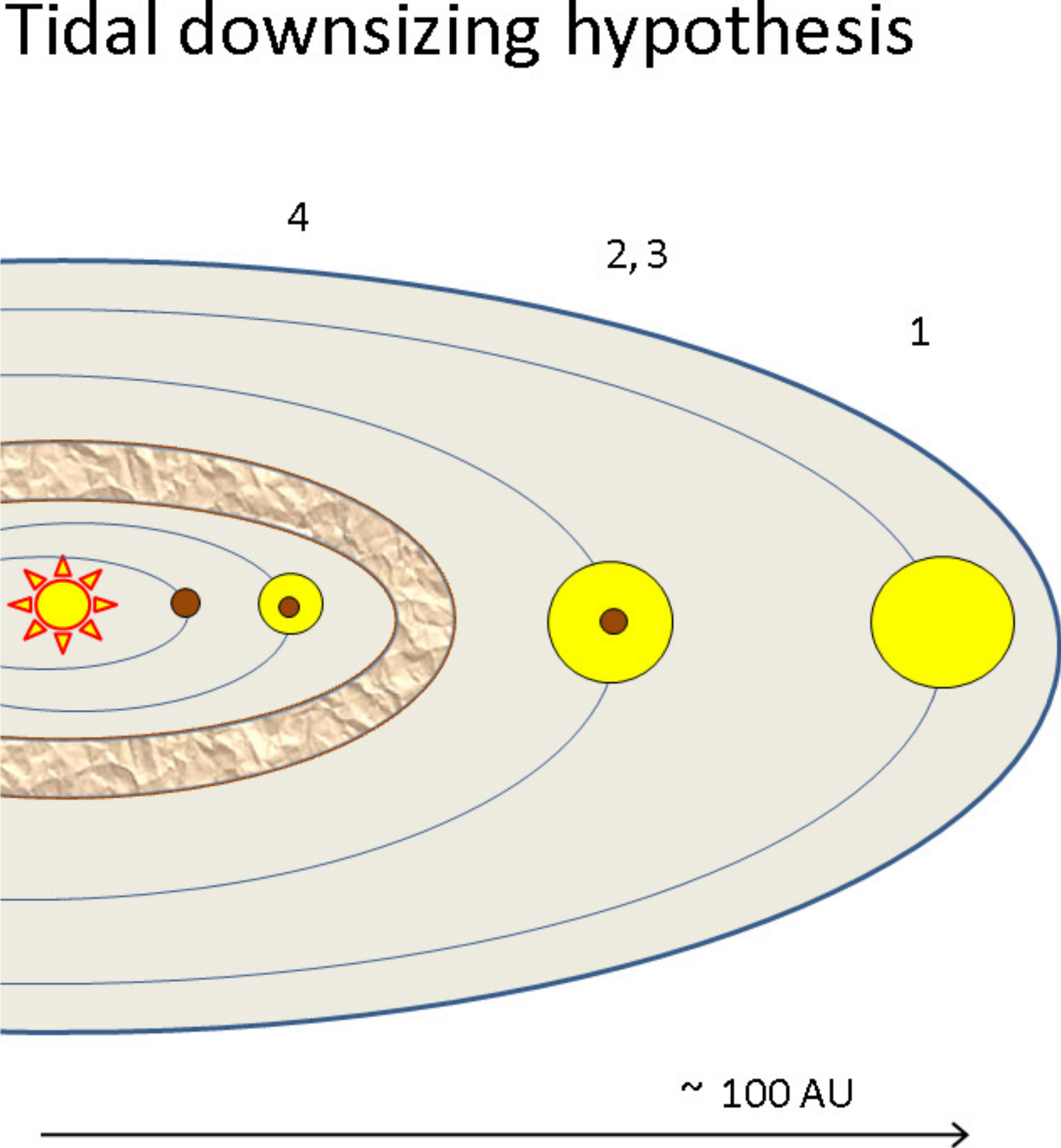}
\caption{Tidal Downsizing hypothesis is a sequence of four steps: (1) gas clump birth; (2) migration; (3) grain sedimentation and core formation; (4) disruption. 
Not all of these steps may occur for a given clump (see \S \ref{sec:TD_basic} for detail).}
\label{fig:sketch}
\end{figure}

Tidal Downsizing hypothesis is a sequence of four steps, illustrated in Fig. \ref{fig:sketch}:

(1) A gas clump of Jovian mass is born  at separation of $\sim 100$ AU from the star in a gravitationally unstable gas disc (see \S \ref{sec:disc_fragm}).

(2) The clump migrates inward rapidly due to torques from the disc, as shown by simulations (\S \ref{sec:rapid}).

(3) A core and solid debris (planetesimals) form in the centre of the clump by grain sedimentation and gravitational instability of the solid component in the centre of the clump (\S\S \ref{sec:dust_inside}, \ref{sec:planetesimals}, \ref{sec:cores}). 

(4A) If the fragment did not contract sufficiently from its initial extended state, it is disrupted by tides from the star \citep[][and \S \ref{sec:term}]{BoleyEtal10}. The core and the debris are released back into the disc, forming debris rings (shown as a brown oval filled with a patern in Fig. \ref{fig:sketch}). The core continues to migrate in, although at a slower rate. 

(4B) If the fragment contracts faster than it migrates then it is not disrupted and becomes a gas giant planet with a core. Note that the latter does not have to be massive.

The planet formation process ends when the gas disc is dissipated away \citep{AlexanderREtal14a}.

\subsection{Key concepts and physical constraints}\label{sec:tscales}

\parg{Pre-collapse gas fragments,} formed by gravitational instability in the disc (see \S \ref{sec:inside} and \S \ref{sec:term})  are initially cool, with central temperatures $T_{\rm c}\sim$ a hundred K, and extended, with the radius of the clump (planet) estimated as \citep{Nayakshin15a}
 \begin{equation}
 R_{\rm p} \approx 0.7 {G M_{\rm p}\mu\over k_b T_{\rm c}} \approx 2 \hbox{AU} \left({M_{\rm p}\over 1 \mj}\right) T_2^{-1}\;,
 \label{rp1}
 \end{equation}
 where $T_2 = T_{\rm c}/100$~K, and $\mu \approx 2.43 m_p$ is the mean molecular weight for Solar composition molecular gas.  Clump effective temperatures are typically of order of tens of K \citep[e.g.,][]{VazanHelled12}. The fragments are expected to contract rapidly and heat up initially; when reaching $T_{\rm c}\sim 1000 $~K their contraction becomes much slower \citep[e.g., Fig. 1 in][]{Nayakshin15a}.

\parg{Second collapse.} If the planet contracts to the central temperature $T_{\rm c}\sim 2,000$~K, it collapses rapidly due to H$_2$ dissociation \citep{Bodenheimer74} into the "second core" \citep{Larson69}, which has $T_{\rm c} \gtrsim 20,000$~K and a radius of only $R_{\rm p} \sim 1\rsun \approx 0.005$~AU (see \S \ref{sec:term}). 

\parg{Super-migration.} Numerical simulations (\S \ref{sec:rapid}) show that gas clumps born by gravitational instability "super-migrate" in, that is, their separation from the star may shrink from $a \sim 100$ AU to arbitrarily close to the star, unless the disc dissipates earlier. The migration time $t_{\rm mig}$ is from a few thousand years to a few $\times 10^5$ years at late times when the disc mass is low.

 \parg{Tidal disruption} of the planet takes place if its radius is larger than the Hill radius of the planet,
 \begin{equation}
 R_{\rm H} = a \left({M_{\rm p}\over 3 M_*}\right)^{1/3} \approx 0.07 a \left({M_{\rm p}\over 1 \mj}\right)^{1/3}\;,
 \label{RH1}
 \end{equation}
 where $a$ is the planet-star separation and $M_*$ was set to $1\msun$. Pre-collapse fragments can be disrupted at $a\sim $ a few to tens of AU whereas post-collapse planets are safe from tidal disruptions except perhaps for very small separations, $a\lesssim 0.1$~AU.
 
 \parg{Exclusion zone.} The smallest separation which a migrating \underline{pre-collapse} gas fragment can reach is found by comparing equations \ref{RH1} and \ref{rp1} for $T_{\rm c} = 2000$ K:
 \begin{equation}
a_{\rm exc} = 1.33 \hbox{ AU } \left({ M_{\rm p}\over 1 \mj}\right)^{2/3}\;.
\label{aex1}
\end{equation}
This implies that there should be a drop in the number of gas giant planets inwards of $a_{\rm exc}$.  Inside the exclusion zone, only the planets that managed to collapse before they were pushed to $a_{\rm exc}$ remain gas giants.

\parg{Grain sedimentation} is possible inside pre-collapse fragments \citep[see \S \ref{sec:cores}, ][]{McCreaWilliams65} as long as the fragments are cooler than $\sim 1500$~K. Grain growth and sedimentation time scales are a few thousand years  (eq. \ref{tsed1}). Massive core ($M_{\rm core} \ge 1 \mearth$) assembly may however require from $10^4$ to a few~$\times 10^5$~years.

\parg{Planetesimals} are debris of disrupted planets in the model, and are born only when and where these disruptions take place (\S \ref{sec:planetesimals} and \ref{sec:hier}). The relation between planets and planetesimals are thus inverse to what it is in the Core Accretion picture.

\parg{Pebble accretion.} 10-cm or larger grains accreting onto the planet may accelerate its collapse by increasing the planet weight (\S \ref{sec:pebbles}). This process leads to distinct testable metallicity correlation signatures.

\parg{Negative feedback by cores} more massive than a few $\mearth$. These cores release so much heat that their host gas clumps expand and may lose their gas completely, somewhat analogously to how red giant stars lose their envelopes. Core feedback can destroy gas clumps at separations as large as tens of AU (\S \ref{sec:feedback}).

 \begin{figure*}
\includegraphics[width=1.9\columnwidth]{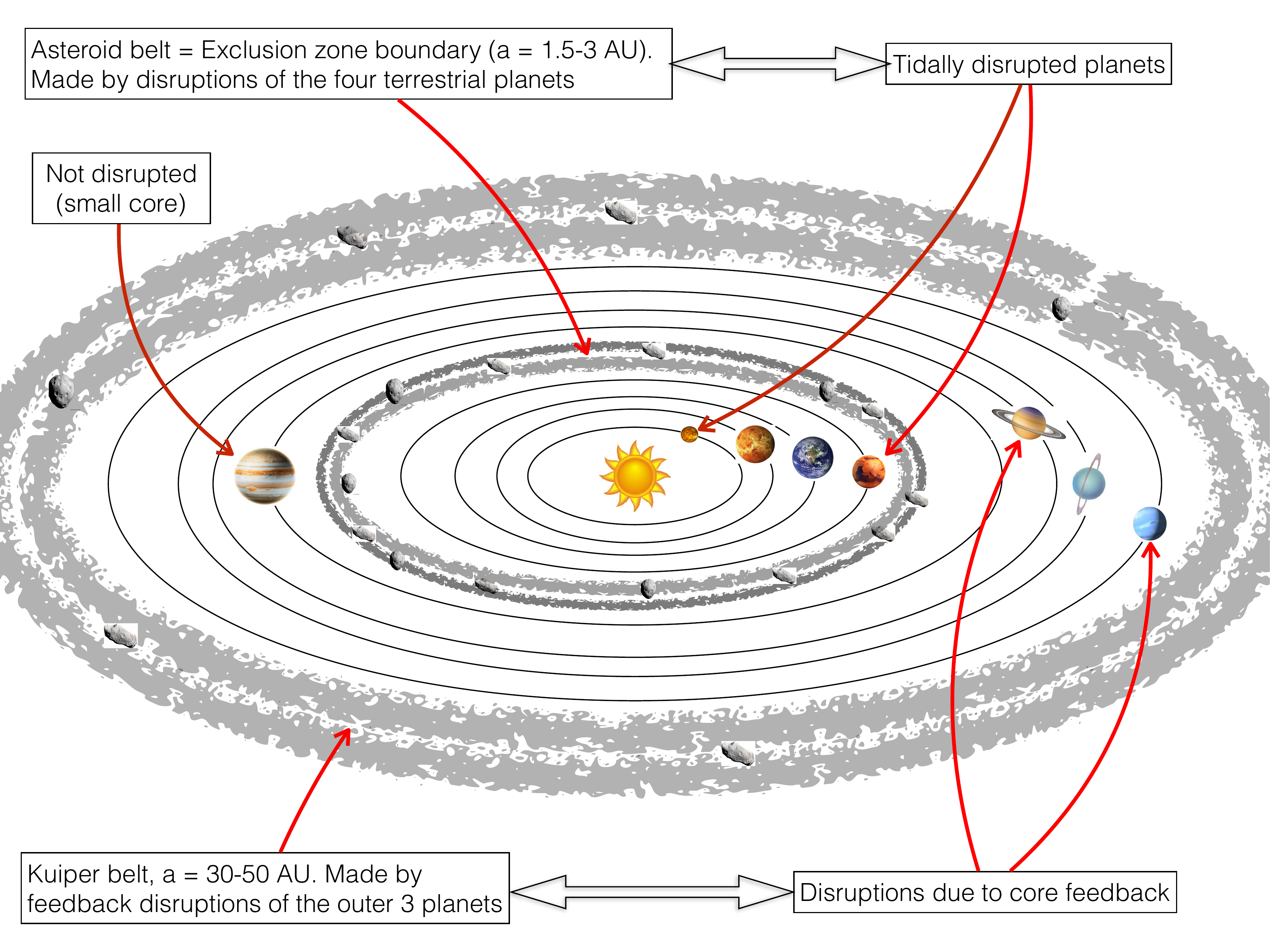}
 \caption{A qualitative model for the Solar System formation in Tidal Downsizing, described in \S \ref{sec:SS_basic}. In this scenario, the Solar System was formed by tidal disruption of the first four gas fragments (Mercury to Mars), survival of the fifth (Jupiter), and disruption of the outer three fragments due to feedback from their very bright cores (Saturn, Uranus and Neptune).}
 \label{fig:SS_sketch}
 \end{figure*}

\subsection{A zeroth order Solar System model}\label{sec:SS_basic}

Figure \ref{fig:SS_sketch} shows a schematic Tidal Downsizing model for the formation of the Solar System. In this picture, the inner four terrestrial planets are the remnants of gas fragments that migrated in most rapidly and lost their gaseous envelopes due to the tides from the Sun at separations $a\gtrsim a_{\rm exc}$, e.g., a few AU (cf. eq. \ref{aex1}), potentially explaining the origin and the location of the Asteroid belt. Since these fragments were made earlier when the disc was likely more massive, they migrated in very rapidly and had little time for core assembly. This may explain qualitatively why the terrestrial planet masses are so low compared to the much more massive cores of the four giants.

Continuing this logic, we should expect that the mass of a core in the planet increases with the distance from the Sun, in general. If the Jupiter's core mass is below $\lesssim 5 \mearth$, that is in between the terrestrial planet mass and the more distant "ice giants" \citep[such a core mass is allowed by the Jupiter's interior models, e.g.,][]{Guillot05}, then Jupiter was not strongly affected by the feedback from its core. It is therefore reasonable that Jupiter kept all or a major fraction of its primordial H/He content at its current location of 5.2 AU. Pebble accretion onto Jupiter, and/or partial H/He mass loss, made its bulk composition metal-rich compared with the Sun.

Even further from the Sun, Saturn, Uranus and Neptune are expected to have even larger cores, which is consistent with Saturn's core \citep[constrained to weigh $5-20\mj$, see][]{HelledG13} most likely being heavier than Jupiter's, and with Uranus and Neptune consisting mainly of their cores, so having $M_{\rm core}\gtrsim 10 \mearth$. At these high core masses, the three outer giants of the Solar System evolved differently from Jupiter. In this model, they would have had their envelopes puffed up to much larger sizes than Jupiter had. Saturn has then lost much more of its primordial H/He than Jupiter, with some of the gas envelope still remaining bound to its massive core. Uranus and Neptune envelopes' were almost completely lost. As with the Asteroid belt, the Kuiper belt is the record of the tidal disruptions that made Saturn, Uranus and Neptune. A more detailed interpretation of the Solar System in the Tidal Downsizing scenario is given in \S \ref{sec:SS}.

The Solar System is not very special in this scenario, being just one of thousands of possible realisations of Tidal Downsizing (see Fig. \ref{fig:sketch2}). 
The main difference between the Solar System and a typical observed exoplanetary system \citep[e.g.,][]{WF14} may be that the proto-Solar Nebula was removed relatively early on, before the planets managed to migrate much closer in to the Sun. The spectrum of Tidal Downsizing realisations depends on many variables, such as the disc metallicity, the timing of the gas disc removal, the number and the masses of the gas clumps and the planetary remnants, and the presence of more massive stellar companions. There is also a very strong stochastic component due to the clump-clump and the clump-spiral arm interactions \citep{ChaNayakshin11a}.

%

\section{Multidimensional gas disc simulations}\label{sec:3D}

\subsection{Disc fragmentation}\label{sec:disc_fragm}

To produce Jupiter at its current separation of $a\approx 5$~AU via disc fragmentation \citep{Kuiper51b}, the protoplanetary disc needs to be very massive and unrealistically hot \citep[e.g.,][]{GoldreichWard73,CassenEtal81,LaughlinBodenheimer94}. Analytical arguments and 2D simulations with a locally fixed cooling time by \cite{Gammie01}  showed that self-gravitating discs fragment only when (1) the \cite{Toomre64} $Q$-parameter is smaller than $\sim 1.5$, and (2) when the disc cooling time is $t_{\rm cool} = \beta \Omega_K^{-1} \lesssim $ a few times the local dynamical time, which is defined as $1/\Omega_K = (R^3/GM_*)^{1/2}$, where $M_*$ is the protostar's mass. The current consensus in the community is that formation of planets any closer than tens of AU via gravitational instability of its protoplanetary disc {\it in situ} is very unlikely \citep[e.g., see][]{Rafikov05,Rice05,DurisenEtal07,RogersWadsley12,HelledEtal13a,YoungClarke16}, although some authors find their discs to be fragmenting for $\beta$ as large as 30 in their simulationd \citep{MeruBate11a,MeruBate12,Paardekooper12a}.

The \cite{Toomre64} $Q$-parameter must satisfy
\begin{equation}
Q = {c_s \Omega\over \pi G\Sigma} \approx {H\over R} { M_* \over M_{\rm d}}\lesssim 1.5\;,
\label{Q1}
\end{equation}
where $c_s$ and $\Sigma$ are the disc sound speed and surface density, respectively. The second equality in eq. \ref{Q1} assumes hydrostatic balance, in which 
case $c_s/H = \Omega$ \citep{Shakura73}, where $H$ is the disc vertical height scale. The disc mass at radius $R$ was defined as $M_{\rm d}(R) = \Sigma \pi 
R^2$. Finally, $\Omega^2 \approx G M_*/R^3$, neglecting the mass of the disc compared to that of the star, $M_*$. Since $H/R\propto T_{\rm d}^{1/2}$, where $T_{\rm d}$ is the disc mid plane temperature, we see that to fragment, the disc needs to be (a) relatively cold and (b) massive. In particular, assuming $H/R \sim 0.2$ \citep{TsukamotoEtal14} at $R\sim 50-100$~AU, the disc mass at fragmentation is estimated as
\begin{equation}
{M_{\rm d}\over M_*} \approx 0.15 \left( {1.5\over Q} \right) \left({H \over 0.2 \; R}\right) \;.
\label{md1}
\end{equation}

\cite{Lin87} argued that effective $\alpha_{\rm sg}$ generated by spiral density waves should saturate at around unity when the Toomre's parameter $Q$ approaches unity from above. Simulations \citep{Gammie01,LodatoRice04,LodatoRice05} show that $\alpha_{\rm sg}$ for {\it non-fragmenting} discs does not exceed $\sim 0.1$.  This constrains the disc viscous time scale as
\begin{equation}
t_{\rm visc} = {1\over \alpha} {R^2 \over H^2} {1\over \Omega_K} \approx 4\times 10^4 \; {\rm years} \; \alpha_{0.1}^{-1} R_2^{3/2}\;,
\label{tvisc1}
\end{equation} 
where $\alpha_{0.1} = \alpha/0.1$, $R_2 = R/100$~AU and $H/R$ was set to 0.2. Thus, gravitationally unstable discs may evolve very rapidly, much faster than the disc dispersal time  \citep[$\sim 3$ Million years][]{HaischEtal01}. However, once the disc loses most of its mass via accretion onto the star,  $\alpha_{\rm sg}$ may drop well below $\sim 0.1$ and the disc then may persist for much longer in a non self-gravitating state.

\subsection{Rapid fragment migration}\label{sec:rapid}

 \cite{Kuiper51b}  {\it postulated} that Solar System planets did not migrate.  The importance of planet migration for Core Accretion theory was realised when the first hot Jupiter was discovered \citep{MQ95,Lin96}, but gravitational instability planets remained "immune" to this physics for much longer.

\cite{VB05,VB06} performed numerical simulations of molecular cloud core collapse and protostar growth. As expected from the previous fixed cooling time studies (\S \ref{sec:disc_fragm}), their discs fragmented only beyond $\sim 100$~AU. However, their fragments migrated inward towards the protostar very rapidly, on time scales of a few to ten orbits ($\sim O(10^4)$ yrs). The clumps were "accreted" by their inner boundary condition at $10$~AU. This could be relevant to the very well known "luminosity problem" of young protostars \citep{HartmannEtal98}: observed accretion rates of protostars are too small to actually make $\sim 1$ Solar mass stars within the typical disc lifetime. The missing mass is believed to be accreted onto the stars during the episodes of very high accretion rate bursts, $\dot M \gtrsim 10^{-4}\msun$~yr$^{-1}$, which are rare.  The high accretion rate protostars are called "FU Ori" sources \citep[e.g.,][]{HK96}; statistical arguments suggest that a typical protostar goes through a dozen of such episodes. Although other possibilities exist \citep{Bell94,ArmitageEtal01}, massive migrating clumps driven into the inner disc and being rapidly disrupted there yield a very natural mechanism to solve the luminosity problem \citep{DunhamVorobyov12} and the origin of the FU Ori sources \citep{NayakshinLodato12}. Future observations of FU Ori outburst sources may give the presence of close-in planets away by quasi-periodic variability in the accretion flow \citep[e.g.,][]{PowellEtal12}. Recent coronagraphic Subaru 8.2 m Telescope imaging in polarised infrared light of several brightest young stellar objects (YSO), including FU Ori, have shown evidence for large scale spiral arms on scales larger than 100 AU in all of their sources \citep{LiuHB16}. The authors suggest that such spiral arms may indeed be widespread amongst FU Ori sources. This would support association of FU Ori with migrating gas clumps.

\begin{figure}
\includegraphics[width=1.\columnwidth]{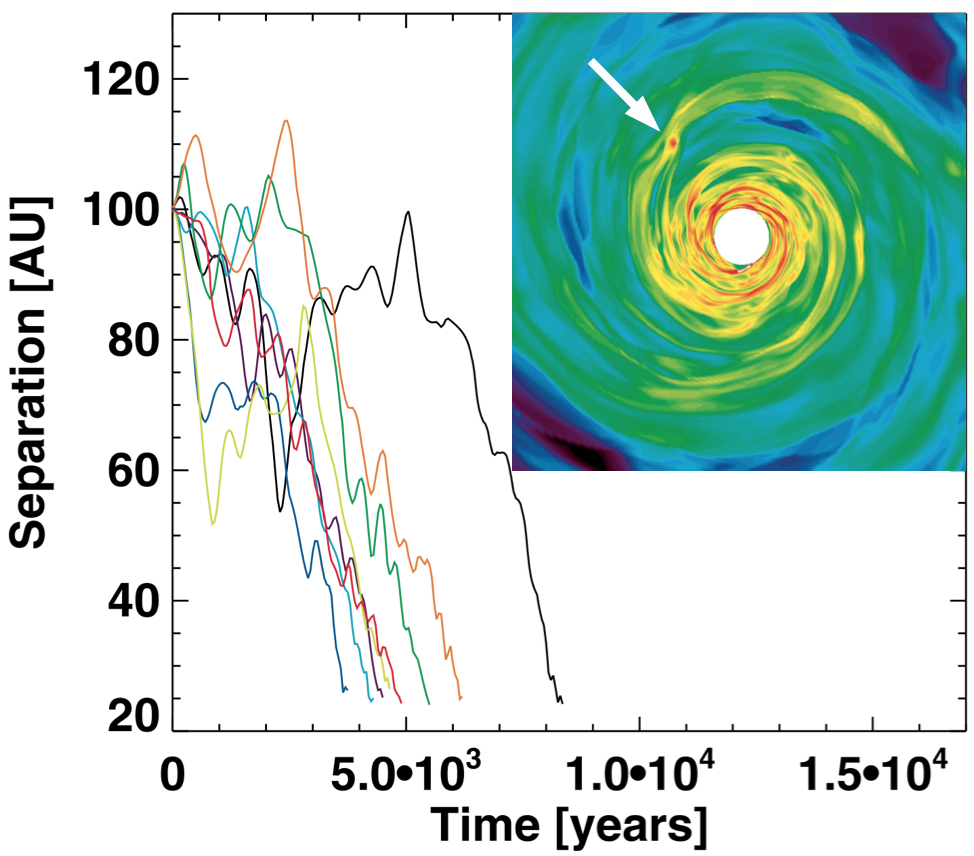}
 \caption{Numerical simulations of a Jupiter mass planet migrating in a self-gravitating protoplanetary disc \citep{BaruteauEtal11}. The planets are inserted in the disc at separation  of $100$ AU, and migrate inward in a few thousand years. Different curves are for the same initial disc model but for the planet starting at 8 different azimuthal locations. The inset shows the disc surface density map.}
 \label{fig:Bar11}
 \end{figure}

In the planet formation literature gas fragment migration was rediscovered by \cite{BoleyEtal10}, who modelled massive and large protoplanetary discs \citep[although the earliest mention of gas fragment migration may have been made by][]{MayerEtal04}. They found that gravitational instability fragments are usually tidally disrupted in the inner disc. Similar rapid migration of fragments was seen by \cite{InutsukaEtal09,MachidaEtal11,ChaNayakshin11a,ZhuEtal12a}. \cite{BaruteauEtal11} (see figure \ref{fig:Bar11}) and \cite{MichaelEtal11} found that gas giants they migrate inward so rapidly because they do not open gaps in {\em self-gravitating} discs. This is known as type I migration regime \citep[see the review by][]{BaruteauEtal14a}. For a laminar disc, the type I migration time scale, defined as $da/dt = - a/t_{\rm I}$ where $a$ is the planet separation from the star,
\begin{equation}
t_{\rm I} = (\Gamma \Omega)^{-1} Q \frac{M_*}{M_{\rm p}}\frac{H}{a} =
3\times 10^4 \;\hbox{yrs}\; a_2^{3/2} \frac{H}{0.2
  a} {Q\over \Gamma} q_{-3}^{-1} ;,
\label{tmig1}
\end{equation}
where $q_{-3} = 1000 M_{\rm p}/M_*$ is the planet to star mass ratio scaled to 0.001, $a_2 = a/100$~AU, and $\Gamma$ is a dimensionless factor that depends on 
the disc surface density profile and thermodynamical properties \citep[$\Gamma$ is the modulus of eq. 6 in][]{BaruteauEtal11}. Simulations show that $\Gamma\sim$ 
a few to ten for self-gravitating discs, typically. 

Due to the chaotic nature of gravitational torques that the planet receives from the self-gravitating disc, planet migration is not a smooth monotonic process. This can be seen from the migration tracks in Fig. \ref{fig:Bar11}, which are for the same disc with cooling parameter $\beta = 15$ and the same $M_{\rm p} = 1\mj$ planet, all placed at $a=100$~AU initially, but with varying azimuthal angles $\phi$ in the disc. The extremely rapid inward migration slows down only when deep gaps are opened in the disc, which typically occur when $q > 0.01-0.03$ at tens of au distances. This is appropriate for brown dwarf mass companions. 

\subsection{Fragment mass evolution}\label{sec:AorM}

Most authors find analytically that initial fragment mass, $M_{\rm in}$, at the very {\it minimum} is $3\mj$ 
\citep[e.g.,][]{Rafikov05,KratterEtal10,ForganRice11,ForganRice13,TsukamotoEtal14}, suggesting 
that disc fragmentation should yield objects in the brown dwarf rather than planetary mass regime \citep[e.g.,][]{SW08}. One exception is \cite{BoleyEtal10}, who found analytically $M_{\rm in} \sim 1-3\mj$. Their 3D simulations formed clumps with initial mass from $M_{\rm in} \approx 0.8\mj$ to $\sim 3\mj$. \cite{ZhuEtal12a} found initial masses larger than $10 \mj$ in their 2D fixed grid simulations, commenting that they assumed a far more strongly irradiated outer disc than \cite{BoleyEtal10}. \cite{Boss11} finds initial fragment mass from $\sim 1 \mj$ to $\sim 5\mj$.

However, $M_{\rm in}$ remains highly uncertain. In the standard accretion disc theory, the disc mid plane density is $\rho_{\rm d} = \Sigma/(2 H)$. Using eq. \ref{Q1}, the initial fragment mass can be estimated as
\begin{equation}
M_{\rm in} = {4\pi\over 3} \rho_{\rm d} H^3 \approx {1 \over 2} M_* \left( {H\over R} \right)^{3} {1.5 \over Q}\;.
\label{m_in1}
\end{equation}
For $H/R = 0.2$ and $M_* = 1\msun$, this yields $M_{\rm in} = 4\mj$, but for $H/R=0.1$ we get approximately ten times smaller value. While the mass of the disc at fragmentation depends on $H/R$ linearly, $M_{\rm in} \propto (H/R)^3$, so the fragment mass is thus much more sensitive to the properties of the disc at fragmentation.

If the clump accretes more gas from the disc then it may move into the brown dwarf or even low stellar mass regime. To become bound to the planet, gas entering the Hill sphere of the planet, $R_{\rm H}$, must lose its excess energy and do it quickly, while it is still inside the Hill sphere, or else it will simply exit the Hill sphere on the other side \citep[cf.][for a similar Core Accretion issue]{OrmelEtal15}. \cite{ZhuEtal12a} used 2D fixed grid hydrodynamical simulations to follow a massive protoplanetary disc assembly by axisymmetric gas deposition from larger scales. They find that the results depend on the mass deposition rate into the disc, $\dot M_{\rm dep}$, and may also be chaotic for any given clump. Out of 13 gas fragments formed in their simulations, most (six) migrate all the way to the inner boundary of their grid, four are tidally disrupted, and three become massive enough (brown dwarfs) to open gaps in the disc.

Even when the gas is captured inside the Hill radius it still needs to cool further. \cite{NayakshinCha13} pointed out that the accretion rates onto gas fragments in most current hydrodynamical disc simulations may be over-estimated due to neglect of planet feedback onto the disc.  It was found that fragments more massive than $\sim 6 \mj$ (for protoplanet luminosity of $0.01 L_\odot$) have atmospheres comparable in mass to that of the protoplanet. These massive atmospheres should collapse under their own weight. Thus, fragments less massive than a few $\mj$ do not accrete gas {\it rapidly} whereas fragments more massive than $\sim 10\mj$ do.
 
\begin{figure}
\includegraphics[width=0.9\columnwidth]{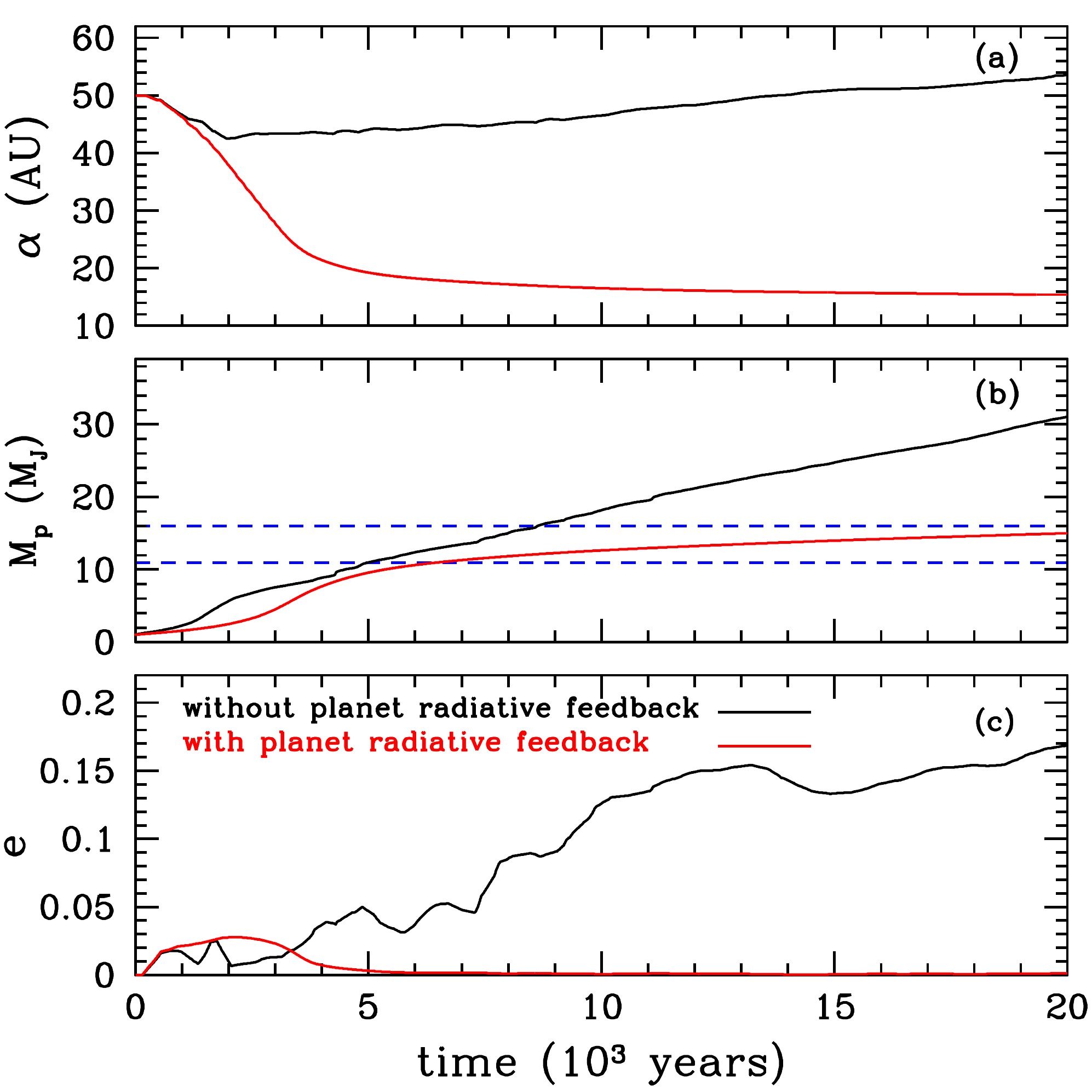}
 \caption{From Stamatellos (2015). The evolution of a fragment in two identical simulations which differ only by inclusion of radiative feedback from accretion onto the planet. Panels (a), (b), (c) show the fragment separation, mass and orbital eccentricity, respectively.}
 \label{fig:ForNot}
 \end{figure}

\cite{Stamatellos15} considered {\it accretion luminosity} feedback for planets after the second collapse.  Figure \ref{fig:ForNot} shows time evolution of the fragment separation, mass, and eccentricity for two simulations that are identical except that one of them includes the radiative pre-heating of gas around the planet (red curves),  and the other neglects it (black curves). Preheating of gas around the fragment drastically reduces the accretion rate onto it, and also encourages it to migrate inward more rapidly, similarly to what is found by \cite{NayakshinCha13}. In addition, Nayakshin (2016, in preparation), finds that gas accretion onto the jovian mass gas clumps depends strongly on dust opacity of protoplanetary disc (which depends on grain growth amongst other things); the lower the opacity, the higher the accretion rate onto the planet.
 
 \subsection{\SNc{The desert of gas giant planets at wide separations}}\label{sec:desert}
 
 Direct imaging observations show that the fraction of stars orbited by gas giant planets at separations greater than about 10 au is 1\% only \citep[see][and also \S \ref{sec:wide} for more references]{GalicherEtal16}. This is widely interpreted to imply that massive protoplanetary discs rarely fragment onto planetary mass objects. However, this is only the simplest interpretation of the data and the one that neglects at least three very important effects that remove gas giant planet mass objects from their birth-place at $a\gtrsim 50$~AU.
 
 A few Jupiter mass gas clump can (1) migrate inward on a time scale of just a few thousand years, as shown in \S \ref{sec:rapid};  (2) get tidally disrupted, that is downsized to a solid core if one was formed inside the clump \citep{BoleyEtal10}; (3) accrete gas and become a brown dwarf or even a low mass secondary star (\S \ref{sec:AorM}). 
 
In Nayakshin (2017, in preparation), it is shown that which one of these three routes the clump takes depends most strongly on the cooling rate of the gas that enters the Hill sphere of the planet. The time scale for the gas to cross the Hill sphere is about the local dynamical time, $t_{\rm cr} \sim 1/\Omega_K$, where $\Omega_K$ is the local Keplerian frequency at the planet's location. The gas gets compressed and heated as it enters the sphere. If the cooling rate is shorter than  $t_{\rm cr}$, then the gas should be able to radiate its excess energy away and get bound to the planet and eventually accreted by it. In the opposite case the gas is unable to cool; its total energy with respect to the planet is positive and thus it leaves the Hill sphere on the other side, never accreting onto the planet.

Both pre-collapse and post-collapse planets (see \S \ref{sec:term} for terminology) were investigated. Simulations are started with a gas clump placed in a massive gas disc at separation of 100 AU. A range of initial clump masses was investigated, from $M_{\rm p} = 0.5 \mj$ to $M_{\rm p} =16\mj$, in step of the factor of 2. The gas radiative cooling was done with prescription similar to the one in \cite{NayakshinCha13} but without including radiating feedback\footnote{inclusion of radiative feedback would tend to stifle accretion of gas onto planets as explained in \S \ref{sec:AorM}, favouring the planetary rather than the brown dwarf outcomes.}. To take into account modelling uncertainties in dust opacities of protoplanetary discs \citep[see, e.g.,][]{SemenovEtal03,DD05}, the interstellar dust opacity of \cite{ZhuEtal09} was multiplied by an arbitrary factor $f_{\rm op} = 0.01$, $0.1$, or 10.

\begin{figure*}
\includegraphics[width=0.99\columnwidth]{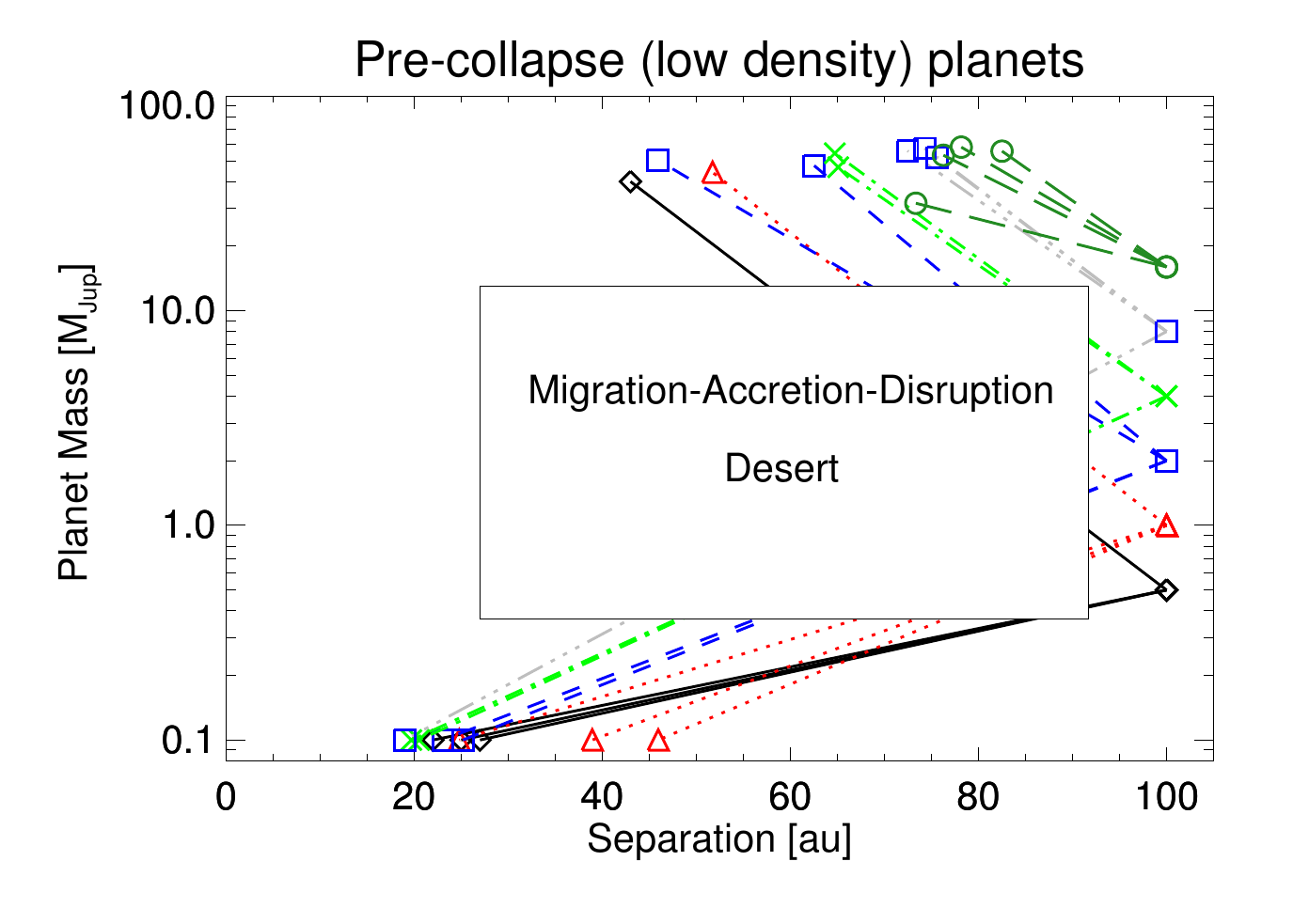}
\includegraphics[width=0.99\columnwidth]{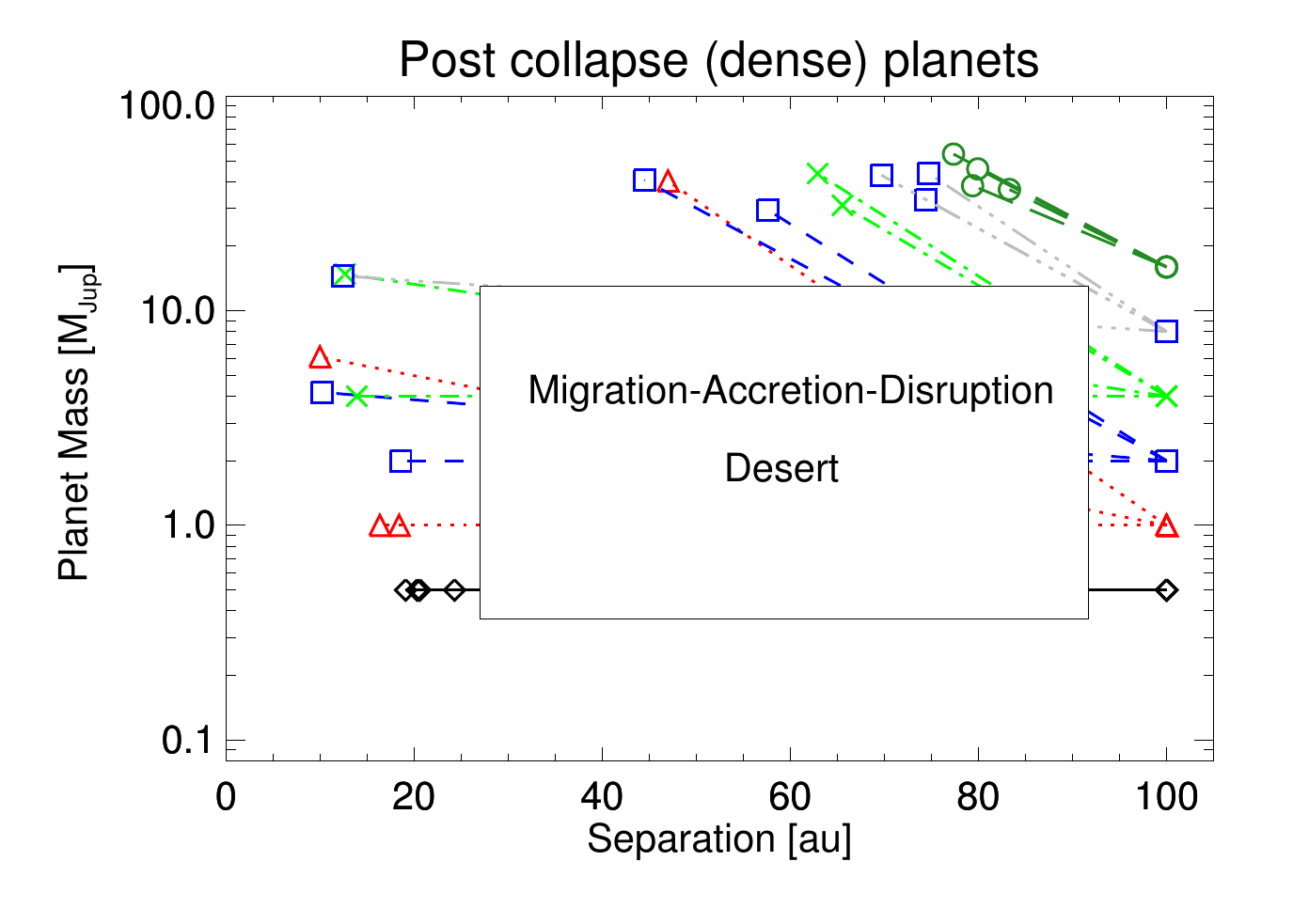}
 \caption{The initial and final positions of planets in the mass versus separation parameter space for planets embedded in massive proto-planetary discs. Note that not a single simulation ended up within the boxed region which is termed a desert. The desert is due to the clumps being taken moved out of that region by the inward migration, gas accretion or tidal disruption of pre-collapse planets. This desert may explain why directly imaged gas giant planets are so rare.}
   \label{fig:MoneyP}
 \end{figure*}

The results of these simulations are presented in Fig. \ref{fig:MoneyP}. For each simulation, only two symbols are shown: the initial planet mass versus the separation, and then the final object mass and separation. These two points are connected by straight lines although the planets of course do not evolve along those lines. For each starting point there are four lines corresponding to the simulations with the four values of $f_{\rm op}$ as detailed above. 

As expected, short cooling time simulations (small $f_{\rm op}$) lead to planets accreting gas rapidly. These objects quickly move into the massive brown dwarf regime and stall at wide separations, opening wide gaps in the parent disc. 

In the opposite, long cooling time (large values for $f_{\rm op}$) case, the planets evolve at almost constant mass, migrating inward rapidly. The final outcome then depends on how dense the planet is. If the planet is in the pre-collapse, low density, configuration, which corresponds to the left panel in Fig. \ref{fig:MoneyP}, then it is eventually tidally disrupted. It is then arbitrary assumed that the mass of the surviving remnant is $0.1\mj$ (this mass is mainly the mass of a core assembled inside the fragment, and will usually be smaller than this). Such remnants migrate slowly and may or may not remain at their wide separations depending on how long the parent disc lasts.
Post-collapse planets, on the other hand, are not tidally disrupted and can be seen on nearly horizontal tracks in the right panel of fig. \ref{fig:MoneyP}. These objects manage to open deep gaps in their parent discs because discs are less vertically extended and are not massive enough to be self-gravitating at $\lesssim 20$~AU. They migrate in in slower type II regime.

For all of the objects in the fig. \ref{fig:MoneyP}, their further evolution dependents on the mass budget of the remaining disc and the rate of its removal by, e.g., photo-evaporation. Since the objects of a few $\mj$ masses migrate most rapidly, it is likely that the objects of that mass that survived in the right panel of the figure will migrate into the inner disc.

 The most important point from the figure is this. The numerical experiments with a single clump embedded into a massive disc show that it is entirely impossible for the clump to remain in the rectangular box termed a desert in the figure. The observed $\sim 1$\% population of gas giant planets at wide separations \citep{GalicherEtal16} must have evolved in an unusual way to survive where they are observed. Either the parent disc was removed unusually rapidly, by, e.g., a vigorous photo-evaporation from an external source \citep{Clarke07} or the rapid inward migration of the planet was upset by N-body effects. The latter may be relevant to the HR 8799 system \citep{MaroisEtal10}.
 
%
%
 %
 %
 %
 %
 %

 \section{Simulations including solids}\label{sec:sim_solids}
 

\subsection{Dynamics of solids in a massive gas disc}\label{sec:gi_solids}

Dust particles in the protoplanetary disc are influenced by the aerodynamical friction with the gas \citep{Weiden77}, which concentrates solid particles in dense structures such as spiral arms \citep{RiceEtal04,RiceEtal06,ClarkeLodato09} and gas clumps.

\begin{figure}
\includegraphics[width=1.05\columnwidth,angle=0]{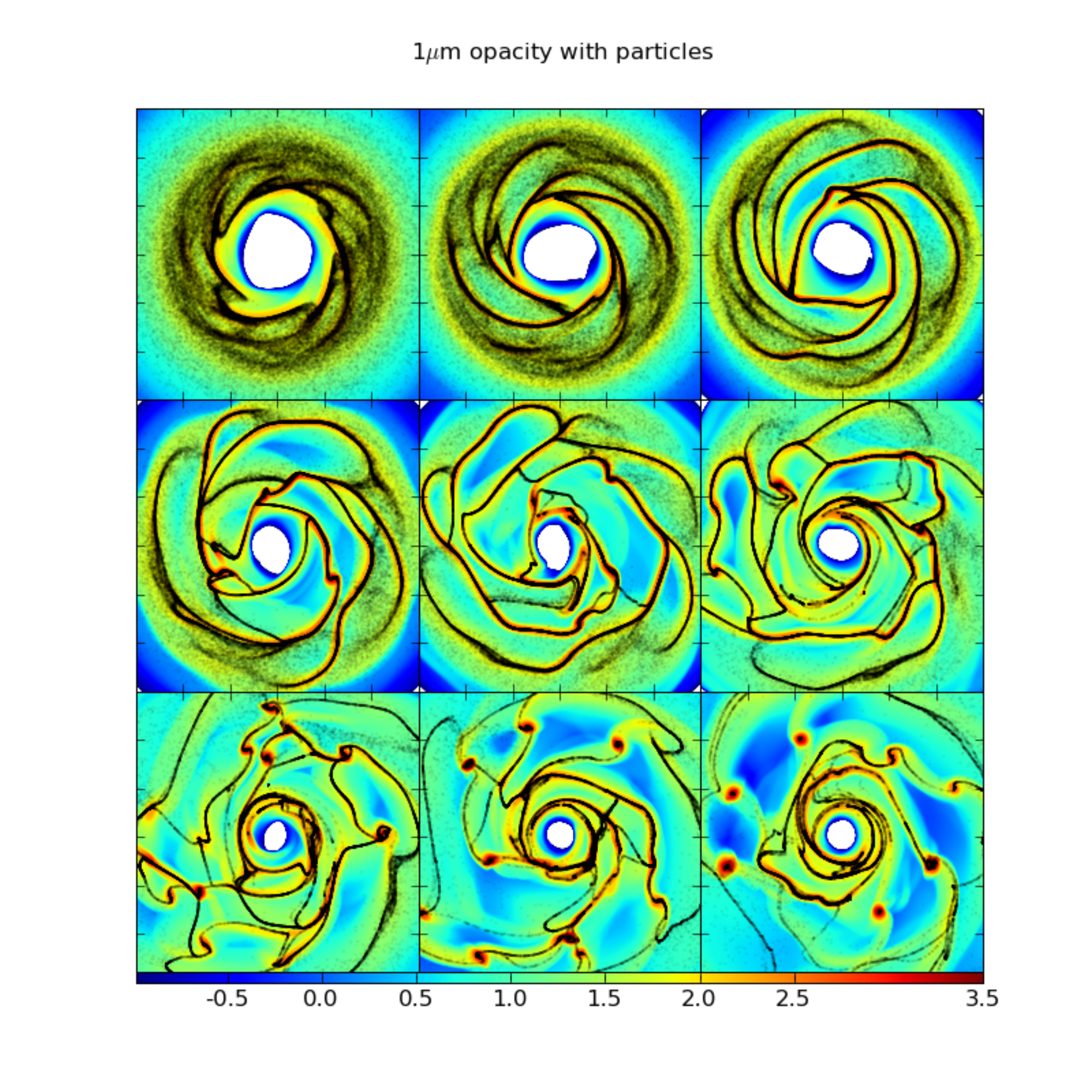}
\includegraphics[width=0.7\columnwidth,angle=-90]{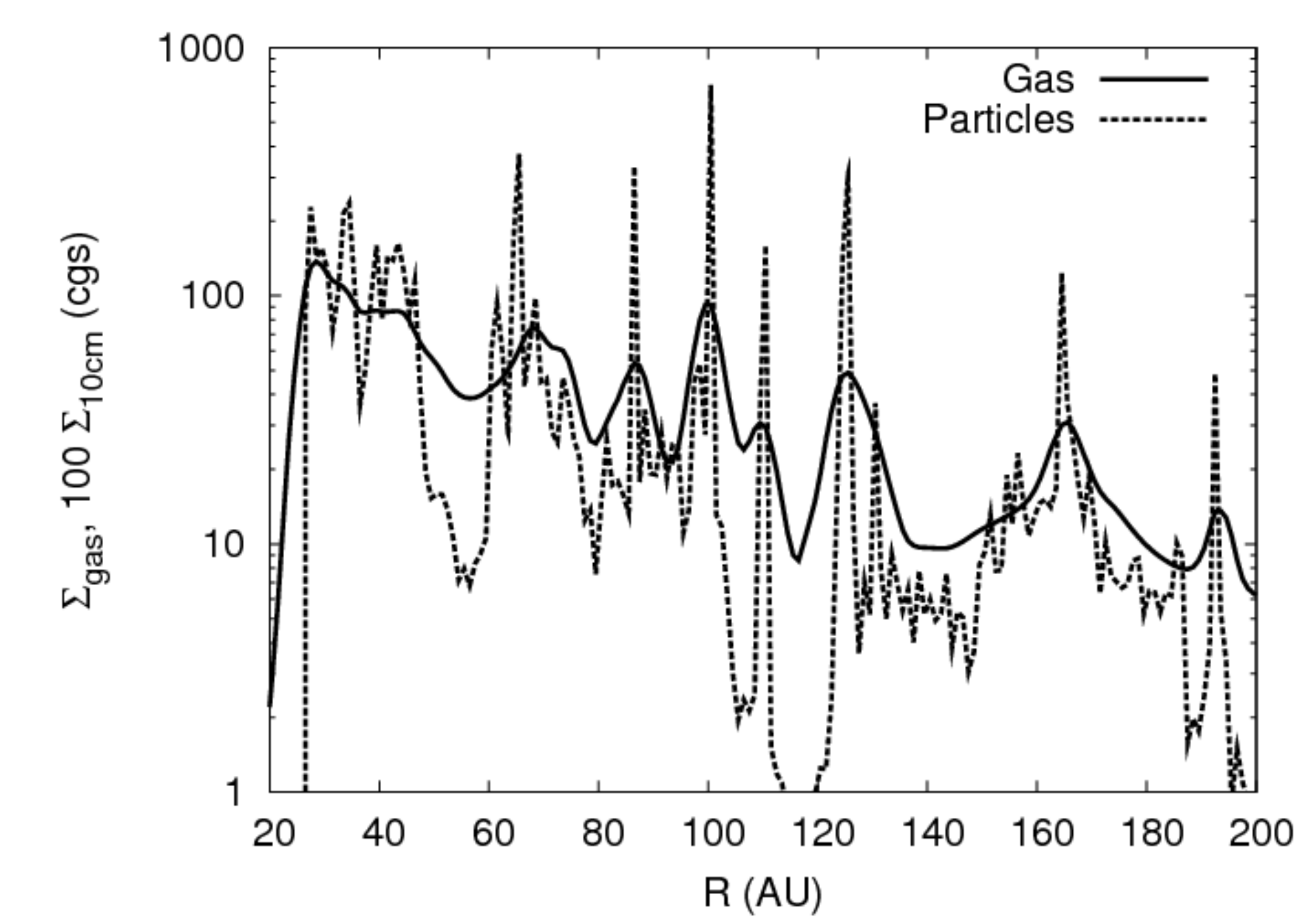}
 \caption{Simulations of Boley \& Durisen (2010). {\bf Top}: the gas disc surface density (colours) and the locations of 10 cm dust grains (black dots) in a simulation of a $0.4\msun$ disc orbiting a $1.5\msun$ star. The snapshots' time increases from left to right and from top to bottom. {\bf Bottom:} 
 Azimuthally averaged gas and dust particles surface densities versus radius in a self-gravitating disc. The peaks in the gas 
surface density correspond to the locations of gas fragments. Note that solids are strongly concentrated in the fragments and are somewhat deficient in between the fragments. }
 \label{fig:BD10}
 \end{figure}

\cite{BoleyDurisen10} performed hydrodynamics simulations of massive self-gravitating discs with embedded 10~cm radius particles. Figure \ref{fig:BD10} shows some of their results. The top panel shows a time sequence of gas disc surface density maps with the grain positions super-imposed. Spiral arms and gas clumps become over-abundant in 10 cm particles compared to the initial disc composition. This is seen in the bottom panel of the figure that presents azimuthally averaged surface densities of the gas and the solid phase. The latter is multiplied by 100. We see that solids tend to  be much stronger concentrated than gas in the peaks of the gas surface density. \cite{BoleyEtal11a} emphasised that composition of the planets formed by gravitational instability may be more metal-rich than that of the parent protoplanetary disc.

\subsection{Core formation inside the fragments}\label{sec:dust_inside}

\cite{ChaNayakshin11a} performed 3D Smoothed Particle Hydrodynamics \citep[e.g.,][]{Price12} simulations of a massive self-gravitating gas disc with dust. Dust particles were allowed to grow in size by sticking collisions with the dominant background population of small grains tightly bound to the gas. In addition, self-gravity of dust grains was included 
as well.  The disc of $0.4\msun$ in orbit around a star with mass of $0.6\msun$ became violently gravitationally unstable and hatched numerous gas fragments, most of which migrated in and were tidally disrupted. Grains in the disc did not have enough time to grow in size significantly from their initial size $a_g = 0.1$~cm during the simulations, but grains inside the gas fragments grew much faster.

\begin{figure}
\includegraphics[width=0.9\columnwidth]{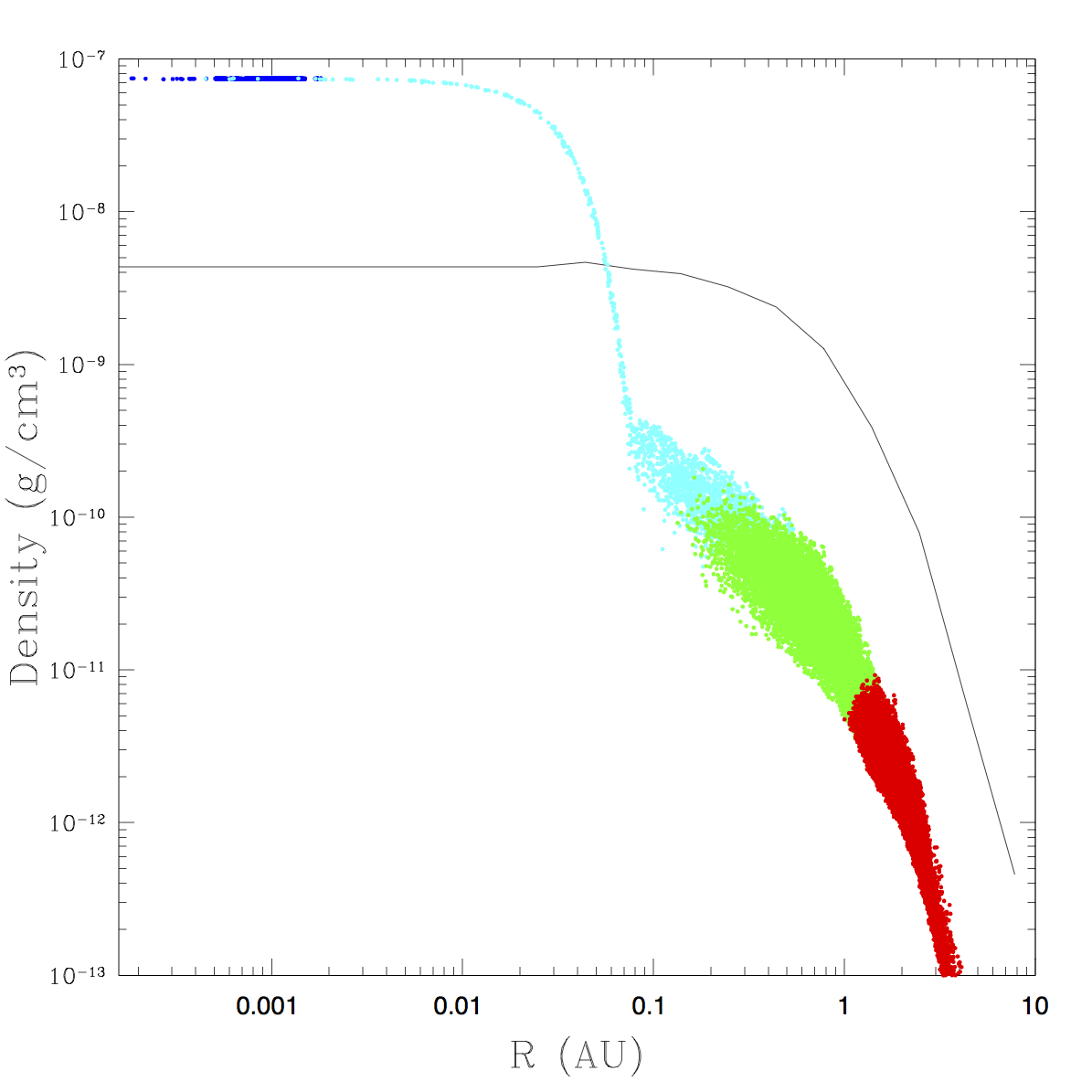}
 \caption{Gas (black) and dust grains (colour) density as a function of distance from the centre of a gas fragment \citep[from][]{ChaNayakshin11a}. The colour of grain 
particles reflects their size. The coloured points show the grain density at the positions of individual grain particles.
 The colours are: red is for $a < 1$ cm grain particles, green for $1 < a <10$ cm, cyan for
  $10< a < 100 $ cm and blue for $a > 1$ m. When the gas is tidally disrupted, the blue and the cyan grains remain self-bound in a core of mass $7.5\mearth$.}
 \label{fig:ChaN11}
 \end{figure}

One of the fragments formed in the outer disc lived sufficiently long so that its grains sedimented and got locked into a {\it self-gravitating 
bound} condensation of mass $\sim 7.5\mearth$. Figure \ref{fig:ChaN11} shows the gas density (black) and the dust density profiles (colours) within this fragment as a function of distance from its centre. There is a very clear segregation of grain particles by their size, as larger grains sink in more rapidly. The dense dust core is composed of particles with $a_g \gtrsim$ 50 cm. 

The linear extent of the dusty core is $\sim 0.05$ AU, which is the gravitational softening length of the 
dust particles for the simulation. This means that gravitational force between the dust particles is artificially reduced if their separation is less than the softening length. The gas fragment shown in Fig. \ref{fig:ChaN11} migrated in rapidly (although not monotonically) and was tidally destroyed at separation $\sim 15$~AU. The self-gravitating condensation of solids (the core) however survived this disruption and remained on a nearly circular orbit at the separation of $\sim 8$~AU. This simulation presents a proof of concept for Tidal Downsizing.

Gas fragments formed in the simulation showed a range of behaviours. More than half migrated in rapidly and were destroyed. Some fragments merged with others. Others did not merge but exchanged angular momentum with their neighbours and evolved onto more eccentric orbits, with either smaller or larger semi-major axes than their original orbits. This indicates that Tidal Downsizing may result in a number of planet and even more massive companions outcomes.

\subsection{Birth of planetesimals in the fragments}\label{sec:planetesimals}

 \cite{BoleyEtal10} concluded that fragments made by gravitational instability and that are tidally disrupted "... will have very different environments from the typical conditions in the outer disk, and they represent factories for processing dust and building {\it large 
solid bodies}. Clump disruption therefore represents a mechanism for processing dust, modifying grain growth, and building large, possibly Earth-mass, objects during 
the first stages of disk formation and evolution." 

In \cite{Nayakshin10b}, \S 7, it was argued that making large solids by grain sedimentation is much more straightforward in Tidal Downsizing than it is in Core Accretion since there is no Keplerian shear that may pump turbulence in the case of the planetesimal assembly in the protoplanetary disc \citep{Weiden80}, the grains are not lost into the star \citep[the famous 1 metre barrier,][]{Weiden77}, and the expected grain sedimentation velocities are below grain material break-up speeds. 
\cite{NayakshinCha12} argued that not only massive cores but also smaller, $\sim 1-1000$~km size bodies can be made inside the fragments. Analytical arguments supporting these ideas will be detailed in \S \ref{sec:hier}. Here we focus on the orbits of these bodies after a fragment is disrupted.

Simulations show that self-gravitating  gas fragments formed in protoplanetary discs always rotate \citep[e.g.,][]{MayerEtal04,BoleyEtal10,GalvagniEtal12}, so that not all solids are 
likely to condense into  a single central core due to the excess angular momentum in the cloud \citep{Nayakshin11a}. At gas densities characteristic of pre-collapse gas fragments,  solids larger than $\sim 1-10$~km in radius decouple from the gas aerodynamically in the sense that the timescale for in-spiral of these bodies into the core is $\gtrsim 10^5$~years, which is longer than the expected lifetime of the host fragments \citep[see Fig. 1 in][]{NayakshinCha12}. 

Neglecting aerodynamical friction for these large bodies, and assuming that they are supported against fall into the core by rotation, we may ask what happens to them once the gas envelope is disrupted. Approximating the fragment density profile as constant in the region of interest, and labelling it $\rho_0$, the mass enclosed within radius $R$ away from the centre of the core is $M_{\rm enc} = M_{\rm core} + (4\pi/3)\rho_0 R^3$. The circular speed of bodies at $R$ is $v_{\rm circ}^2 = G M_{\rm enc}/R$. Bodies circling the core at distances such that $M_{\rm enc} \gg M_{\rm core}$ will be unbound when the gas leaves, whereas bodies very near the core remain strongly bound to it. It is thus convenient to define the core influence radius,
\begin{equation}
R_{\rm i} = \left[ {3 M_{\rm core}\over 4 \pi \rho_0}\right]^{1/3}\;.
\label{ri}
\end{equation}
For central fragment density an order of magnitude larger than the mean density, eq. (10) of \cite{NayakshinCha12} shows that $R_{\rm i} \sim 0.1 R_{\rm f}$, where $R_{\rm f}$ is the fragment radius. Since the fragment is denser than the tidal density $\rho_{\rm t} = M_*/(2\pi a^3)$, where $a$ is the fragment separation from the host star, $R_{\rm i}$ is also considerably smaller than the Hill radius {\it of the core}, $R_{\rm i}/R_{\rm H, core} \approx (\rho_{\rm t}/\rho_0)^{1/3} \ll 1 $, hence the bodies inside $R_{\rm i}$ are not disrupted off the core via stellar tides.

\begin{figure}
\includegraphics[width=0.99\columnwidth]{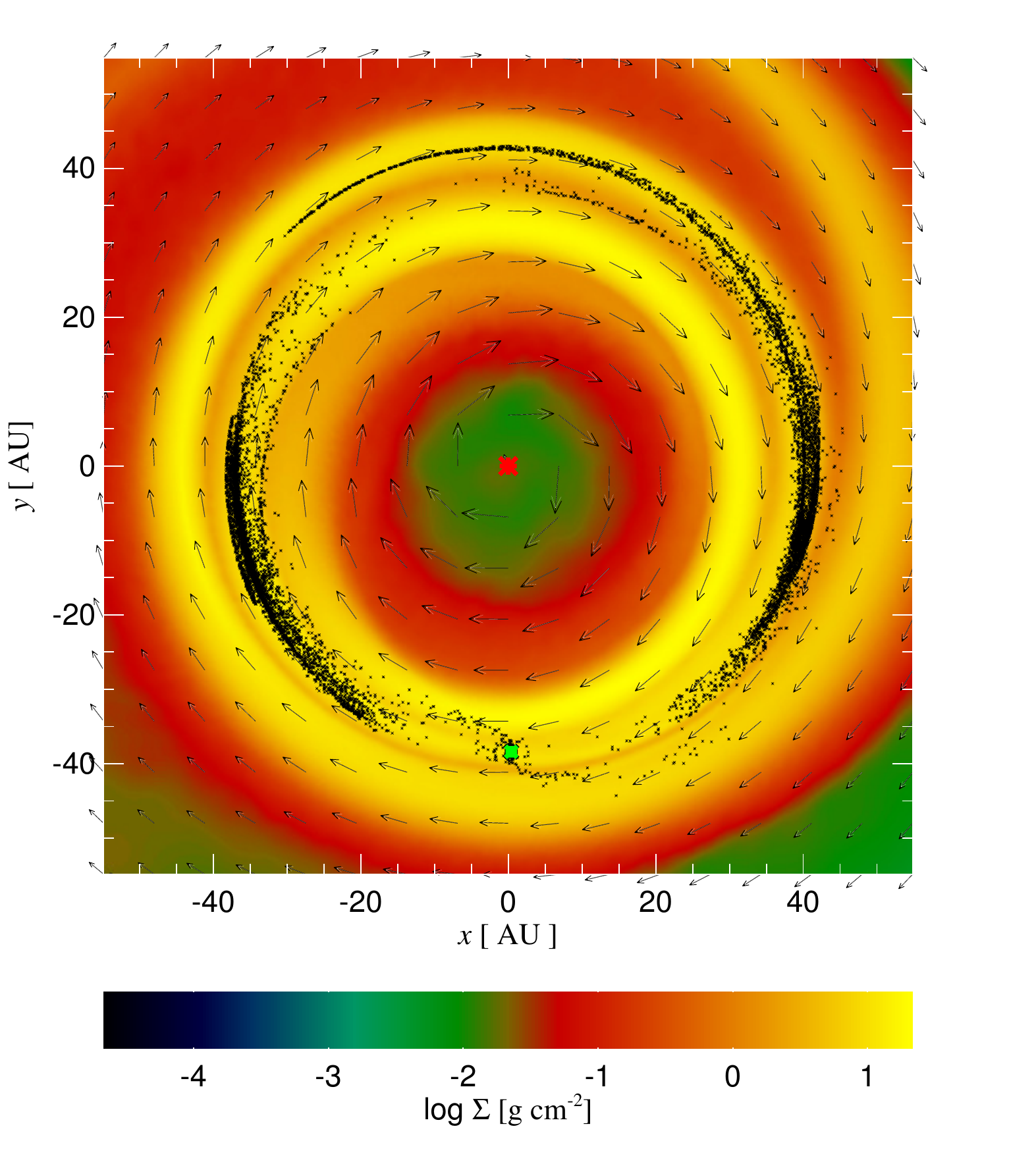}
 \caption{Gas (colour) surface density map after a tidal disruption of a gas fragment at $a\sim 40$ AU from the host star \citep[from][]{NayakshinCha12}. Black dots show positions of large solid bodies ("planetesimals") that initially orbited the central core of mass $M_{\rm core} = 10\mearth$, marked with the green asterisks at the bottom of the figure. }
 \label{fig:NCha12}
 \end{figure}

\cite{NayakshinCha12} used the 3D dust-SPH code of \cite{ChaNayakshin11a} to simulate the disruption of a gas fragment in orbit around the star. It was assumed for simplicity that planetesimals orbit the central core on circular orbits in a disc inside the gas fragment. No protoplanetary disc was included in the simulation. Figure \ref{fig:NCha12} shows the gas and the solids shortly after the fragment of mass $5\mj$ is tidally disrupted \citep[this figure was not shown in the paper but is made using the simulations data from][]{NayakshinCha12}. The core mass in the simulation is set to $10\mearth$, and its position is marked with the green cross at the bottom of the figure at $(x,y)\approx (0,-40)$. The gas (all originating from the clump) is shown by the diffuse colours. The position of the central star is shown with the red asterisk in the centre.  The black dots show the planetesimal particles. 

Solid bodies closest to the core remain bound to it even after the gas envelope is disrupted. These may contribute to formation of satellites to the massive core, as needed for Neptune and Uranus. Bodies farther out are however unbound from the core when the gas
is removed and are then sheared into debris rings with kinematic properties (e.g., mild eccentricities and inclinations) resembling the Kuiper and the Asteroid belts in the Solar System. The debris ring widens to $\Delta R\sim 20$~AU at later times in the simulation \citep[see Fig. 3 in][]{NayakshinCha12}.

This shows that if planetesimals are formed inside pre-collapse fragments, then debris rings made after their disruptions may look very much the same as the "bona fide" planetesimal discs postulated by \cite{Safronov72}, implying that we should look for observational tests that could distinguish between the two scenarios for planet debris formation (see \S \ref{sec:Z_debris}).

\subsection{Igneous materials inside fragments}\label{sec:igneous}
 

\begin{figure}
\includegraphics[width=1\columnwidth]{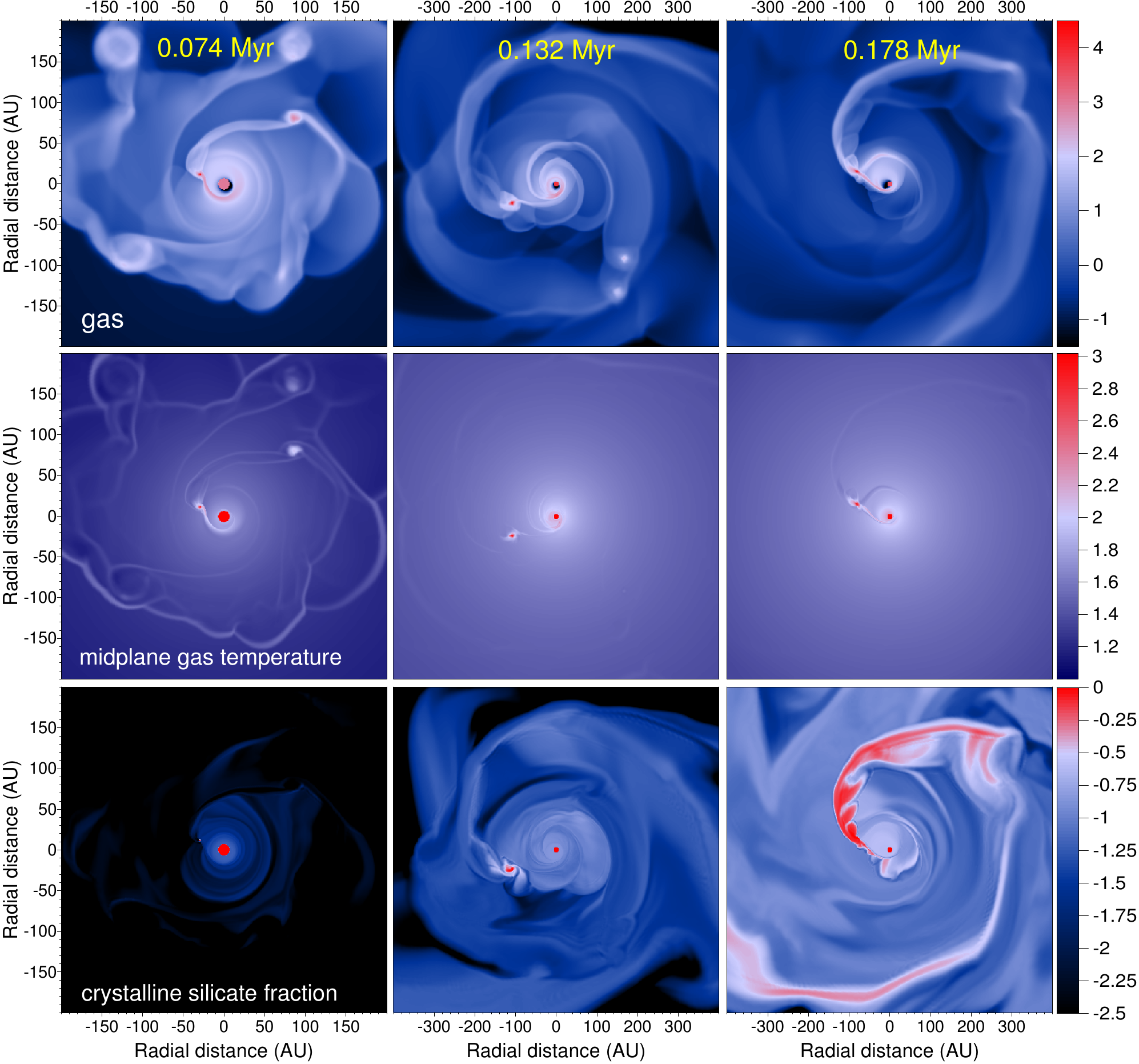}
 \caption{Snapshots from 2D simulations by Vorobyov (2011). Formation of crystalline silicates in fragments formed by gravitational collapse of a young and massive protoplanetary disc. Note the migration and disruption of the fragments along with their high gas temperatures (middle panel). This naturally creates igneous materials {\it in situ} in the disc at $\sim 100$~AU where the background disc has temperature of only $\sim 10-20$~K, and may explain why comets represent a mix of materials made at tens and $\sim 1000-2000$~K.}
 \label{fig:Vor2011}
 \end{figure}

Solar System mineralogy shows importance of high temperature $T \ge 1000 - 2000 $K processes even for very small solids called chondrules and crystalline
silicates. Chondrules are 0.1 to a few mm igneous spherules found in abundance in most unmelted stony meteorites (for example, chondrites). Roughly 85\% of meteorite falls are ordinary chondrites, which can be up to 80\% chondrules by volume. Therefore, chondrules are a major component of the solid material in the inner Solar System \citep{MorrisDesch10}. Chondrules are likely to form individually from precursors that were melted and then rapidly cooled and crystallised. The puzzle here is that high temperatures needed for formation of chondrules in the disc directly are not available beyond $a\sim 1$~AU.

A similar composition problem exists for comets. They are icy bodies a few km across that leave vaporised tails of material when they approach the inner Solar System. The composition of comets is bewilderingly diverse.  Some of the materials in cometary nuclei have not
\citep{KawakitaEtal04} experienced temperatures greater than $\sim 30 - 150$ K. Crystalline silicates, e.g. olivine, require temperatures of at least 1000 K to make \citep{WoodenEtal07}. It was thus suggested \citep[e.g.,][]{Gail01} that igneous materials were made inside 1~AU region and then were transported to tens of AU regions. However, crystalline silicates in comets may account for as much as $\sim 60$\% of weight, requiring surprising efficiency for such large scale outward transport of solids \citep{WestphalEtal09}.
 
\cite{NayakshinEtal11a}, \cite{Vorobyov11a} and \cite{BridgesEtal12a} noted that high-temperature processed materials could be made inside pre-collapse gas fragments because these are appropriately hot $500 \lesssim T_{\rm c} \le 2000$ K. Grains of less than $\sim 1$ cm in size sediment towards the centre of the fragment slowly, being impeded by convective gas motions \citep{HB11,Nayakshin14b}. When the fragment is disrupted, the grains are released back into the surrounding gas disc and will then be mixed with amorphous materials made in the main body of the disc, requiring no global outward grain transport.

Fig. \ref{fig:Vor2011} shows \cite{Vorobyov11a}'s calculations that employ a model for the formation of crystalline silicates as a function of the surrounding gas density and temperature. The top, the middle and the bottom rows of the snapshots show maps of the gas projected density, temperature and the crystalline silicates fraction, respectively, for three consecutive snapshots from the same simulation. Note that the gas temperature is high only inside the gas fragments and thus all high-T solid processing occurs inside these fragments at large distances from the star.  Repeated fragment disruption events like the one shown in the figure may be able to build up a significant reservoir of annealed igneous materials in both the outer and the inner disc.

\begin{figure*}
 \includegraphics[width=0.7\columnwidth]{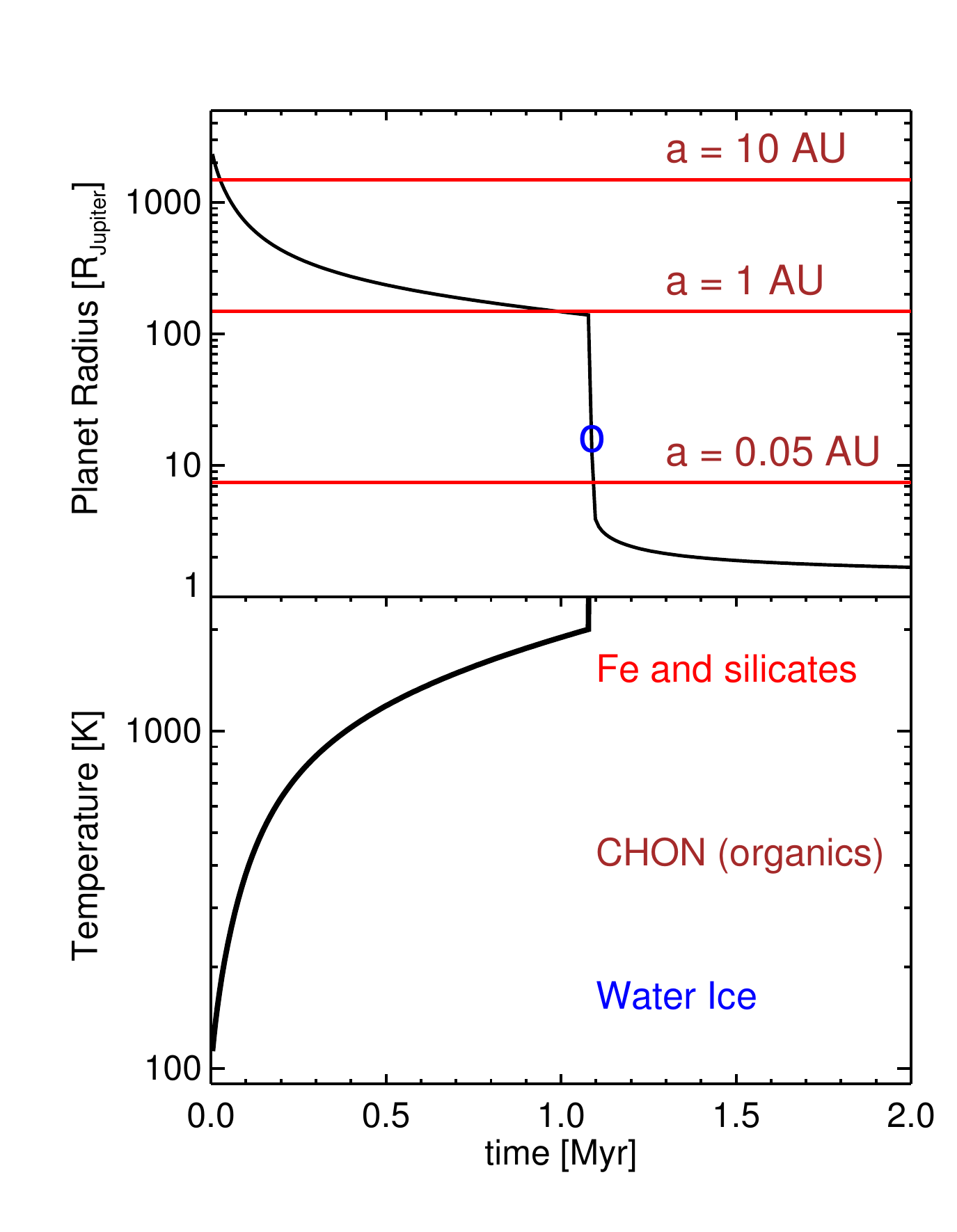}
 \includegraphics[width=0.7\columnwidth]{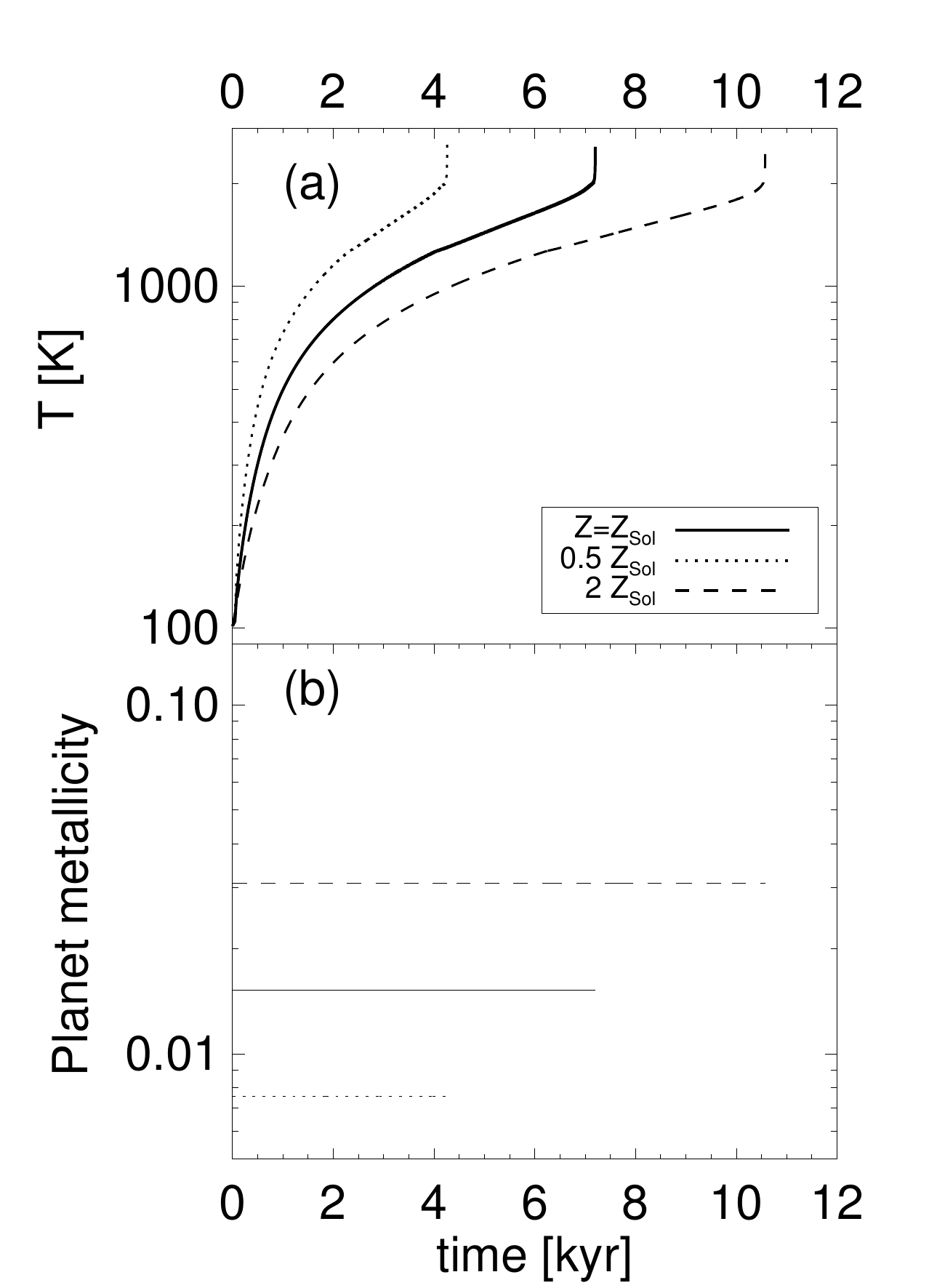}
  \includegraphics[width=0.7\columnwidth]{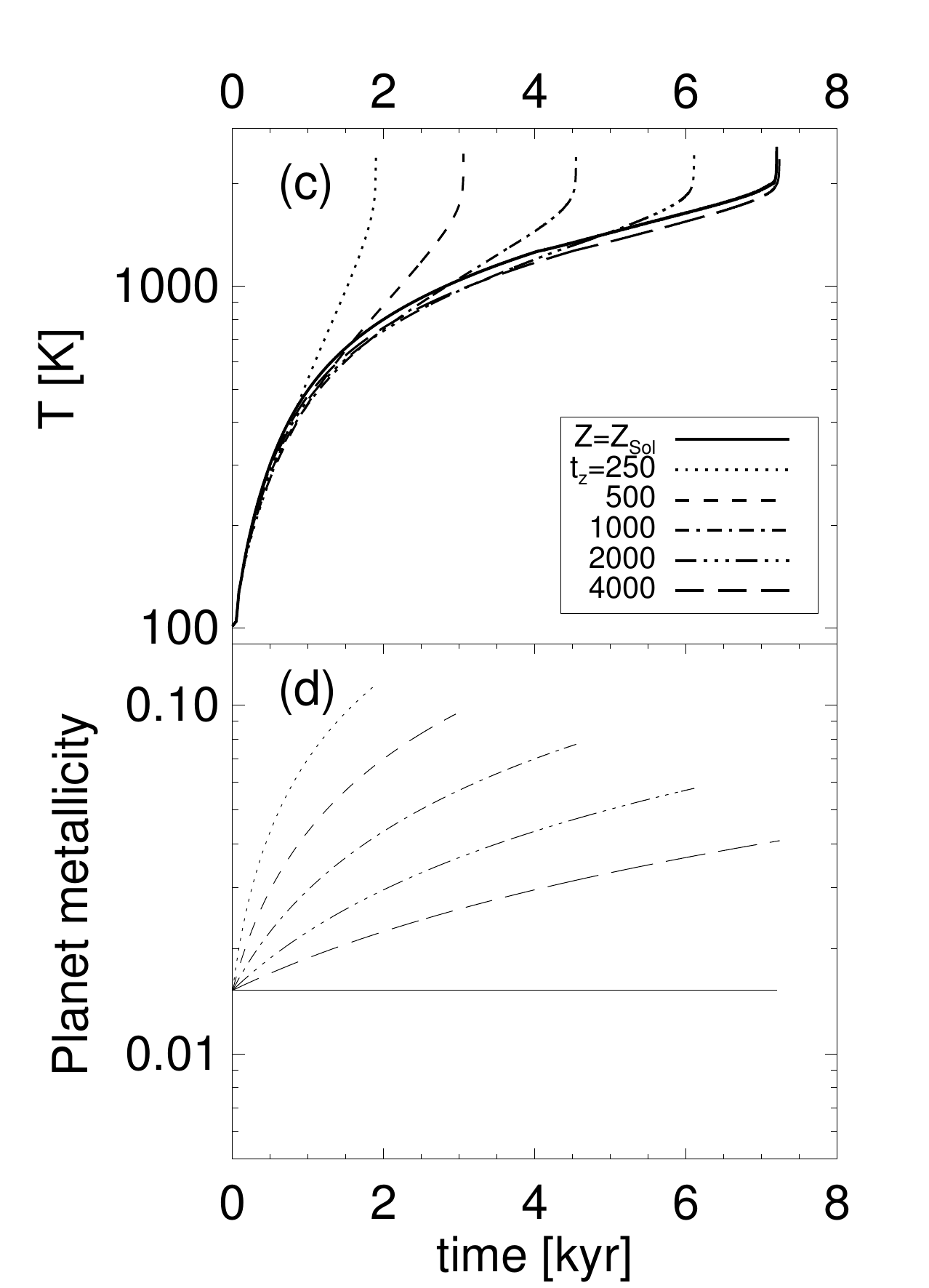}
  \caption{{\bf Left:} From \cite{Nayakshin15a}. Radiative contraction of an isolated gas fragment of mass $M_{\rm p} = 1\mj$. See \S \ref{sec:term} for detail. {\bf Middle and right:} Contraction of a gas fragment at constant or increasing metallicity, discussed in \S \ref{sec:pebbles}. Panel (a): evolution of the central temperature versus
    time for constant metallicity planets of $4 \mj$ masses; 
    panel (b) shows the (constant in time) metallicity, $z$, of the
    planets. Panels (c) and (d): same but for planets loaded by grains
    at constant rates parameterised by the metallicity doubling time
    $t_z$. Note that the faster the metals are added to the planet, the
    quicker it collapses.}
 \label{fig:exp1}
\end{figure*}

%
%
%
%
%
%
%
%

\section{Survival of fragments}\label{sec:inside}
\subsection{Terminology: pre-collapse and hot start}\label{sec:term} 

Contraction of an isolated gas clump of mass $M_{\rm p} = 1\mj$ to the present day Jupiter proceeds in two stages \citep{Bodenheimer74}. In the first, the pre-collapse stage, the fragment is dominated by molecular H, its temperature is in hundreds to 2000 K, the radius $R_{\rm p}$ is from a fraction of an AU to $\sim 10$ AU,  and its density is between $10^{-12}$ to $\sim 10^{-7}$ g~cm$^{-3}$ \citep[][]{Nayakshin10a}. This stage is analogous to the first core stage in star formation \citep{Larson69}. First cores {\it of stars} accrete gas rapidly and so contract and heat up almost adiabatically \citep{Masunaga00}, reaching the second core stage in some $\sim 10^3-10^4$ years, depending on the core gas accretion rate. For the problem at hand, however, we assume that gas accretion is not important (cf. \S \ref{sec:AorM}). 

The left panel of Fig. \ref{fig:exp1} shows radius $R_{\rm p}$ and central temperature $T_{\rm c}$ of an isolated $M_{\rm p} = 1 \mj$ planet, cooling radiatively at the interstellar dust opacity, versus time. It takes 1~Myr for the fragment to contract to temperature $T_{\rm c} = 2000$ K, at which point H$_2$ molecules dissociate. The process requires $\approx 4.5$~eV of energy per molecule to break the bonds, presenting a huge energy sink for the fragment. Robbed of its thermal energy, the fragment then collapses dynamically to much higher densities. When densities of order $\rho\sim 10^{-3}$ g~cm$^{-3}$ in the centre are reached, the collapse stops.  The post-collapse stage is called the second core in star formation; it is analogous to the "hot start" models \citep[e.g.,][]{MarleyEtal07}. The initial radius of the planet in the hot start configuration is as large as a few $R_\odot$, but  the planet is very luminous and contracts quickly to smaller radii \citep[e.g.,][]{SpiegelBurrows12}. In Fig. \ref{fig:exp1}, the beginning of the second core stage is marked by the blue open circle in the bottom left panel.

The red horizontal lines in the top left panel show the Hill radii (eq. \ref{RH1}) for several values of planet-star separation $a$, assuming $M_*=1\msun$. When  $R_{\rm p}$ approaches $R_{\rm H}$ from below, mass loss from the planet commences. \cite{NayakshinLodato12} showed that the planet mass loss can be stable or unstable depending on the planet mass-radius relationship. For a molecular hydrogen planet with polytropic index $n=5/2$,  $\zeta_p = -3$ in equation 26 in the quoted paper, and the mass transfer is unstable. Physically, the planet expands rapidly ($R_{\rm p} \propto M_{\rm p}^{-3}$ for this $n$) as it loses mass. This expansion and mass loss is a runaway process until the core starts to dominate the mass of the planet, at which point the planet radius-mass relation changes. The mass loss then slows down and eventually stops. In the coupled disc-planet models below (\S \ref{sec:dp_code}),  a simplifying assumption that mass transfer begins when $R_{\rm p}$ exceeds $R_{\rm H}$ and instantaneously unbinds the planet is made.

The top left panel of Fig. \ref{fig:exp1} shows that pre-collapse planets can be disrupted at separations from $a\sim 1$ to tens of AU from the host star. Survival of a gas fragment as a giant planet at separations of $\lesssim$ a few AU requires the fragment to undergo second collapse {\it before} it migrates into the inner disc.

\subsection{Radiative contraction}\label{sec:rad_collapse}

Given that migration times of gas fragments can  be as short as $t_{\rm mig} \sim 10^4 $ years (\S \ref{sec:rapid}), survival of any Jupiter mass gas clumps that cools radiatively, as in  Fig. \ref{fig:exp1},  in the inner few AU disc appears very unlikely. This is confirmed by \cite{ForganRice13b}, see \S \ref{sec:FR13}. Furthermore, \cite{VazanHelled12} considered a more realistic setup in which pre-collapse planets are embedded in a protoplanetary disc at selected distances from the star. They found that disc irradiation of the planet further slows down the contraction and may even reverse it, heating the planet up, puffing it up and eventually unbinding it \citep[see also][]{CameronEtal82}. This "thermal bath" effect makes the challenge of having {\it any} moderately massive gas fragments, $M_{\rm p}\lesssim $ a few $\mj$, to collapse in the inner $\sim 10$~AU via radiative cooling nearly impossible.

Finally, \cite{HB11} pointed out that, {without grain growth and sedimentation}, gas giant planets formed by gravitational instability and cooling radiatively would {\it anti-correlate} with metallicity of the parent star, [M/H], which contradicts the observed positive correlation \citep{FischerValenti05}.  Assuming 
that dust opacity is proportional to metal mass in the planet, they found that higher dust opacity pre-collapse fragments naturally take 
longer to cool radiatively. 

However, the full picture may be more complex if grain opacity is significantly reduced by grain growth, see \cite{HB11}. \SNc{For example, it is not impossible that grain opacity in high metallicity gas clumps would be actually smaller since grain growth time scales are shorter. If that were the case then gas clumps would contract and collapse more rapidly in high metallicity environments and that could give rise to a positive metallicity correlation, perhaps similar to the one observed. As pointed out in \cite{Nayakshin15c} this scenario is however disfavoured for a number of reasons.}

\subsection{Pebble accretion}\label{sec:pebbles}

As already discussed in \S \ref{sec:gi_solids}, grains that are moderately weakly coupled to gas via aerodynamical friction (a few mm to a few cm in size) are captured by a dense body or fragment embedded into the disc \citep{RiceEtal06,JohansenLacerda10,OrmelKlahr10,BoleyDurisen10}.


\cite{Nayakshin15a} studied contraction of {\it coreless} gas fragments of different metallicities, i.e., the limit when grains do not get locked into the core because the fragment is too hot or when the sedimentation process is too long. It was found that if $Z =$~const within the fragment, then fragments of higher metallicity collapse slower, confirming results of \cite{HB11}. However, if the fragment metallicity was increased gradually, by adding grains to the fragment, then the larger the pebble accretion rate, the faster the fragment was found to contract.  

The panels (a) and (c) of figure \ref{fig:exp1} show the central temperature of gas fragments of initial mass $M_{\rm p0} = 4\mj$, with an initial $T_{\rm c} = 100$~K and the dust opacity reduced by a factor of 10 from the interstellar values \citep{ZhuEtal09}. Panels (b) and (d) show metallicity evolution of the fragments.

In the figure, the constant $Z$ cases are presented in panels (a,b), whereas panels (c,d) show the cases where metals are added to the planet at a constant rate, parameterised by parameter $t_{\rm z}$: $\dot M_{\rm Z} = dM_{\rm Z}/dt = Z_\odot M_{\rm p0}/t_{\rm z}$, where $M_Z$ is the mass of metals inside the planet, and $M_{\rm p0}$ is the mass of the planet at time $t=0$. The initial metallicity for all the cases on the right is Solar, $Z = Z_\odot$. Grain growth and settling into the core is turned off, so that fragments keep uniform composition. The full problem with grain growth and settling into the core is non-linear and is considered in \S \ref{sec:feedback}.

Physically, addition of pebbles to the fragment may be likened to addition of "dark" mass to the planet. The total energy of the fragment, $E_{\rm tot}$, evolves in time according to
\begin{equation}
{d E_{\rm tot}\over dt} = - L_{\rm rad}  - L_{\rm peb} \;,
\label{etot1}
\end{equation}
where $L_{\rm rad}$ and $L_{\rm peb}$ are respectively, the radiative luminosity of the planet, and the potential energy gain due to pebble accretion, defined as a luminosity: 
\begin{equation}
L_{\rm peb} = {G M_{\rm p}\dot M_z \over R_{\rm p}}\;,
\end{equation}
This term is negative since the potential energy change of the fragment as pebbles are added is negative. For moderately
massive fragments, $M_{\rm p} \lesssim$ a few $\mj$, radiative luminosity is small, as we have seen, and so pebble accretion is the dominant {\it effective} cooling mechanism \citep{Nayakshin15a}.

In reality the fragment does not cool -- it just becomes more massive without a gain in kinetic or thermal energy, and hence must contract. Assuming the planet to be a polytropic sphere of gas with adiabatic index $n$ with an admixture of grains treated as dark mass not contributing to pressure or entropy, it is possible to obtain an analytic solution for how the central 
temperature of the sphere evolves when its metallicity is increased \citep{Nayakshin15a}:
\begin{equation}
T_{\rm c} = T_0 \left( {M_{\rm p}\over M_{\rm p0}}\right)^{6\over 3-n} = T_0
\left[{1-Z_0\over 1-Z}\right]^{6\over 3-n}\;,
\label{tc1}
\end{equation}
where $Z_0$ and $T_0$ are initial metallicity and central temperature of the planet.
In the limit $Z_0 < Z \ll 1$ it can be further
simplified.  $(1-Z_0)/(1-Z) \approx 1 + (Z-Z_0)$, and using the
identity $(1+x)^b \approx \exp(bx)$ valid for $x\ll 1$:
\begin{equation}
T_{\rm c} = T_0  \exp\left[ {6\Delta Z\over 3-n}\right]\;,
\label{tc_vs_z3}
\end{equation}
where $\Delta Z = Z - Z_0$. Clearly, if $6/(3-n) \gg 1$ then the planet heats up (contracts) very rapidly with addition of grains. In particular, for di-atomic molecules of H$_2$, $\gamma=7/5$, or $n= 5/2$, so
\begin{equation}
T_{\rm c} = T_0  \exp\left[ {12 \Delta Z}\right] = T_0  \exp\left[ {0.18 {\Delta Z\over Z_\odot}}\right] \;.
\label{tc_vs_z}
\end{equation}
This predicts that increasing the metallicity of the fragment by the factor of $\sim 6$ increases its central temperature by factor of $e$, taking the pre-collapse fragment much closer to second collapse. 


\subsection{Metallicity correlations as function of $M_{\rm p}$}\label{sec:transition}

The time it takes for an isolated gas fragment of mass $M_{\rm p}$ to reach central temperature of $T_{\rm c} \gtrsim 2000$ K and collapse via H$_2$ dissociation is (very approximately)
\begin{equation}
t_{\rm rad} \sim 1 \hbox { Myr } \left({1\mj \over M_{\rm p}}\right)^2 \left( {Z\over Z_\odot}\right)\;,
\label{trad}
\end{equation}
where the interstellar grain opacity is assumed \citep[e.g., see Fig. 1 in][]{Nayakshin15a}. This equation neglects energy release by the core, which is justifiable as long as the core is less massive than a few Earth masses (\S \ref{sec:feedback}).

The migration time in the type I regime is as short as $\sim 10^4$ years (cf. eq. \ref{tmig1}). When the planets reach the inner $\sim 10$~AU disc, where the disc is usually not self-gravitating, with Toomre's $ Q \gg 1$, more massive planets tend to open gaps and migrate in the slower type II regime. The migration time in that regime is typically $\gtrsim 10^5$ years.

Thus, radiative collapse is too slow to beat migration, and hence pebble accretion is needed to speed it up, for gas fragments of a moderate mass, $M_{\rm p}\lesssim 3\mj$. Since more pebbles are bound to be present in higher metallicity discs, the moderately massive gas giants are expected to correlate with [M/H] of the host positively. For planets more massive than $\sim 5\mj$, the radiative cooling time is comparable or shorter than the migration time. This suggests that massive gas giant planets may collapse radiatively at low [M/H] before they migrate in and are tidally disrupted.
At even higher masses, $M_{\rm p} \gtrsim10 \mj$, including the brown dwarf regime,  fragments always collapse more rapidly via radiation than they migrate in, whatever the metallicity of the host disc. 

This predicts that metallicity correlations of giant planets should undergo a fundamental change around the mass of $\sim 5\mj$.

\subsection{Second disruptions at $a\lesssim 0.1$~AU}\label{sec:2nd_dis}

Post-collapse (second core stage) planets are denser than pre-collapse planets by a few orders of magnitude, so they are much less likely to be tidally compromised. However, as seen from the left panel of Fig. \ref{fig:exp1}, there is a brief period of time when a contracting post-collapse gas giant planet may be disrupted at 
separation $a\lesssim 0.1$~AU. In \cite{Nayakshin11a}, a toy model for both the disc and the planet was used to argue that many massive cores found by the {\it Kepler} satellite in abundance at separation of $\sim 0.1$ from  their host stars could be made via such "second" disruptions. Based on the toy model, it was shown that post-collapse planets migrating early on, when the disc accretion rate is large, $\dot M \gtrsim 10^{-7} \msun$~yr$^{-1}$, may be disrupted at characteristic distance of $a\lesssim 0.1$~AU, whereas planets migrating later, when the disc accretion rate is much smaller are more likely to be sufficiently compact to avoid the disruption.

\cite{NayakshinLodato12} improved on this calculation by using a realistic 1D time dependent disc model, although still using a very simple (constant effective temperature) cooling model for the planet. A rich set of disc-planet interaction behaviour was found, which is not entirely surprising since the disc can exchange with the planet not only the angular momentum but also mass. The disc may be also switching between the cold molecular H and the hot ionised H stable branches of the disc \citep{MM81,MM84,Bell94}, resulting in large increases or decreases in the accretion rate. This may lead to the planet's migration type changing from type II to type I or vice versa. Importantly, if the planet mass loss proceeds mainly via the Lagrangian L1 point and the migration type is II, then the planet migrates outward during the intense mass loss phases. 


\begin{figure}
\includegraphics[width=0.99\columnwidth]{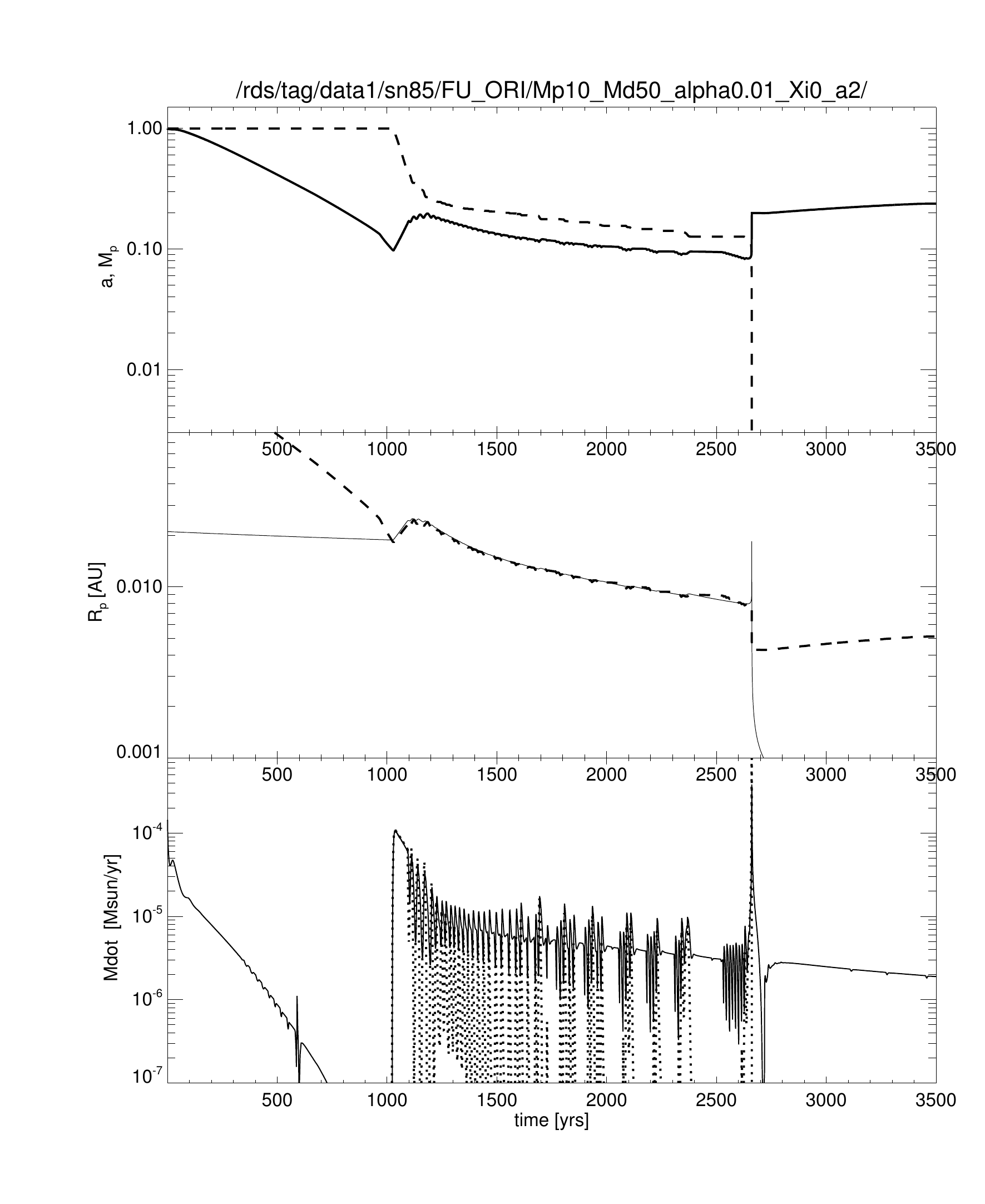}
 \caption{A coupled evolution of the disc and the migrating planet from Nayakshin \& Lodato 2012. {\bf Top panel}: planet separation from the star (solid) and planet's mass in units of $10 \mj$ (dashed). {\bf Middle:} Planet radius ($R_{\rm p}$, solid) and planet Hills radius (dashed). {\bf Bottom}: Accretion rate onto the star (solid) and the mass loss rate of the planet (dotted).}
 \label{fig:NL12}
 \end{figure}

Figure \ref{fig:NL12} shows an example calculation from \cite{NayakshinLodato12} in which a second collapse fragment of mass $M_0 = 10 \mj$ is inserted into a protoplanetary disc at $a_0 =1$~AU. Initially the planet is much smaller than its Hills radius, so the mass loss rate is zero.  The planet opens a very deep gap in the disc, cutting off mass supply to the inside disc, which empties onto the star. This creates a gas-free hole inside the planet orbit. As the planet migrates inward, both $R_{\rm p}$ and $R_{\rm H}$ shrink with time, but the planet contraction time is far longer than its migration time of $\sim 10^3$~years (this is the case of a very massive disc). Therefore, the Hill radius catches up with $R_{\rm p}$ when the planet-star separation $a\sim 0.1$~AU.

When $R_{\rm H} - R_{\rm p}$ becomes comparable with the planetary atmosphere height-scale, the planet starts to lose mass rapidly via L1 point. This fills the disc inward of the planet with material lost by the planet, and accretion onto the star resumes at a very high rate. Since the viscous time is short at such small distances from the star, accretion rate onto the star matches the mass loss rate by the planet (except for very brief periods of time). An FU Ori like outburst commences which is powered by the star devouring the material shaved off the planet. At the beginning of the outburst, a quasi equilibrium is established: the star accretes the planet material at exactly the rate at which it is lost by the planet. The mass of the planet starts to decrease rapidly (see the dashed curve in the top panel of the figure). The equilibrium is however soon destabilised as rapid transitions between the low and the  high temperature states in the disc occur {\it in the gap region of the disc}, and hence the disc switches between the two states much more rapidly than could be expected, leading to the complex quasi-periodic behaviour seen in the lower panel of Fig. \ref{fig:NL12}. Such rapid transitions may be related to the less violent and shorter duration outbursting sources known as EXORs \citep{Herbig89,SAEtal08,LorenzettiEtal09}. The long duration outbursts seen in other examples in \cite{NayakshinLodato12} may correspond to the high luminosity long duration classical FU Ori events, as suggested earlier by \cite{VB05,VB06,BoleyEtal10}.

The planet eventually loses so much mass that the gap closes; this triggers an even faster mass loss rate, producing the large spike in the accretion rate at $t\approx 2600$~years in the bottom panel of Fig. \ref{fig:NL12}. The second disruptions also leave behind solid cores assembled within the planets during pre-collapse stage. This may lead to a metallicity signature in the period distribution of small planets (see \S \ref{sec:MM_valley}).

%
%
%
%
%
%
%

\section{Cores in Tidal Downsizing scenario}\label{sec:cores}

\subsection{Grain sedimentation inside the fragments}\label{sec:grandest}

Grain sedimentation time scales can be made assuming for simplicity constant density within the gas fragment 
\citep{Boss98}. Combining the Epstein and the Stokes drag regimes, it is possible to derive \citep[eq. 41 in ][]{Nayakshin10a} the sedimentation velocity for a spherical grain of radius $a_g$ and material density $\rho_a$:
\begin{equation}
v_{\rm sed} = \frac{4 \pi G \rho_a a_g R}{3 c_s} {\lambda + a_g \over \lambda} \left(1 + f_{\rm g}\right)\;,
\label{u_epstein}
\end{equation}
where $\lambda = 1/(n \sigma_{\rm H2})$ is the mean free path for hydrogen molecules, $n$ and $\sigma_{\rm H2}\approx 10^{-15}$cm$^{2}$ are the gas density and collision cross section, $R$ is the distance from the centre of the fragment, and $c_s$ is the sound speed. The dimensionless factor $f_{\rm g}$ is the mass fraction of grains in the fragment interior to radius $R$; it is initially small, $f_{\rm g}\sim 0.01$, but may become greater than unity when grains sediment to the fragment centre.

For a reference, at $a_g = 1$~cm, $v_{\rm sed} \approx 1.2$~m/s in the Epstein's regime ($a_g \ll \lambda$) for $R=1$~AU and fragment temperature of 300 K. Note that $v_{\rm sed} \propto a_g$, so that large grains fall to the centre faster. Sweeping smaller grains in their path as they fall, larger grains are grow by accretion of the smaller ones \citep[see, e.g., ][]{DD05}. The time to reach the centre from radius $R$ is independent of $R$:
\begin{equation}
t_{\rm sed}= \frac{R}{v_{\rm sed}} 
\approx 5 \times 10^3 \;\hbox{yrs}\; \left({3\; {\rm g \; cm}^{-2}\over \rho_a a}\right) {\lambda \over \lambda + a_g} 
\label{tsed1}
\end{equation}
for $f_{\rm g} \ll 1$. We observe that this time scale is shorter than the planet migration time for grains with size $a_g \gtrsim 1$~cm. This opens up the possibility of making solid cores within the fragment prior to its tidal disruption \citep{McCreaWilliams65,DecampliCameron79,Boss98,BoleyEtal10}.
Numerical modelling shows that convection presents a significant hurdle to grain sedimentation \citep[][and \S \ref{sec:g_and_c}]{HelledEtal08,HS08}. 


\subsection{Gravitational collapse of the "grain cluster"}\label{sec:core_collapse}

The main difficulty in forming planets by a direct gravitational collapse of the solid component {\it in the protoplanetary 
disc} is the differential shear \citep{GoldreichWard73} and turbulence in the disc \citep{Weiden80}. Just a tiny fraction of the circular motion of the protoplanetary disc, $v_{\rm K} = 30$~km/s at 1~AU, transferred into gas turbulent motions is sufficient to result in the maximum mass made by the gravitational collapse being negligibly small compared to a planet mass \citep[see \S 7.3 in][]{Nayakshin10b}. 

In Tidal Downsizing making planetary mass cores by direct collapse of the grain component inside a gas fragment may be simpler. Once a significant fraction of the fragment grains sediment into the central region of the fragment, grains start to dominate the mass density there, so that $f_{\rm g} \gg 1$ in the central region \citep[see Fig. \ref{fig:ChaN11} here, and also figs. 2 or 4 in][]{Nayakshin10a}. Gas fragments found in simulations of self-gravitating discs usually rotate approximately as solid bodies, making rotational velocities in their centres rather small \citep{Nayakshin11a}; thus rotation is not likely to prevent gravitational collapse of the grain cluster (the region where $f_{\rm g}\gg 1$) entirely. In \cite{Nayakshin10a}, \S 3.6.2, evolution of a single size grain population within a constant density gas background was considered. If was shown that when the fragment grains sediment within the radius 
\begin{equation}
R_{\rm gc} \approx 0.1 R_{\rm p}  \left({f_{\rm g}\over 0.01}\right)^{1/2}\;, 
\label{Rgc1}
\end{equation}
where $f_{\rm g}$ is the {\it initial} grain mass fraction in the fragment, and $R_{\rm p}$ is the planet radius, gas pressure gradient is no longer able to counteract the collapse. The grain cluster may then collapse into a dense core.

\subsection{Hierarchical formation of smaller bodies}\label{sec:hier}

Many astrophysical systems follow the hierarchical fragmentation scenario first suggested for galaxies by \cite{Hoyle53}. In his model, as a very massive gas cloud contracts under its own weight, smaller and smaller regions of the cloud become self-gravitating. The Jeans mass in the cloud is $M_{\rm Jeans} \sim c_s^3/(G^3 \rho)^{1/2}$, where $c_s$ and $\rho$ are the gas sound speed and density, respectively. The Jeans mass is originally equal to that of the cloud (galaxy). Provided that $c_s$ remains roughly constant due to cooling, increasing $\rho$ during the collapse implies that smaller sub-parts of the cloud start to satisfy the condition $M < M_{\rm Jeans}$, where $M$ is mass of the sub-part. These regions can then collapse independently from the larger system. This process continues, eventually making star clusters, groups of stars, individual stars, and perhaps even gas giant planets on the smallest scales where the hierarchical collapse stops because gas can no longer cool effectively below the opacity fragmentation limit  \citep{Rees76}.

Is there a similar hierarchy of collapse scales for the grains sedimenting down inside the gas fragments? 

Consider an off-centre spherical region with radius $\Delta R$ and gas density $\rho$ somewhat higher than the background density. Grains inside the region will sediment towards the centre of that region on a
time scale $\Delta t$ {\it independent} of $\Delta R$:
\begin{equation}
\Delta t \approx \frac{3 c_s \mu }{4 \pi G \rho_a a_g^2 \sigma_{\rm H2}} \; {1 \over \rho (1 + f_{\rm g})}\;,
\label{tsed3}
\end{equation}
where $f_{\rm g} > 1 $ is the local grain concentration and it is assumed that $\lambda  \ll a_g$.
From this we see that if the total density in the perturbed region, $\rho(1+f_{\rm g})$, is greater than that of the surroundings, it will collapse more rapidly than the whole grain cluster considered in \S \ref{sec:core_collapse}. The collapse accelerates with time: $\Delta t$ is inversely proportional to density and the density increases as the perturbation collapses. Thus the grains in this region are able to collapse into an {\it independent} solid body before the whole grain cluster collapses.

This argument suggests that perturbations of {\it all} sizes can collapse. A very small $\Delta R$ region collapses slowly since the collapse velocity, proportional to $\Delta R$, is quite small. However the collapse time is as short as that for a much more extended perturbation. Taken at face value, this would imply that even tiny solid bodies, with final post-collapse radius $a_{\rm fin}$ as small as $\lesssim 1 $~m could form via this process. However, in practice there is another limit to consider. A small body born by collapse of a small perturbation is very likely to be inside of a larger perturbation (which in itself may be a part of a yet bigger one). Therefore,  the small body will be incorporated into a larger collapsing system unless the body can decouple dynamically from the larger system. 

Consider now a post-collapse body of radius $a_b$ , and material density $\rho_b \sim 1$ g~cm$^{-3}$. Since the body is inside the region where $f_{\rm g} > 1$, we can neglect aerodynamical friction with the gas and consider only interaction of the body with grains in the region. The body may be able to decouple from the bulk of the grains collapsing into the core if the stopping distance of the body is larger than $R_{\rm gc}$. This requires that the column depth of the body
\begin{equation}
\Sigma_b = \rho_b a_b > \rho_{\rm gc} R_{\rm gc} = \rho_0 R_{\rm f} \approx {M_{\rm f}\over \pi R_{\rm f}^2}
\label{min_ast1}
\end{equation}
is larger than the column depth of the parent gas fragment. Introducing a mean temperature of the fragment as $T_{\rm p} \approx GM_{\rm p}\mu/(3 k_B R_{\rm p})$, we obtain the minimum size of an object that can separate itself out of the core:
\begin{equation}
a_{\rm min} = 3.7 \hbox{ km } T_3^2  {1\mj \over M_{\rm p}} \rho_b^{-1}\;.
\label{min_ast2}
\end{equation}
This is in the asteroid size range. Finally, we should demand that the body is able to resist gas drag for a long enough time after the core is formed (when the grains in the collapsing grain cluster are mainly incorporated into the core). This problem has been examined in \cite{NayakshinCha12}, also leading to a minimum size in the range of $1-10$~km.

\begin{figure*}
\includegraphics[width=0.47\textwidth]{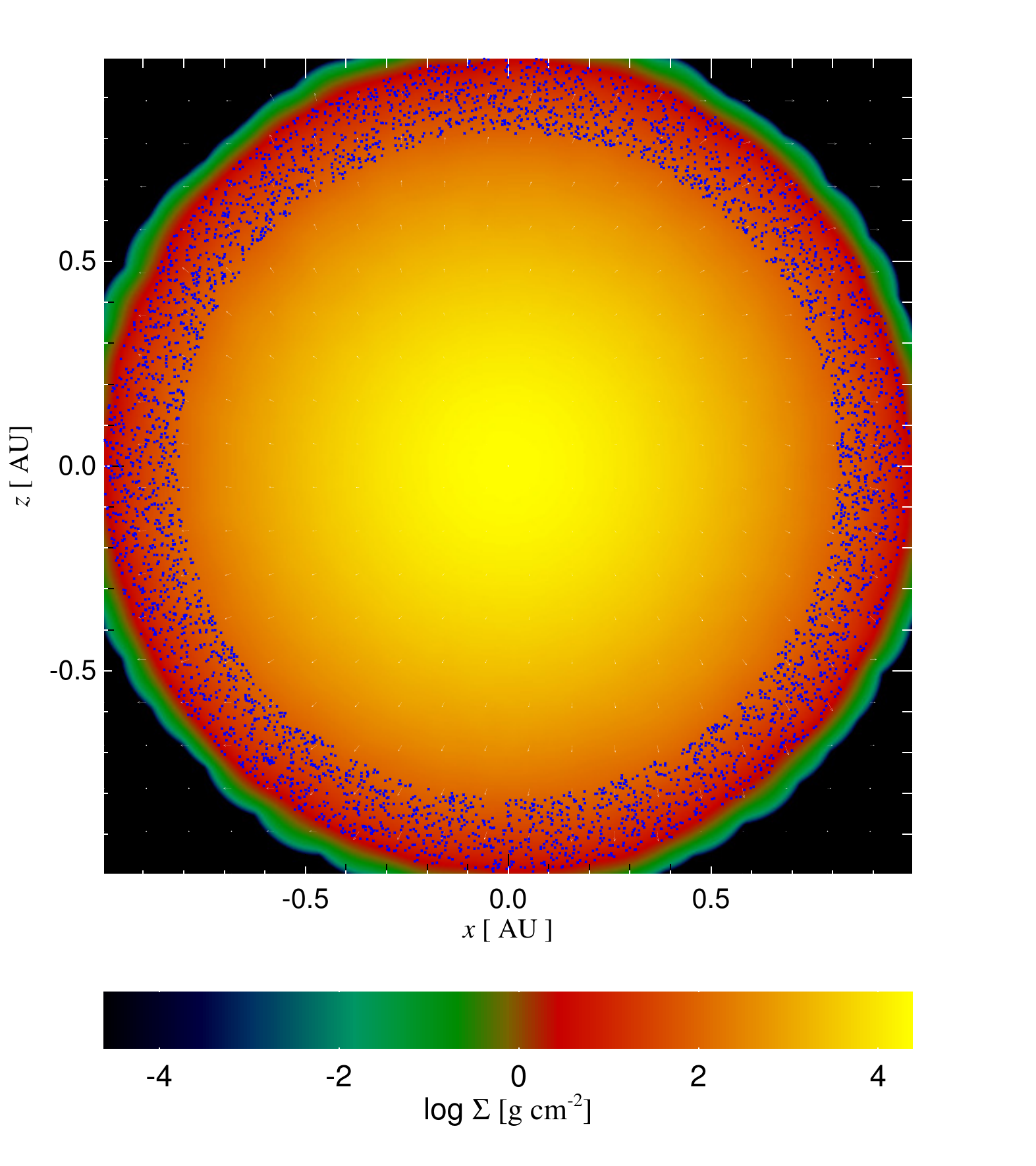}
\includegraphics[width=0.47\textwidth]{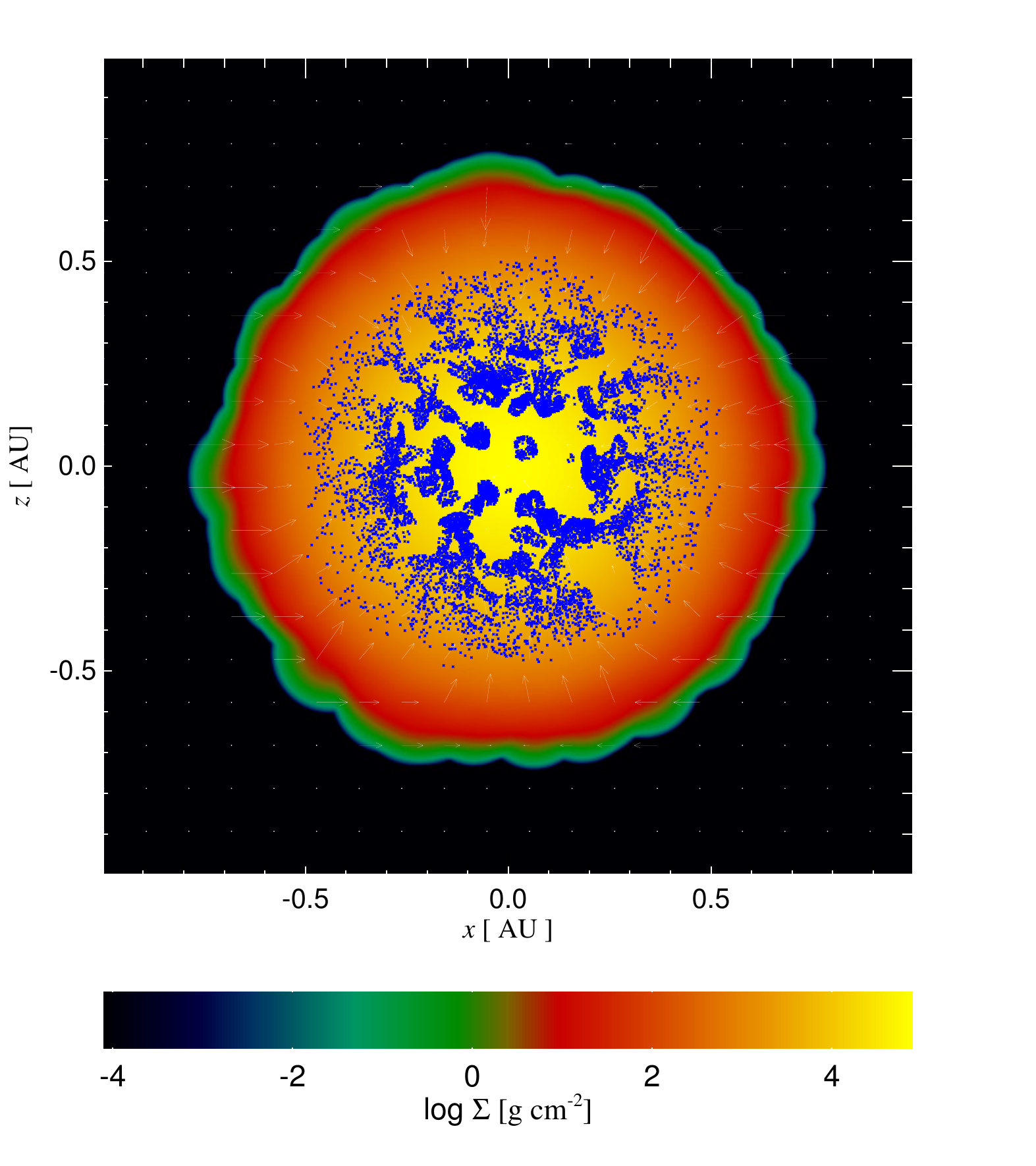}
 \caption{Simulations of a polytropic gas clump (colours) of mass $3\mj$ instantaneously loaded with $0.3 \mj$ worth of 10 cm sized grains (blue dots) distributed in a spherical outer shell. Left and right panels show the initial condition and time $t=7$~years, respectively. Note the development of Raileigh-Taylor instability in which high grain concentration fingers sediment non-spherically. See text in \S \ref{sec:hier} for more detail.}
 \label{fig:num3D}
 \end{figure*}

Fig. \ref{fig:num3D} shows two snapshots from a simulation (Nayakshin 2016, in preparation) of grain-loaded polytropic clump. The figure shows gas surface density (colours) for a slice between -0.1~AU $< y < $ 0.1 AU and $(x,z)$ as shown.  The blue squares on top of the gas mark positions of individual grains. The simulation is started with a relaxed polytropic gas clump of mass $3\mj$, adiabatic index $\gamma=7/5$, and central temperature $T_c = 500$~K. The clump is instantaneously loaded with grains of size $a_{\rm g} = 10$~cm of total mass of $10$\% of the clump mass, uniformly spread inside a spherically symmetric shell between radii of $0.8 R_{\rm p}$ and $R_{\rm p}$, where $R_{\rm p}$ is the planet radius. The initial configuration is displayed in the left panel of Fig. \ref{fig:num3D}.

The right panel of the figure shows what happens with the planet and grains at time $t=7$ years (which corresponds to about 3 dynamical times for the initial clump). Importantly, grain sedimentation process is not spherically symmetric, with "fingers" of higher grain concentration materials protruding inwards.  Undoubtedly, the development of the infalling filaments is driven by the Rayleigh-Taylor instability. These preliminary results indicate that there may be additional physical reasons for development of many rather than one grain concentration centres, lending support to the hypothesis that pre-collapse gas fragments may be sites of both core and planetesimal formation. Also note that the fragment is contracting as predicted by the spherically symmetric model \citep{Nayakshin15a}, although its latter evolution strongly depends on whether a massive core is formed in the centre.

\subsection{Core composition}\label{sec:composition}

A gas fragment of Solar composition \citep{Lodders03} contains 
\begin{equation}
M_{\rm Z} = 0.015 M_{\rm f} \approx 4.5 \mearth \; \frac{Z}{0.015} \; {M_{\rm p}\over \mj}
\label{MZ1}
\end{equation}
 of total mass in astrophysical metals. A third of this mass is in water which is very volatile -- vaporisation temperature $T_{\rm vap}\sim 150-200$~K for the relevant range in gas pressure.
 
Furthermore,  another third of the grain mass is in volatile organics, commonly referred to as "CHON", which is a mnemonic acronym for the
  four most common elements in living organisms: carbon, hydrogen, oxygen, and
  nitrogen. For this review, CHON is  organic material other than water. CHON is a frequently used component in planet
  formation models \citep[e.g.,][]{PollackEtal96,HelledEtal08}. The
  composition of CHON is set to be similar to that of the grains in Comet Halley's coma \cite{Oberc04}. 
  CHON vaporisation temperature is higher than that of water but is still rather low, $T_{\rm vap}\sim 350 - 450$~K for the range of gas pressures appropriate for the interiors of pre-collapse fragments.\footnote{\cite{IaroslavitzP07} note that CHON composition is poorly known, so our results remain dependent on the exact properties of this material.}
 
 Given the fact that fragments migrate in as rapidly as $\sim 10^4$ years, the core must form similarly quickly or else the fragment will either collapse and become a 
second core  or be disrupted, at which point core growth terminates. In practice, a rapid core formation 
requires that gas fragments are compact and dense, but this also means that water and ice and CHON are unlikely to be able to sediment into the centre because the 
fragments are too hot \citep{HS08}. {\it Cores made by Tidal Downsizing are hence likely to be rock-dominated}\footnote{The feedback by the core may puff up contracting host fragment, cooling its centre and making it possible for some late volatile accretion onto the core (see \S \ref{sec:rhd}). However, creating of ice-dominated cores via this mechanism would appear too fine tuned. It would require the fragment to expand significantly to allow ices to sediment and yet not too strongly as to completely destroy it.}. This is significantly different from the classical Core Accretion where massive cores are most naturally assembled beyond the ice line and are thus ice-dominated \citep{PollackEtal96,ColemanNelson16}. In \S \ref{sec:core_comp} we shall discuss current observations of core compositions in light of these differences between the two theories.

A Solar composition Jupiter mass fragment could only make a rocky core of mass $M_{\rm core} \sim 1.5 \mearth$ if all refractory grains 
sediment to its centre. More massive gas fragments could be considered \citep[as done by][]{Nayakshin10a,Nayakshin10b} but such fragments contract radiatively 
very rapidly, making sedimentation of even refractory grains difficult. Thus, to make a massive solid core, $M_{\rm core}\gtrsim 10 \mearth$, metal enrichment of 
fragments, such as pebble accretion or metal enrichment at birth \citep{BoleyDurisen10,BoleyEtal11a}, is necessary.

 \subsection{Core feedback and maximum mass}\label{sec:feedback}
 
 As the core is assembled, some of its gravitational potential energy is radiated into the surrounding gas envelope. How much exactly is difficult to say since 
the opacity, equation of state, and even the dominant means of energy transport  for hot massive planetary cores are not well understood yet \citep{StamenovicEtal12}. The problem is also highly non-linear since the overlying gas envelope structure may modify the energy loss rate of the core, and the temperature of the surrounding gas in turn depends on the luminosity of the core \citep{HoriIkoma11,NayakshinEtal14a}. 

\subsubsection{Analytical estimates}

Nevertheless, assuming that a fraction $0 < \xi_c \lesssim 1$ of core accretion energy, $E_{\rm core} \sim G M_{\rm core}^2/R_{\rm core}$, is released into the fragment and that the latter cannot radiate it away quickly, the core mass is limited by the following order of magnitude estimate:
\begin{equation}
\xi_c {G M_{\rm core}^2\over R_{\rm core}} \lesssim {G M_{\rm p}^2 \over R_{\rm p}}\;.
\label{fb1}
\end{equation}
Defining the escape velocity as $v_{\rm esc} = \sqrt{G M_{\rm p}/R_{\rm p}}$, 
\begin{equation}
{M_{\rm core}\over M_{\rm p}} \lesssim {v_{\rm esc, p}^2 \over \xi_c v_{\rm esc, c}^2}
\end{equation}
Since $v_{\rm esc, p}\sim $ 1 km/s and $v_{\rm esc, c} \gtrsim 10$ km/s, this yields $M_{\rm core}/M_{\rm p} \lesssim 0.01 \xi_c^{-1} $.  A more careful calculation, in 
which the fragment is treated as a polytropic sphere with index $n = 5/2$ yields the following maximum "feedback" core mass \citep{Nayakshin16a}:
\begin{equation}
M_{\rm core} \le M_{\rm fb} = 5.8 \mearth \left({T_3 M_{\rm p} \over 1 \mj}\right)^{3/5}
\rho_{\rm c}^{-1/5} \xi_c^{-1}\;,
\label{Mcrit}
\end{equation}
where $T_3 = T_c/(1000$~K) is the central temperature of the fragment and $\rho_{\rm c}$ is the core mean density in units of g/cm$^3$. 
$T_3$ cannot exceed $\approx 1.5$ because at higher temperatures grains vaporise and the core stops growing via their sedimentation anyway. 
Also, although not necessarily clear from the analytic argument, 
fragments with masses higher than a few $\mj$ are not normally able to hatch massive cores because they contact quickly radiatively \citep[cf. Fig. 18 in][and also \S \ref{sec:rad_collapse}]{NayakshinFletcher15}. Therefore, the factor in the brackets in eq. \ref{Mcrit} cannot actually exceed a few, leading to the maximum core mass of $\sim 10-20\mearth$.

\subsubsection{Radiative hydrodynamics calculation}\label{sec:rhd}

Numerical calculations are desirable to improve on these estimates. In \cite{Nayakshin16a}, a 1D radiative hydrodynamics (RHD) code of \cite{Nayakshin14b} is employed to study the evolution of a fragment accreting pebbles. Unlike the earlier study of core-less fragments in \cite{Nayakshin15a}, grain growth and sedimentation onto the core are allowed. The energy equation for the fragment (see eq. \ref{etot1}),  now taking into account the energy release by the core, reads
\begin{equation}
{d E_{\rm tot}\over dt} = - L_{\rm rad}  + L_{\rm core} - L_{\rm peb} \;,
\label{etot2}
\end{equation}
where the new term on the right hand side, $L_{\rm core}$, is the core luminosity. This term is positive because energy release by the core injects energy into the gas envelope (the fragment). 

\begin{figure}
\includegraphics[width=0.99\columnwidth]{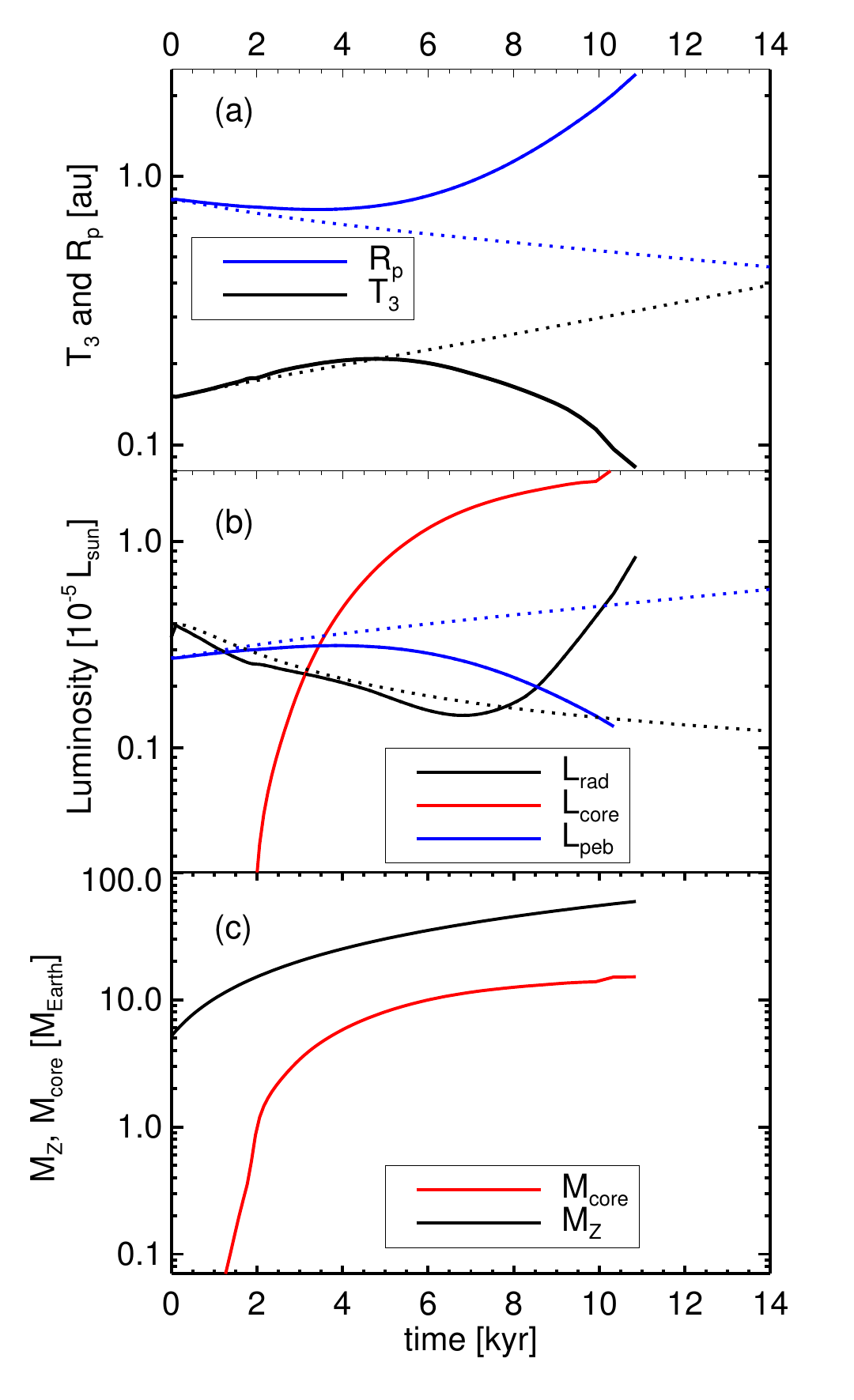}
 \caption{Panel (a) shows the gas fragment central temperature $T_3 = T_{\rm c}/10^3 K$, and planet radius, 
   $R_{\rm p}$, versus time for simulations with (solid curves) and without (dotted) core formation, as described in \S \ref{sec:rhd}. 
   Panel (b) shows core luminosity,
   $L_{\rm core}$, pebble luminosity, $L_{\rm peb}$, and the radiative luminosity of the fragment as labelled. Panel (c): The core mass, $M_{\rm core}$, and the total metal content of the fragment.}
   \label{fig:rhd}
 \end{figure}
 
 In the experiments shown in this section, the initial cloud mass, metallicity and central temperature are $M_{\rm p}=1\mj$, $Z= 1 Z_\odot$ and $150$~K, respectively. The metal loading time scale is set to $t_{\rm z} = 2000$~years. Figure \ref{fig:rhd} compares two runs, one without grain growth and without core formation \citep[so identical in setup to][]{Nayakshin15a}, and the other with grain growth and core formation allowed. Panel (a) of the figure shows in black colour the evolution of $T_3$, the central fragment temperature measured in $10^3$~K, and the planet radius, $R_{\rm p}$ [AU], shown with blue curves. The solid curves show the case of the fragment with the core, whereas the dotted ones correspond to the core-less fragment. 
 
 Panel (b) of Fig. \ref{fig:rhd} presents the three luminosities in eq. \ref{etot2}.  The dust opacity for this calculation is set to 0.1 times the interstellar opacity\footnote{This is done for numerical convenience rather than a physical reason. The RHD code of \cite{Nayakshin14b} uses an explicit integration technique and so becomes very slow as the fragment contracts. For the case at hand, setting the opacity to lower values allows faster execution times without compromising the physics of feedback. For the sake of future coupled disc-planet evolution calculations, it is appropriate to note that the RHD code is impractical to use generally, and this is why the "follow adiabats" approach is used later on in \S \ref{sec:1disc}. }  at $Z=Z_\odot$ \citep{ZhuEtal09}. This increases the importance of $L_{\rm rad}$ term by a factor of $\sim 10$; for the nominal grain opacity, $L_{\rm rad}$ would be completely negligible. Finally, panel (c) of Fig. \ref{fig:rhd} shows the total metal mass in the planet and the core mass with the black and red curves, respectively.
 
Consider first the core-less fragment. As the fragment contracts, $L_{\rm rad}$ quickly becomes negligible compared to $L_{\rm peb}$. This is the pebble accretion dominated no-core regime studied in \S \ref{sec:pebbles} and in \cite{Nayakshin15a}. The fragment contracts as it accretes pebbles.

In the case with the core, panel (a) shows that the fragment collapse reverses when $L_{\rm core}$ exceeds $L_{\rm peb}+L_{\rm rad}$. By the end of the calculation the gas envelope is completely unbound, with the final $M_{\rm core} = 15.2 \mearth$, consistent with equation \ref{Mcrit}.  It is worth emphasising that the appropriate fragment disruption condition is not the luminosity of the core, which first exceeds the sum $L_{\rm peb}+L_{\rm rad}$ when $M_{\rm core}\approx 10\mearth$, but the total energy released by the core. On the other hand, for a migrating planet, the fact that the fragment stopped contracting when the core reached $\approx 8\mearth$ may be sufficient to change the fate of the fragment as is it is more likely to be disrupted when it stops contracting.

\subsubsection{Comparison to Core Accretion}\label{sec:fb_comp}

In Core Accretion theory, the core is more massive and much more compact than the envelope in the interesting stage, that is before the atmosphere collapses \citep{Mizuno80,PollackEtal96,PT99}. Therefore, in this theory $L_{\rm core} \gg L_{\rm peb}$ always, and so one can neglect $L_{\rm peb}$ in equation \ref{etot2}. The luminosity of the core is an obstacle that needs to be overcome in Core Accretion before the atmosphere collapses. It is thought that grain growth reduces the opacity in the atmosphere by factors of $\sim 100$, so that the atmosphere can re-radiate the heat output of the core and eventually collapse \citep{PollackEtal96,Mordasini13}.

 In Tidal Downsizing, there are two regimes in which the pre-collapse gas clump (planet) reacts to pebble accretion onto it differently. While the mass of the core is lower than a few $\mearth$, the gas clump contracts because $L_{\rm core} \ll L_{\rm peb}$. The latter is large because the gas envelope mass is very much larger than that of a pre-collapse Core Accretion planet. This is the regime studied in \cite{Nayakshin15a,Nayakshin15b}, where pebble accretion was shown to be the dominant effective contraction mechanism for moderately massive gas giants.
 
 The second regime, when core mass exceeds $\sim 5\mearth$, is analogous to Core Accretion. Here the core luminosity is large and cannot be neglected. This effect was studied recently in \cite{Nayakshin16a} and is equally key to Tidal Downsizing. Due to this, massive cores are not simply passive passengers of their migrating gas clumps (\S \ref{sec:feedback}).


The roles of massive cores in Tidal Downsizing and Core Accretion are diagonally opposite.

\subsection{Gas atmospheres of cores}\label{sec:atm}


\cite{NayakshinEtal14a} studied formation of a dense gas envelope around the core. This effect is analogous to that of Core Accretion, although the envelope (called atmosphere here) is attracted not from the disc but from the surrounding gas fragment.  Assuming hydrostatic and thermal equilibrium for the envelope of the core, the atmospheric structure was calculated inward starting from $r_i = G M_{\rm core}/c_\infty^2$, where $c_\infty$ is the sound speed in the first core sufficiently far away 
from the core, so that its influence on the gas inside the fragment may be approximately neglected.

It was then shown that for given inner boundary conditions (gas pressure and temperature at $r_i$), there exists a maximum core mass, $M_{\rm crit}$, for which the hydrostatic solution exists. For core masses greater than $M_{\rm crit}$, the atmosphere weight becomes comparable to $M_{\rm core}$, and the iterative procedure with which one finds the atmosphere mass within radius $r_i$ runs away to infinite masses. $M_{\rm crit}$ was found to vary from a few $\mearth$ to tens of $\mearth$.

In \cite{NayakshinEtal14a}, it was suggested that the fragments in which the mass of the core reached $M_{\rm crit}$ will go through the second collapse quickly and hence become young gas giant planets. However, the steady state assumptions in \cite{NayakshinEtal14a} may not be justified during collapse. 
Experiments (unpublished) with {\it hydrodynamic} code of \cite{Nayakshin14b} showed that when the atmospheric collapse happens, there is a surge in the luminosity entering $r_i$ from the inner hotter regions. This surge heats the gas up and drives its outward expansion. This reduces gas density at $r_i$, causing the pressure at $r_i$ to drop as well, halting collapse. 


If the fragment is sufficiently hot even without the core, e.g., $2000 - T_{\rm c} \ll T_{\rm c}$, then the presence of a massive core may be able to 
accelerate the collapse by compressing the gas and increasing the temperature in the central regions above 2000 K. However, if the fragment managed to reach the near collapse state {\it without the core being important}, then it would seem rather fine tuned that the fragment would then need the core to proceed all the way into the 
second collapse. The fragment is already close to collapse, so presumably it can collapse without the help from the core.

Therefore, at the present it seems prudent to discount the atmospheric collapse instability as an important channel for gas fragment collapse. While this conclusion on the importance of bound gas atmospheres near the solid cores differs from that of \cite{NayakshinEtal14a},  their calculation of 
the atmosphere structure and the mass of the gas bound to the core is still relevant. If and when the fragment is disrupted, the atmosphere remains bound to the core. 
This is how Tidal Downsizing may produce planets with atmospheres composed of volatiles and H/He (cf. \S \ref{sec:atmo}).

%
%
%
%
%

\section{Population synthesis}\label{sec:dp_code}


Detailed numerical experiments such as those presented in the previous sections are very computationally expensive and can be performed for only a limited number of cases. This is unsatisfactory given the huge parameter space and uncertainties in the initial conditions and microphysics, and the fact that observations have now moved on from one planetary system to $\sim $ a thousand.

A more promising tool to confront a theory with statistics of observed planets is population synthesis modelling \citep[PSM; see][]{IdaLin04a}. A widely held opinion "with enough free parameters everything can be fit" could be justifiable only perhaps a decade ago. Now, with with $\sim O(100)$ observational constraints from the Solar System and exoplanets, population synthesis is becoming more and more challenging. A balanced view of population synthesis is that it cannot ever prove that a model is right, but experience shows that it can challenge theories strongly. It can also highlight differences between planet formation theories and point out areas where more observations and/or theory work is needed. 

There is much to borrow from Core Accretion population synthesis \citep{IdaLin04b,MordasiniEtal09a,MordasiniEtal09b}. It is quite logical to follow the established approaches to modelling the protoplanetary disc, but then differ in planet formation physics. A planet formation module of the population synthesis should evolve the planet-forming elements of the model, integrating their internal physics, and interaction with the disc via grains/gas mass exchange and migration. The outcome of a calculation is the mass, composition, location, and orbit of one or more planets resulting from such a calculation. By performing calculations for different initial conditions (e.g., disc mass or radial extent) one obtains distributions of observables that can then be compared to the observations.



\subsection{Galvagni \& Meyer model}\label{sec:GM}

 \cite{GalvagniMayer14} study was focused on whether hot Jupiters could be accounted for by gas fragments rapidly migrating from the outer self-gravitating disc. This  (pre-pebble accretion) study was based on 3D SPH simulations  of pre-collapse gas fragment contraction and collapse by \cite{GalvagniEtal12}, who used a prescription for radiative cooling of the fragments, and found that gas fragments may collapse up to two orders of magnitude sooner than found in 1D  \citep[e.g.,][]{Bodenheimer74,HelledEtal08}.  \cite{GalvagniMayer14} concluded that many of the observed hot Jupiters could actually be formed via Tidal Downsizing. The model did not include grain growth and sedimentation physics, thus not addressing core-dominated planets.

\subsection{Forgan \& Rice model}\label{sec:FR13}

\cite{ForganRice13b} solved the 1D viscous time dependent equation for the disc, and introduced the disc photo-evaporation term. Their protoplanetary disc model is hence on par in complexity with some of the best Core Accretion population synthesis studies \citep[e.g.,][]{MordasiniEtal09a,MordasiniEtal12}. Both icy and rocky grains were considered to constrain the composition of the cores assembled inside the fragments. Fragments were allowed to accrete gas from the protoplanetary disc. For the radiative cooling of gas fragments, analytical formulae from \cite{Nayakshin10c} were employed, which have two solutions for dust opacity scaling either as $\kappa(T) \propto T$ or as $\kappa(T) \propto T^2$, where $T$ is the gas temperature. \cite{ForganRice13b} also allowed multiple gas fragments per disc to be followed simultaneously.

\begin{figure}
\includegraphics[width=0.99\columnwidth]{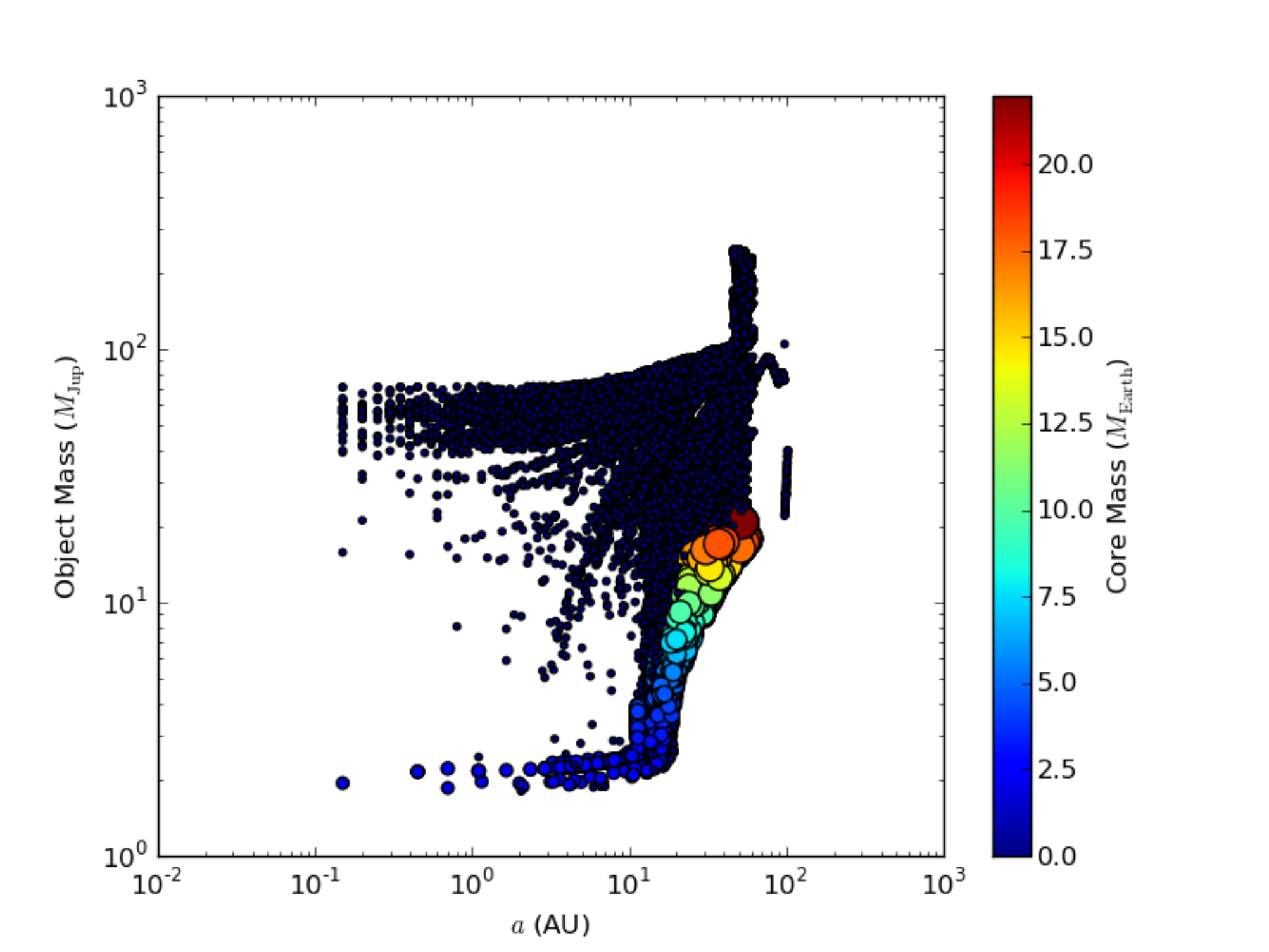}
 \caption{Population synthesis results from Forgan \& Rice (2013b; the right panel of their figure 10), showing the mass of the fragment versus its separation from the host star. Colours show the mass of the cores assembled inside the fragments.}
 \label{fig:FR13}
 \end{figure}

\cite{ForganRice13b} made four different population synthesis calculations, varying the opacity law, the disc migration rate and the assumptions about what happens with the disc beyond 50 AU after it produces fragments. Results of one of such population synthesis experiments are presented in Fig. \ref{fig:FR13}, showing the fragment mass at time $t=1$ Million years versus its separation from the star. The colour of  the circles shows the core mass within the fragments. 

The authors conclude that the model falls way short of explaining the data.
Gas fragments are either disrupted well before they are able to enter the central few AU region, producing hardly any hot Jupiters, or accrete gas rapidly, becoming brown dwarfs (BDs) and even more massive stellar companions to the host star. No massive cores are released into the disc because the fragments that are disrupted do not manage to make massive cores, and the fragments that do make massive cores are in the brown dwarf regime and are not disrupted. 

\subsection{Nayakshin (2015) model}\label{sec:1disc}

In  \cite{Nayakshin15c,Nayakshin15d}, pebble accretion onto precollapse gas fragments was added to population synthesis for the first time. The disc model is similar to that of \cite{ForganRice13b},  but also includes the interaction of the planet with the disc as in \cite{NayakshinLodato12}. The disc not only influences the planet but also receives the back torques from the planet, so that a gap and even a large inner hole can be self-consistently opened. If the planet is disrupted, its gas is deposited in the disc around the planet location. The disc photo-evaporation rate is a Monte-Carlo variable and the limits are adjusted such as to ensure that the disc fraction decays with the age of the system as observed, e.g., $\propto \exp(-t/t_l)$, where $t_l = 3$~Myr \citep{HaischEtal01}. 
\subsubsection{Grains and cores in the model}\label{sec:g_and_c}

The internal physics of the fragments is modelled numerically rather than analytically. The fragments are strongly convective \citep{HelledEtal08,HS08}, which implies that a good approximation to the gaseous part of the fragment is obtained by assuming that it is in a hydrostatic balance and has a constant entropy. The entropy however evolves with time as the fragment cools or heats up. This is known as "follow the adiabats" approach \citep[e.g.,][]{HenyeyEtal64,MarleauCumming14}. The irradiation of the planet by the surrounding disc \citep[the thermal "bath effect", see][]{VazanHelled12} is also included.

The gas density and temperature profiles within the fragment are solved for numerically. The dust evolution module of the code considers three grain species: rocks (combined with Fe), organics (CHON) and water. Grain growth, sedimentation and convective grain mixing are included. Grains are shattered in fragmenting collisions when the sedimentation velocity is too high \citep[e.g.,][]{BlumMunch93,BlumWurm08,BeitzEtal11}. Finally, grains are  vaporised if gas temperature exceeds vaporisation temperature for the given grain species.

Grains reaching the centre accrete onto the solid core. The core initial mass is set to a "small" value ($10^{-4} \mearth$). Growing
core radiates some of its gravitational potential energy away, but a self-consistent model for energy transfer within the core is not yet possible
due to a number of physical uncertainties \citep[e.g.,][]{StamenovicEtal12}. For this reason the energy release by the core is parameterised
 via the Kelvin-Helmholtz contraction time of the core, $t_{\rm kh}$, which is set to be  of order $t_{\rm kh}\sim 10^5 - 10^6$~years. The
luminosity released by the core is injected into the fragment. 

\begin{figure}
\includegraphics[width=0.99\columnwidth]{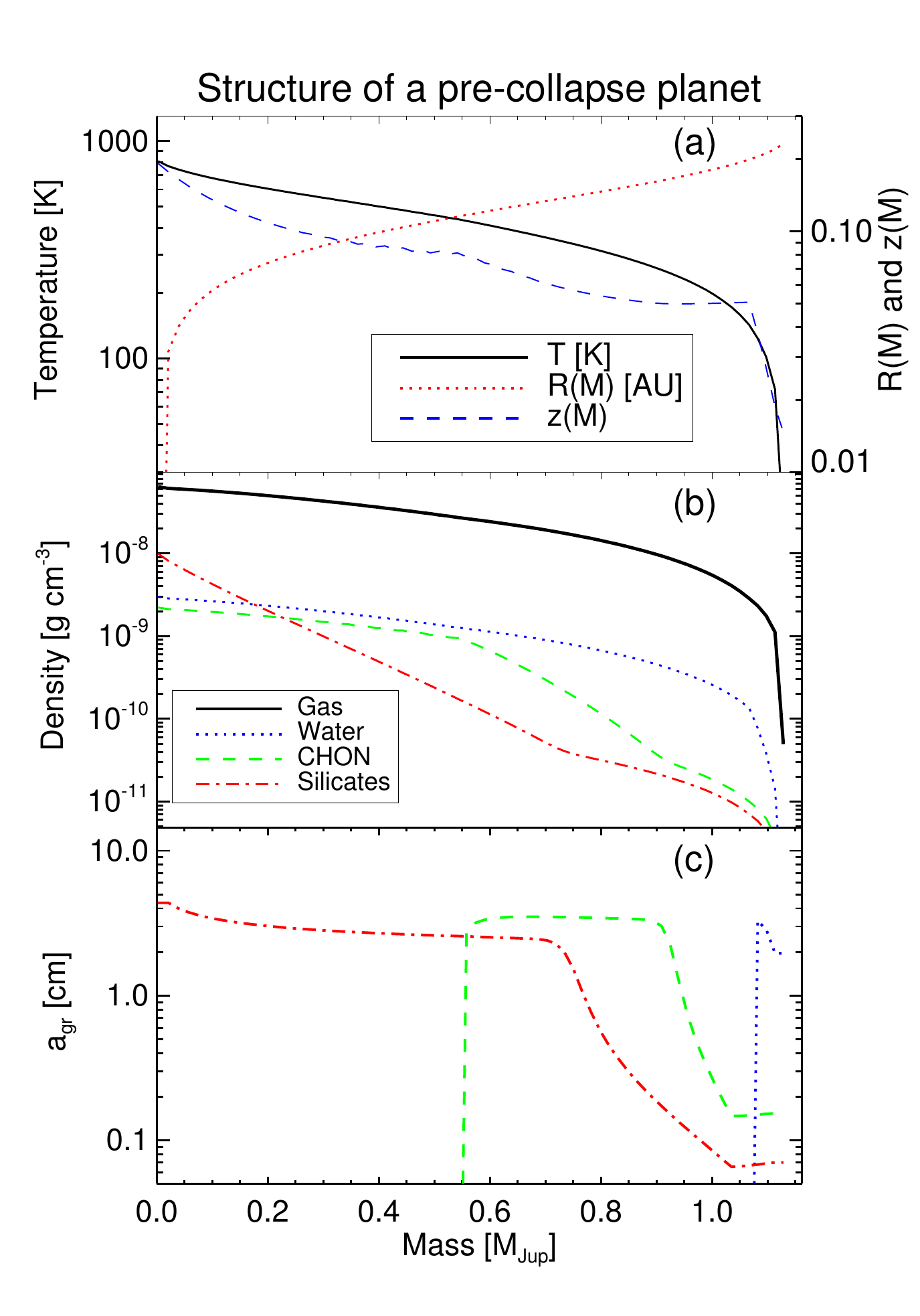}
\caption{Internal structure of a planet \citep[at time $t=24450$~years in simulation M1Peb3 from][]{Nayakshin15c}
as a function of total (gas plus metals, including the core) enclosed
  mass. Panel (a) shows the temperature, Lagrangian radius (in units of AU), and local
  metallicity, $z(M)$. Panel (b) shows gas (solid) and the three grain metal species
  density profiles, while panel (c) shows the species' grain size, $a_{\rm  gr}$. }
\label{fig:structure}
\end{figure}

Figure \ref{fig:structure} shows an example calculation of the internal structure of a gas fragment from population synthesis modelling by \cite{Nayakshin15c}. Since the gas is hot in the inner part and cool in the outer parts, volatile grains (ice and CHON) are able to settle down only in the outer parts of the fragment. In contrast, rocky grains can sediment all the way into the core. This is best seen in the bottom panel (c) of the figure: water ice grains are only large in the outermost $\sim 5$\% of the fragment. Interior to this region, the planet is too hot so that water ice vaporises. Strong convective mixing then ensures that the ratio of the water volume density to the gas density is constant to a good degree (compare the blue dotted and the black solid curves in panel b of the figure) in most of the cloud. Similarly, CHON grains can grow and sediment in the outer $\sim$ half of the fragment only. Note that in the region where CHON grains are large and can sediment, their density shows a significant concentration towards the central parts of the fragment. 

The density of rocky grains is very strongly peaked in Fig. \ref{fig:structure}, cf. the red dash-dotted curve in panel (b). In fact, most of the silicates are locked into the central core, and only the continuing supply of them from the protoplanetary disc via pebble accretion keeps rock grain densities at non-negligible levels.

Also note that the relative abundance of the three grain species varies strongly in the fragment due to the differences in sedimentation properties of these species, as explained above. The outer region is very poor in rocks and very rich in water ice. The innermost region is dominated by rocks. The results of population synthesis planet evolution module are in a very good qualitative agreement with earlier more detailed stand-alone pre-collapse planet evolution calculations \citep{Nayakshin10b,Nayakshin14b}.

\subsubsection{The combined disc-planet code}\label{sec:combi}

\begin{figure*}
\includegraphics[width=0.99\columnwidth]{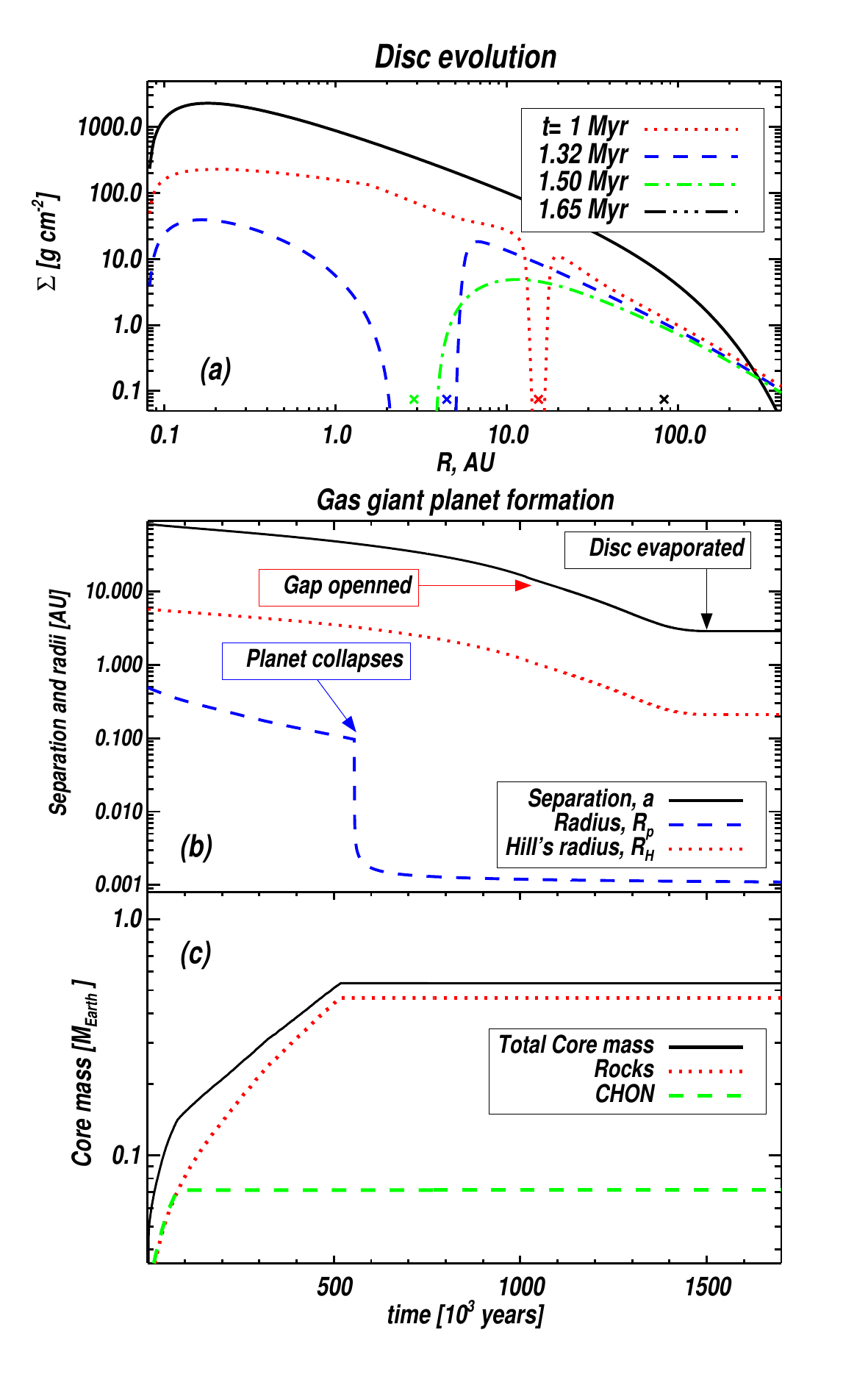}
\includegraphics[width=0.99\columnwidth]{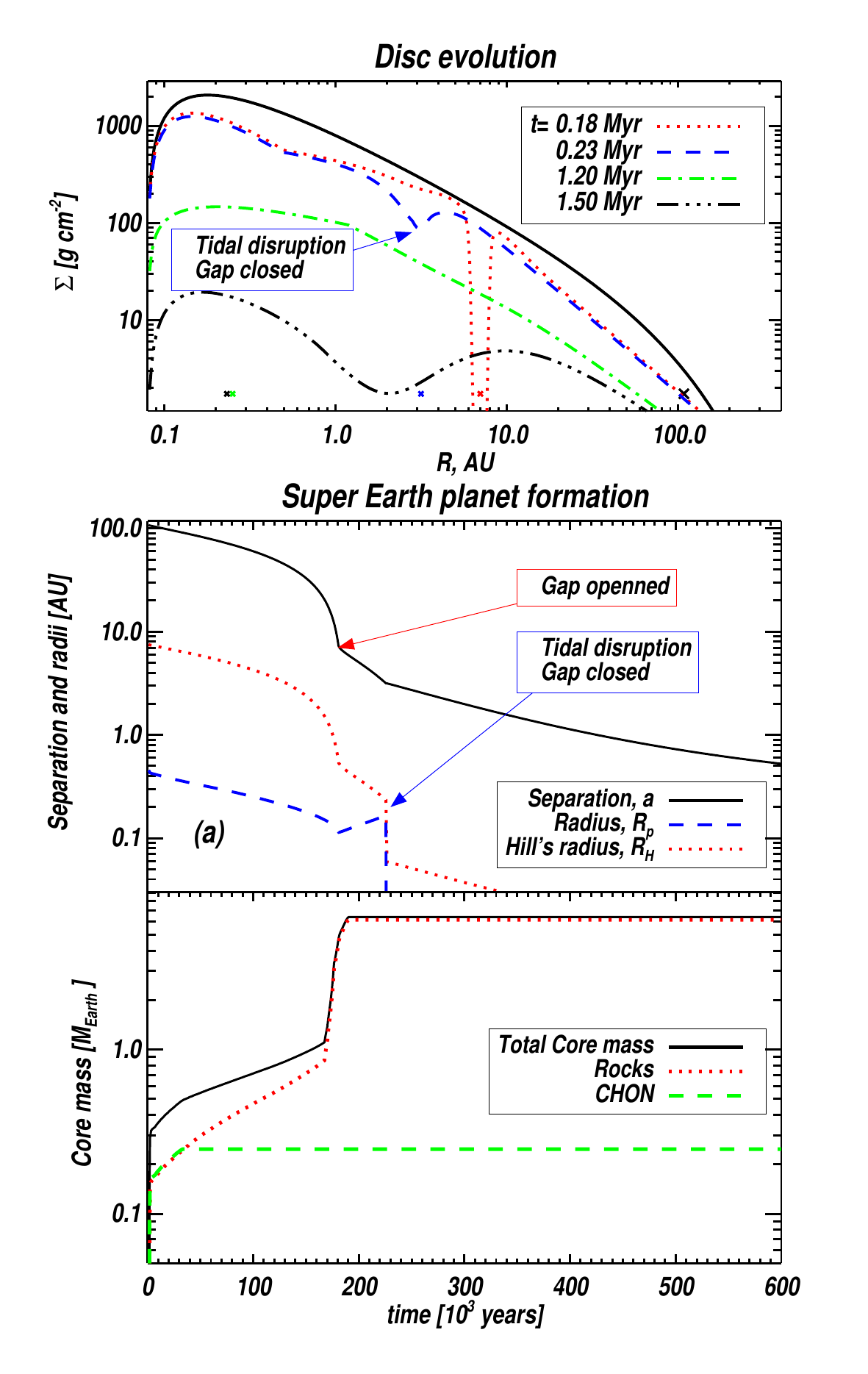}
\caption{Evolution of the protoplanetary disc (panel a) and the embedded fragment (b
  and c). The fragment survives to become a gas giant planet. Panel (a) shows disc surface density profiles at times $t=0$ (solid
  curve) plus several later times as labelled in the legend. The position of
  the planet at corresponding times is marked by a cross of same collor at the
  bottom of the panel. Panel (b) shows the planet separation, radius and the
  Hills radius, whereas panel (c) shows the mass of the core versus time.}
\label{fig:disc_planet}
\end{figure*}

The disc and the fragment evolutionary codes are combined in one, with interactions between them occurring via (a) gravitational torques that dictate the planet migration type and rate, and the structure of the disc near the planet and downstream of it; (b) via pebble accretion that transfers the solids from the disc into the fragment; (c) energy exchange via the disc irradiating the outer layers of the planet, and the planet heating the disc up due to the migration toques close to its location \citep[e.g.,][]{LodatoEtal09}.

One significant shortcoming of the present population synthesis \citep{Nayakshin15c} is limiting the numerical experiments to one fragment per disc, unlike \cite{ForganRice13b} who were able to treat multiple fragments per disc. Numerical simulations show that fragments form rarely in isolation \citep[e.g.,][]{VB06,BoleyEtal11a,ChaNayakshin11a} and so this limitation should be addressed in the future.

Since planet migration is stochastic in nature in self-gravitating discs \citep[e.g.,][ see Fig. \ref{fig:Bar11}]{BaruteauEtal11}, a migration time scale multiplier, $f_{\rm migr}> 0$, is introduced. This parameter is fixed for any particular run but is one of the Monte Carlo variables \citep[for example, in][ $f_{\rm migr}$ is varied between 1 and 4]{Nayakshin16a}.Further details on the population synthesis code are found in \cite{Nayakshin15c,Nayakshin15d,NayakshinFletcher15}.



\subsubsection{Two example calculations}\label{sec:two_ex}

Figure \ref{fig:disc_planet} presents two example calculations from \cite{Nayakshin15d} which show how Tidal Downsizing can produce a warm jupiter and a hot super-Earth. The two calculations have same initial fragment mass, $M_{\rm p0} = 1\mj$. The main distinction is the migration factor $f_{\rm migr} = 8 $ and $1.3$ for the left and the right panels in Fig. \ref{fig:disc_planet}, respectively.

The top panels show the disc surface density evolution sampled at several different times as indicated in the legend. The initial disc mass is similar in both runs, $M_{\rm d0}\sim 0.07\msun$. The crosses on the bottom of the panels depict the planet position at the same times as the respectively coloured surface density curves. The initial surface density of the discs is shown with the solid curve. The red dotted curves show the disc surface density at the time when a deep gap in the disc is first opened. Since the planet on the right migrates in more rapidly, the surrounding disc is hotter when it arrives in the inner ten AU, so that the gap is opened when the planet is closer in to the host star than in the case on the right. 

The contraction of both fragments is dominated by pebble accretion from the disc (\S \ref{sec:pebbles}). The major difference between the two calculations is the amount of time that the two planets have before they arrive in the inner disc. The slowly migrating fragment on the left has a much longer time to contract, so that it manages to collapse at time $t=1.32$ Million years. The other fragment, however, is disrupted at time $\sim 0.2$~Million years. On detailed inspection, it turns out that the fragment would also collapse if it continued to accrete pebbles. However, when the gap is opened, pebble accretion shuts down. The fragment in fact expands (note the upturn in the blue dashed curve in the middle panel on the right) due to the luminosity of the massive $M_{\rm core} \approx 6.4\mearth$ core assembled inside. The fragment continues to migrate after opening the gap, a little slower now in type II regime. Nevertheless, this continuous migration and puffing of the fragment up by the internal luminosity of the core is sufficient to disrupt it tidally just a little later.  After the disruption, the core continues to migrate and arrives in the inner disc at $a=0.23$~AU by the time the disc is dissipated.

%
%
%
%
%
%
%
%
%


\subsection{Overview of population synthesis results}\label{sec:bird}

\begin{figure*}
\includegraphics[width=0.95\columnwidth]{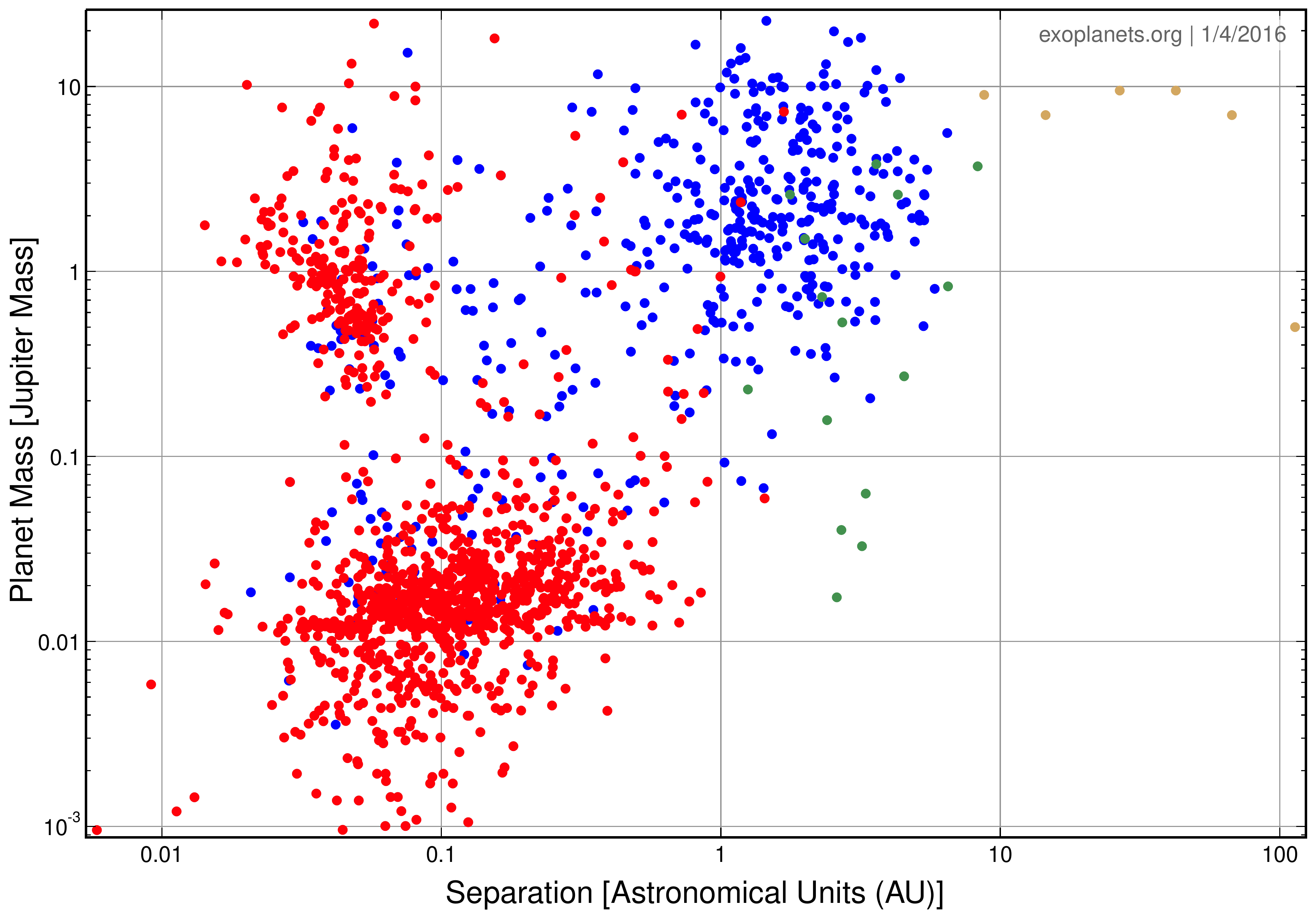}
\includegraphics[width=1.05\columnwidth]{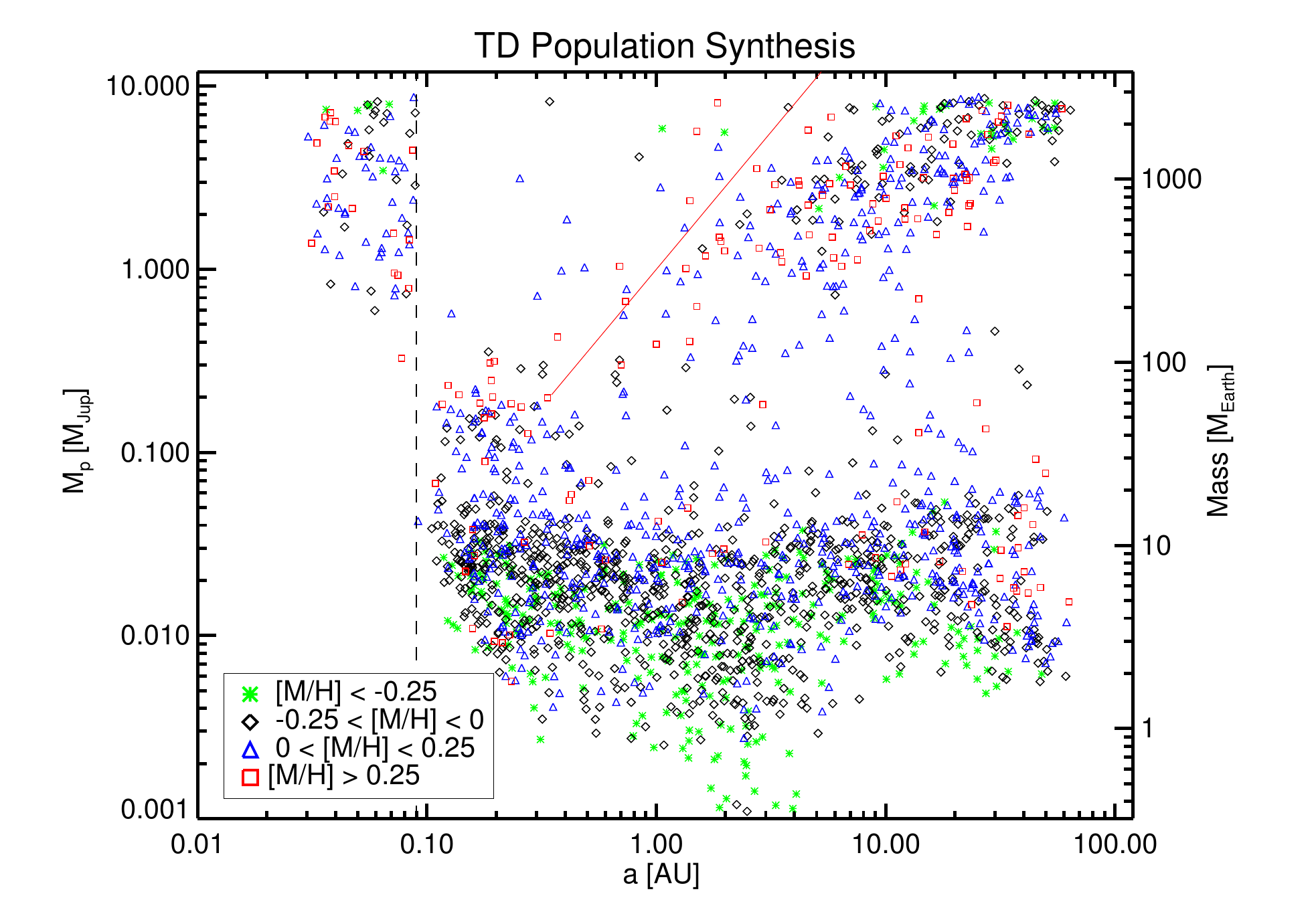}
 \caption{{\bf Left:} Planet mass versus separation from "exoplanets.org" database as of January 4 2016. Red, blue, green and yellow symbols correspond to planets 
detected by transit, RV, microlensing and direct detection methods, respectively. {\bf Right:} Same plot but showing results from a Tidal Downsizing population synthesis calculation from Nayakshin (2016a), colour-coded by metallicity of the host star.}
 \label{fig:scatter}
 \end{figure*}
 
The left panel of figure \ref{fig:scatter} shows planetary mass versus separation from the host star taken from the "exoplanets.org" catalogue \citep{HanEtal14}. 
The colours of the points depict which one of the four exoplanet detection techniques were used to discover the plnet, as 
described in the caption. The lower right hand corner of the figure is almost certainly empty of planets only due to observational selection biases. This region is difficult to observe because the planets are too dim or too low mass and also have very long periods. It may well be teeming with planets. 

In addition to this bias, there is also a strong tendency towards detecting massive planets while missing lower mass ones at a given orbital period or separation. Due to these selection biases, the figure seems to indicate that massive gas giants at small separation are quite abundant. In reality, however, hot Jupiters -- gas giants at $a\lesssim 0.1$~AU -- are over 10 times less frequent than gas giants at $a \gtrsim 1$~AU \citep{SanterneEtal15}. Gas giants at any separation are about an order of magnitude less frequent than planets with size/mass smaller than that of Neptune \citep{MayorEtal11, HowardEtal12}.

The right panel of the figure shows population synthesis from \cite{Nayakshin16a}. Only 10\% of the 30,000 population synthesis runs are shown in this figure to improve visibility. The 
colours on this plot refer to four metallicity bins as explained in the legend. The vertical dashed line at 0.09 AU is set close to the inner boundary of the protoplanetary 
disc in the population synthesis, $R_{\rm in} = 0.08$~AU. Since population synthesis is not modelling the region inside $R_{\rm in}$, it is not quite clear what would actually happen to the planets that migrated all the way to $R_{\rm in}$. It may be expected that the radius of the inner boundary of real protoplanetary discs spans a range of values from very close to the stellar surface to many times that, and that some of the planets inside our $R_{\rm in}$ will actually survive to present day\footnote{For example, \cite{ColemanNelson16} argue that the inner boundary of the disc is at $R\approx 0.05$~AU due to magnetospheric torques for a typical T Tauri star. In cases when the disc has created only one significant planet, and it migrated all the way to the inner disc edge, they find that the planet may survive at a separation somewhat smaller than $R_{\rm in}$. However, if the disc created several large planets, then the planets inside $R_{\rm in}$ interact via resonant torques with the ones migrating in next to them in the resonant planet "convoy". The inner planets are then usually pushed further in and perish in the star completely.}. Without further modelling it is not possible to say which planets will survive inside $R_{\rm in}$ and which would not. Therefore, we simply show only 1\% of the planets that went inside 0.09 AU in the right panel of Fig. \ref{fig:scatter}. They are randomly selected from the total pool of planets that arrived in the region. Their position in the figure
is a random Monte Carlo variable uniformly spread in the $\log$ space between $a=0.03$~AU  and $a=0.09$~AU.

The red line in the right column of Fig. \ref{fig:scatter} shows the "exclusion zone" created by the Tidal Downsizing process (equation \ref{aex1}), which is the region forbidden for pre-collapse gas fragments. Such fragments are tidally disrupted when reaching the exclusion zone (see further discussion in \S \ref{sec:Pvalley}).  Migration of post-collapse 
fragments dilutes the sharpness of the exclusion boundary somewhat. Also, the exclusion zone arguments of course do not apply to low mass planets (cores) that were already 
disrupted. For this reason the red line in the figure is not continued to lower planet masses.

There are some similarities and some differences between the observed (left panel in Fig. \ref{fig:scatter}) and the simulated (right panel) planets. On the positive side, (a) both population synthesis and observations are dominated by 
the smaller, core-dominated planets; (b) simulated planets cover the whole planet-star separation parameter space, without a need to invoke different models for close-in and far out planets; 
(c)  there is a sharp drop in the planet abundance for planets more massive than $\sim 0.1\mj$ in both simulations and observations; 
(d) gas giants at separations $0.1 < a < 1$~AU are relatively rare in both observations and population synthesis. Further analysis (\S \ref{sec:Z}) will show that correlations between planet presence 
and host star metallicity in the model and observations are similar.

However, (a) there is an over-abundance of massive planets at tens of AU in the models compared to observations; (b) the mass function of hot Jupiters is centred on $\sim 1\mj$ in the observation but is dominated by more massive planets in population synthesis; (c) there is no small planets in the population synthesis at $a\lesssim 0.1$~AU. 

%
%
%
%
%
%
%

\section{Metallicity correlations}\label{sec:Z}

\subsection{Moderately massive gas giants}\label{sec:giants_Z}

A strong positive correlation of giant planet frequency of detection versus host star metallicity, [M/H], is well known
\citep{Gonzalez99,FischerValenti05,MayorEtal11,WangFischer14}. \cite{IdaLin04b} found in their population synthesis that if massive cores, $M_{\rm core} \sim 10\mearth$, appear in the disc only after $\sim 3$ Million years for a typical Solar metallicity protoplanetary disc, then metal-poor systems will tend to make massive cores only after the gas disc is dissipated. Metal-rich systems make cores earlier, before the gas disc is dissipated. Therefore, Core Accretion predicts a strong preference for gas giant planet presence around [M/H]~$> 0$ hosts. This argument is based on the assumption that planetesimals are more abundant at high [M/H] hosts (\S \ref{sec:Z_debris}). 

\begin{figure}
 \includegraphics[width=0.95\columnwidth]{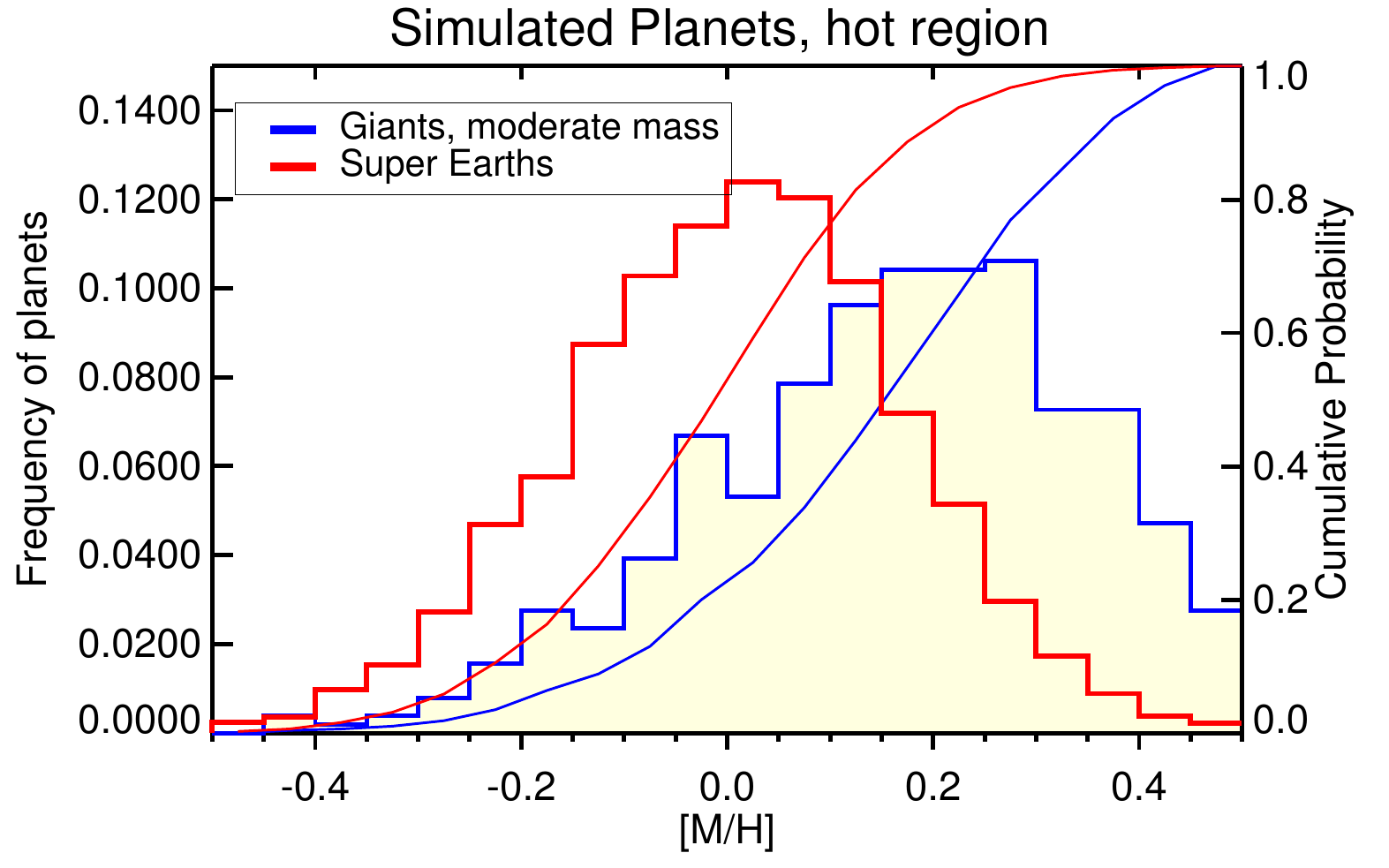}
 \vskip 1 cm
 \caption{Distribution of host star metallicity for planets survived in the inner 5 AU region from Nayakshin \& Fletcher (2015). Gas giant planets correlate strongly with [M/H], whereas
 sub-giant planets do not. See text in \S \ref{sec:giants_Z} and \S \ref{sec:sub-giants} for detail.}
 \label{fig:Z_giants}
 \end{figure}

Since gas fragments collapse more rapidly when accreting pebbles at higher rates (\S \ref{sec:pebbles}), a positive correlation with host star metallicity is also expected in Tidal Downsizing. Figure \ref{fig:Z_giants} shows the host star metallicity distribution for gas giants with mass $0.3\mj < M_{\rm p} < 5 \mj$ from population synthesis of \cite{NayakshinFletcher15} with the blue filled histogram. Only planets that end up at separations less than 5 AU are shown in the figure. The red histogram is for massive cores (see \S \ref{sec:sub-giants}). The continuous 
curves show the corresponding cumulative distributions. The initial metallicity distribution of fragments in this calculation is a gaussian centred on [M/H]=0 with dispersion $\sigma = 0.22$. Survived gas giants are strongly skewed toward metal-rich hosts, as expected, and qualitatively as observed.

Luckily, the similarity in predictions of Core Accretion and Tidal Downsizing essentially ends with the $\sim 1$ Jupiter mass planets inside the inner few AU. 
 
\subsection{Sub-giant planets}\label{sec:sub-giants}

 Observations show that massive core-dominated planets are abundant at all metallicities \citep[e.g.][]{MayorEtal11,BuchhaveEtal14,BuchhaveLatham15}, in contrast to the results for the gas giant planets. More qualitatively, the recent analysis of data by \cite{WangFischer14} shows that gas giants are $\sim 9$ times more likely to form around  [M/H]$ >0$ hosts than they are around [M/H]$<0$ hosts. For sub-Neptune planets the ratio is only around 2. 

The red histogram in Fig. \ref{fig:Z_giants} shows the metallicity distribution from \cite{NayakshinFletcher15} of hosts of "super-Earth" planets defined here as planets with mass in the range $2 \mearth < M_{\rm p} < 15 \mearth$. This distribution is much more centrally peaked than it is for gas giants, in qualitative consistency with the observations.  As already explained in \S \ref{sec:giants_Z}, at low [M/H], most gas fragments migrating inward from their birth place at tens of AU are tidally disrupted. This would in fact yield an anti-correlation between the number of cores created per initial fragment and [M/H] of the star in the context of Tidal Downsizing. However, low metallicity gas fragments contain less massive cores on average (many of which are less massive than $2 \mearth$). Thus, while there are more cores at low [M/H] environments, the more massive cores are found at higher metallicity. The net result is an 
absence of a clear correlation in Tidal Downsizing between the core-dominated planet and the metallicity of their hosts, unlike for gas giants. 

This result is not due to "cherry picking" parameters for population synthesis and is very robust at least qualitatively.  Same physics -- the fact that gas fragments are disrupted more frequently at low [M/H] environments -- explains {\it simultaneously} why gas giants correlate and sub-giants do not correlate with metallicity.

A weak correlation of massive cores with [M/H] of the host star in Core Accretion was explained as following. Cores grow in gas-free environment in discs of low metallicity stars \citep[e.g.,][]{IdaLin04b,MordasiniEtal09b}. These cores are then not converted into gas giants because they had no gas to accrete to make gas-dominated planets. However, this scenario does not tally well with the fact that many of close-in sub-giant planets reside in multi-planet systems, and these are by large very flat \citep[have mutual inclinations $i \lesssim 2^\circ$, see][]{FabryckyEtal14} and have low eccentricities ($e\sim 0.03$).  Such systems are best explained by assembly via migration of planets (made at larger distances) in a {\it gaseous} protoplanetary disc which naturally damps eccentricities and inclinations away \citep{Paardekooper13,HandsEtal14}. 

\subsection{Gas giants beyond a few AU}\label{sec:cold_giants_Z}

The exclusion zone shown with the red line in the right panel of Fig. \ref{fig:scatter} divides the Tidal Downsizing gas giant population in two. Inwards of the line, gas giants must have collapsed into the second cores before they entered this region. Since this is more likely at high metallicities of the host disc, there is a positive [M/H] correlation for the inner gas giants as explored in \S \ref{sec:giants_Z}. Outside the  exclusion zone, however, gas giants may remain in the pre-collapse configuration and still survive when the disc is dispersed. Thus, higher pebble accretion rates do not offer survival advantages at such relatively large distances from the star. This predicts that gas planets beyond the exclusion zone may not correlate with the metallicity of the  host \citep[see Fig. 11 in][]{NayakshinFletcher15}.

Core Accretion is likely to make an opposite prediction. Observations show that protoplanetary discs are dispersed almost equally quickly at small and large distances \citep[see the review by][]{AlexanderREtal14a}. Since classical core assembly takes longer at larger distances, one would expect gas giants at larger distances to require even higher metallicities to make a core in time before the gas disc goes away. Exact separation where this effect may show up may however be model dependent.

While statistics of gas giant planets at distances exceeding a few AU is far less complete than that for planets at $a < 1$~AU, \cite{AdibekyanEtal13} reports that planets orbiting metal-rich stars tend to have longer periods than planets orbiting metal-rich stars (see Fig. \ref{fig:Vardan}).

\begin{figure*}
  \includegraphics[width=0.95\columnwidth]{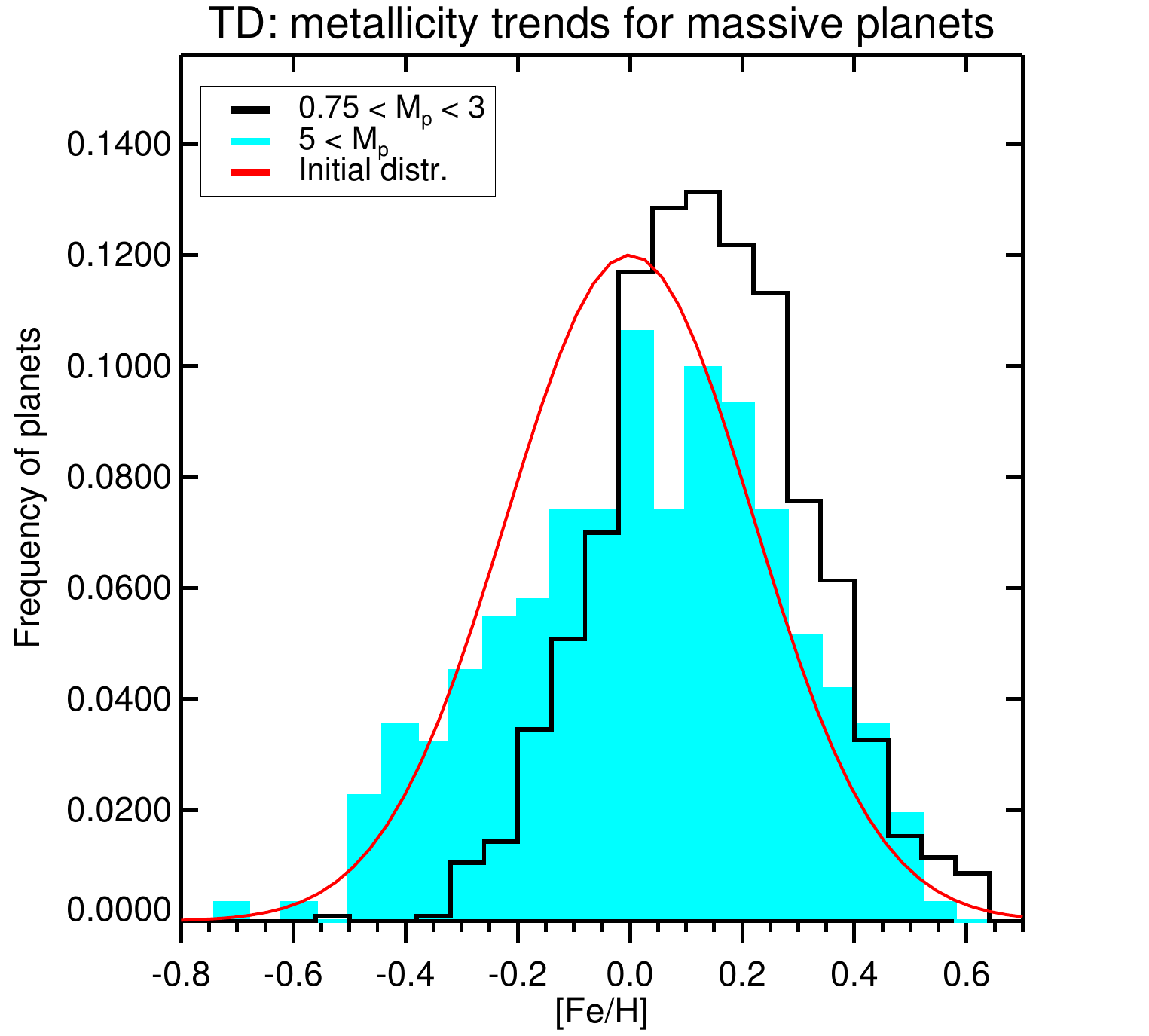}
  \includegraphics[width=1.05\columnwidth]{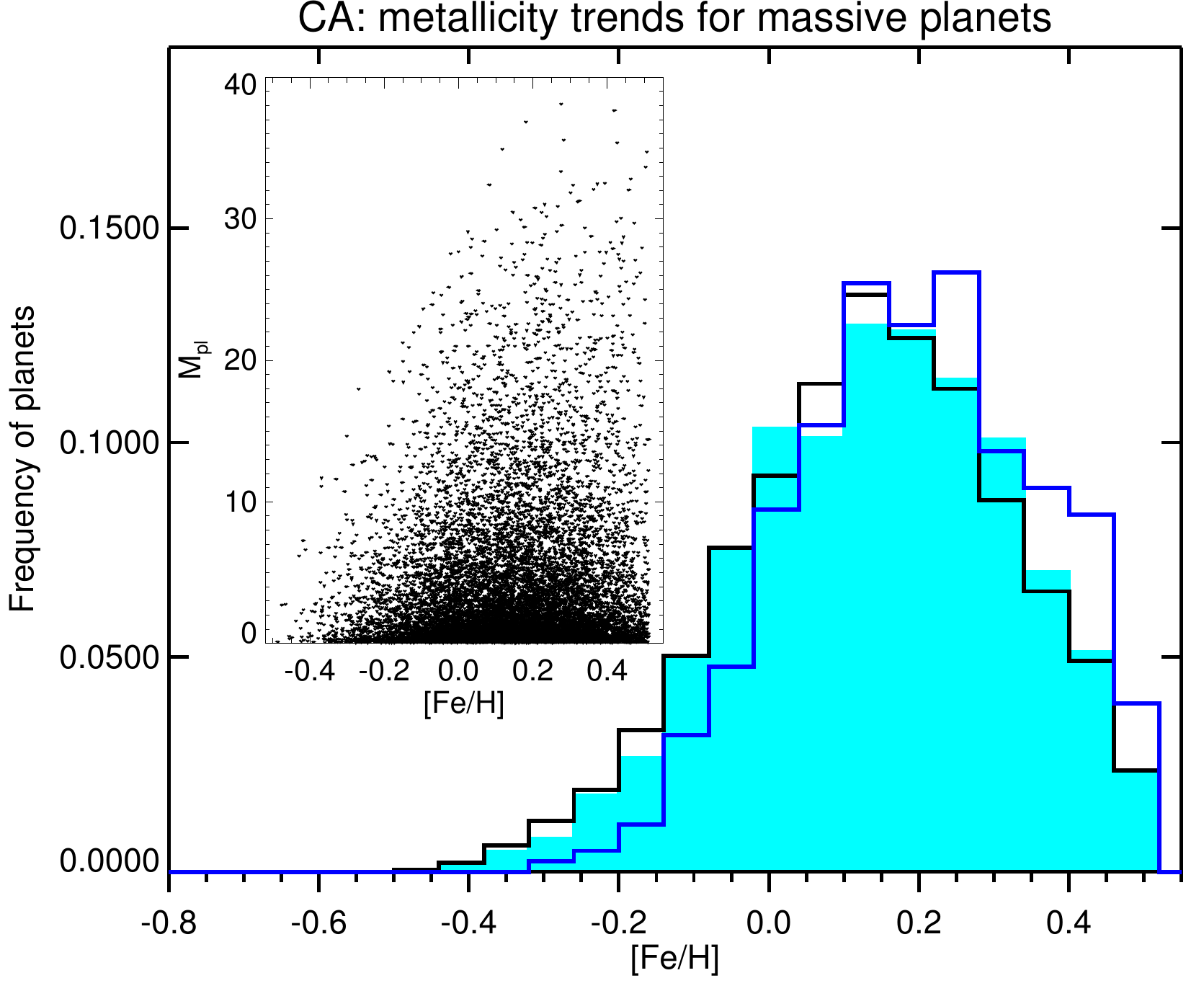}
  \medskip
    \includegraphics[width=1.02\columnwidth]{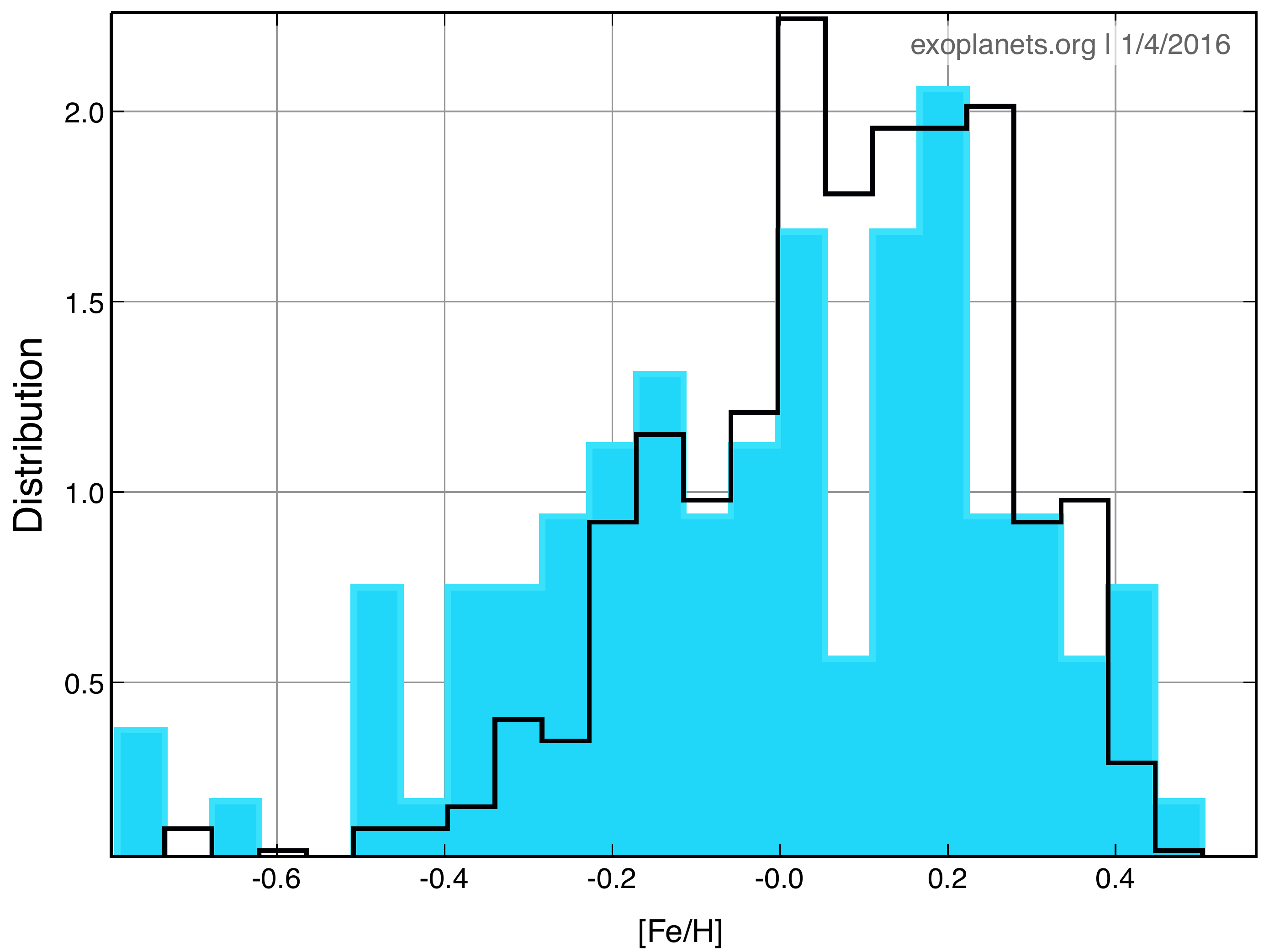}
  \includegraphics[width=0.97\columnwidth]{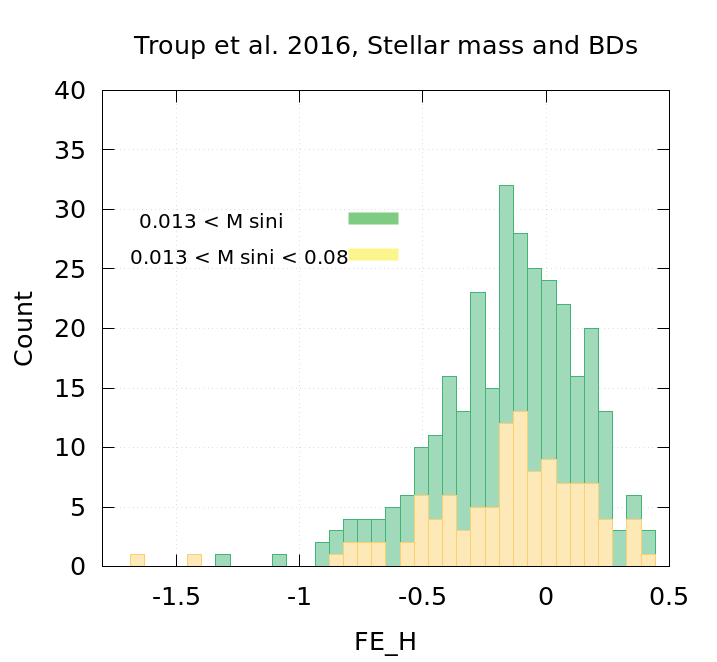}
 \caption{{\bf Top:} Theoretical predictions of population synthesis models. {\bf Left} is distribution of gas giants over host star metallicity in two ranges of planet masses from population synthesis model of \cite{Nayakshin16a}. Black shows planets with mass  $0.75 \mj \le M_{\rm p} \le 3 \mj$, whereas the filled cyan histogram is for $M_{\rm p} > 5\mj$.  {\bf Top right} is Fig. 4 from Mordasini et al (2012), showing planet mass versus host metallicity in their simulations. Tidal Downsizing predicts that the more massive is the planet, the more likely it is to be metal poor; CA makes an opposite prediction (cf. \S \ref{sec:massive_giants_Z}). {\bf Bottom}: Observations. {\bf Bottom left:} Host metallicity distribution for gas giant planets from "exoplanets.org", divided into two mass bins as in the panel above.  {\bf Bottom right:} Similar distribution but for substellar mass companions from Troup et al (2016) with $M_{\rm p}\sin i > 0.013 \msun$ (green) and the brown dwarf sub-sample (yellow; $0.013 < M_{\rm p}\sin i < 0.08 \msun$).}
 \label{fig:Z_massive}
 \end{figure*}

\subsection{Very massive gas giants}\label{sec:massive_giants_Z}

As explained in \S \ref{sec:transition}, Tidal Downsizing makes a robust prediction for how planet-host metallicity correlation should change for more massive planets. 
For planets $M_{\rm p}\gtrsim 5\mj$, the radiative cooling time is comparable to $10^4$ years, implying that fragments of such a mass may collapse before they reach the exclusion zone. High mass planets may therefore avoid tidal disruption simply by radiative cooling. 

Accordingly, we should expect that high mass planets and brown dwarfs should be found with roughly equal frequency around metal rich and metal poor stars, in stark contrasts to Jupiter-mass planets. Fig. \ref{fig:Z_massive}, the top left panel, shows the host metallicity distribution for planets ending up at $a< 15$~AU in simulation ST from \cite{Nayakshin15d}. The figure shows two mass bins, $0.75 \mj \le M_{\rm p} \le 3 \mj$ (black) and $M_{\rm p} \ge 5\mj$ (cyan). The red curve shows the initial (gaussian, centred on [M/H] = 0 and with dispersion $\sigma = 0.22$) distribution of host disc [M/H]. It is seen that moderately massive giants are shifted towards significantly higher metallicities, as previously found (Fig. \ref{fig:Z_giants}). In particular, only 20\% of the planets in the black distribution have [M/H]~$< 0$. Planets more massive than $5\mj$ are distributed more broadly, with 45\% of the planets having negative [M/H].

The Core Accretion model makes an opposite prediction. The inset in the top right panel of Fig. \ref{fig:Z_massive} reproduces\footnote{I thank Cristoph Mordasini very much for providing me with the data from his simulations.} Fig. 4 from \cite{MordasiniEtal12}, whereas the black and the cyan histograms show the host metallicity distribution for planets in the same mass ranges as for the top left panel. The blue histogram shows the metallicity distribution for brown dwarfs. It is easy to see from the figure that the more massive a gas giant planet is, the more metal rich the parent star should be to make that planet by Core Accretion. This result is probably quite robust since it relies on the key physics of the model. It takes a long time to make massive cores and planets in Core Accretion scenario \citep{PollackEtal96,IdaLin04b}. The more massive the planet is to be, the earlier it must start to accrete gas to arrive at its final mass or else the gas disc dissipates away. More metal rich hosts make massive cores more rapidly, so most massive planets should be made in most metal rich discs.

These predictions can be contrasted with the data. The bottom left panel of Fig. \ref{fig:Z_massive} shows the observed metallicity distributions for hosts of gas giant planets that are currently on the "exoplanets.org" database. Planets more massive than $M_{\rm p} = 5 \mj$ are shown with the filled cyan histogram, whereas the moderately massive giants correspond to the black histogram, selected by $0.75 \mj \le M_{\rm p} \le 3 \mj$. The mass cut is the only selection criterion applied to the data. Both histograms are normalised on unit area. The massive group of planets is comprised of 96 objects and has a mean metallicity of $-0.014$, whereas the less massive group is more populous, with 324 objects and the mean metallicity of $0.066$.

While the statistics of exoplanetary data remains limited, we can see that the trend towards lower [M/H] hosts at higher $M_{\rm p}$ is definitely present in the data. We can also confidently conclude that there is no shift towards {\it higher} [M/H] for the more massive planets. The bottom right panel of the figure shows [M/H] correlations for brown dwarf mass companions to stars from \cite{TroupEtal16} which are discussed in the next section.

No fine tuning was done to the population synthesis parameters to achieve this agreement. One physical caveat here is that gas accretion onto the planet is entirely neglected for both pre-collapse and post-collapse configurations. If some post-collapse fragments do accrete gas, then some of the massive planets, $M_{\rm p} \gtrsim 5 \mj$, could have started off as less massive planets. These planets would then remain sensitive to the metallicity of the host. Therefore, if the observed massive $M_{\rm p} > 5 \mj$ planets are a mix of accreting and non-accreting populations, then there would remain some preference for these planets to reside in metal rich systems, but this preference should be weaker than that for the moderately massive gas giants.

\subsection{Close brown dwarf companions to stars}\label{sec:Z_BD}

\subsubsection{Are brown dwarfs related to planets?}\label{sec:BD_vs_planets}

It is often argued \citep[e.g.,][]{WF14} that brown dwarfs and low mass stellar companions must form in a physically different way from that of planets because (a) the frequency of brown dwarf occurrence around Solar type stars is an order of magnitude lower than that for gas giant planets at periods less than a few years \citep[e.g.,][]{SahlmannEtal11,SanterneEtal15}; (b) Gas giant planets correlate strongly with metallicity of the host star, whereas for brown dwarfs the metallicity  distribution is very broad with no evidence for a positive correlation \citep{RaghavanEtal10}; (c) Gas giant planets are over-abundant in metals compared to their host stars \citep{Guillot05,MillerFortney11} whereas brown dwarfs have compositions consistent with that of their host stars \citep{LeconteEtal09}.

These arguments are not water tight, however. 
\SNc{The differences quoted above could be explainable in terms of a single scenario if predictions of that scenario are significantly different for objects of different masses. For example, gas giant planets are an order of magnitude less frequent than sub-giant planets, and their host metallicity correlations are significantly different \citep{MayorEtal11,HowardEtal12}, yet there is no suggestion that these two populations are not related. Both Core Accretion \citep{MordasiniEtal09b} and Tidal Downsizing \citep{NayakshinFletcher15} may be able to explain the sub-Neptune mass planets and gas giants in a single unifying framework.}

\SNc{To be more specific with regard to a Tidal Downsizing origin of close-in brown dwarfs, these objects may be hatched less frequently than gas giants via gravitational instability of the disc. Alternatively, brown dwarfs may be migrating into the inner few au disc from their birth place less efficiently than gas giants do \citep[the migration rate in the type II regime is inversely proportional to the object mass, see, e.g.,][]{LodatoClarke04}. }

\SNc{A trademark of two completely different formation scenarios would be a clean break (discontinuity) between gas giants and brown dwarfs in any of the above observational characteristics. There does not appear to be an observational evidence for such a break.} The occurrence rate of gas giants drops with planet mass towards the brown dwarf regime monotonically \citep[e.g., Fig. 13 in][]{CummingEtal08}; the host metallicity correlation of very massive gas giants becomes weak towards masses of $\sim 10\mj$, before hitting the brown dwarf regime, as discussed in \S \ref{sec:massive_giants_Z}; and the metallicity of gas giants also continuously drops with $M_{\rm p}$ increasing towards brown dwarfs \citep[e.g.,][and also \S \ref{sec:Zpl_giants} and Fig. \ref{fig:Zpl} below]{MillerFortney11}.

 Based on the continuity of the transition in all of these properties, it is \SNc{therefore possible} to consider gas giant planets and brown dwarfs as one continuous population that forms in the same way. \cite{ReggianiEtal16} argue that the observed companion mass function at wide orbits around solar-type stars can be understood by considering giant planets and brown dwarfs a part of the same population as long as a cutoff in planet separation distribution is introduced around $\sim 100$~AU.

A physically similar origin for planets and brown dwarfs is allowed by both planet formation scenarios. In Tidal Downsizing, brown dwarfs were either born big or managed to gain more gas. In Core Accretion, brown dwarfs are over-achieving gas giant planets \citep{MordasiniEtal12}.

\subsubsection{Metallicity correlations of brown dwarfs}\label{sec:BD_z_corr}

\SNc{If planets and brown dwarfs are a continuous population, then it appears that data favour Tidal Downsizing over Core Accretion as a formation route for these objects}.

\cite{RaghavanEtal10} showed that brown dwarf companions to solar mass stars are very broadly distributed over host [M/H]. For low mass {\it stellar} companions, it is the low metallicity 
hosts that are more likely to host the companion. Very recent observations of \cite{TroupEtal16} detail the picture further. These authors presented a sample of 382 close-in stellar and sub-stellar companions, about a quarter of which are brown dwarfs at separations between $\sim 0.1$ to $\sim 1$~AU. Out of these brown dwarfs, 14 have [M/H]$< -0.5$. To put this in perspective, out of many hundreds of planets with mass $0.5 \mj < M_{\rm p} < 5\mj$ on "exoplanets.org" \citep{HanEtal14}, only 4 have [M/H]$<-0.5$. 

The bottom right panel of Fig. \ref{fig:Z_massive} shows the host star metallicity distribution for brown dwarfs (yellow) and for all companions more massive than $0.013 \msun$ (green) from the \cite{TroupEtal16} data. As authors note, their observations strongly challenge Core Accretion model as an origin for the brown dwarfs in their sample.

Indeed, \cite{MordasiniEtal12} in their \S 4.3 state: "While we have indicated in Sect. 4.1 that metallicity does not significantly change the distribution of the mass for the bulk of the population, we see here that the metallicity determines the maximum mass a planet can grow to in a given disk, in particular for subsolar metallicities. There is an absence of very massive planets around low-metallicity stars". To emphasise the point, the authors look at the maximum planet mass in their models at metallicity [M/H]$< -0.4$. For their nominal model, the resulting maximum  planet mass of the low [M/H] tail of the population is only $7\mj$. This is at odds with the observations \citep{RaghavanEtal10,TroupEtal16}.

\subsection{Debris discs}\label{sec:Z_debris}

There is another checkpoint we can use to compare theoretical models of host metallicity correlations with observations: the debris discs \citep{Wyatt08}.

Detailed calculations of planetesimal formation \citep[e.g.,][]{JohansenEtal07,JohansenEtal09}, suggest that planetesimal formation efficiency is a strong function of metallicity of the parent disc. It is therefore assumed that higher [M/H] discs have more abundant supply of planetesimals. This is in fact required if Core Accretion is to explain the positive gas giant correlation with the host star metallicity  \citep[e.g.,][]{IdaLin04b,MordasiniEtal09b}.

As detailed in \S \ref{sec:planetesimals}, Tidal Downsizing scenario offers a different perspective on formation of minor solid bodies. The very central parts of the self-gravitating gas fragments may be producing solid bodies greater than a few km in size by self-gravitational collapse mediated by gas drag (\S \ref{sec:hier}). Observable planetesimals are however created only when the parent gas fragment is disrupted; in the opposite case the planetesimal material is locked inside the collapsed gas giant planet.

Debris discs are detected around nearby stars \citep{Wyatt08} via thermal grain emission in the infra-red
\citep{OudmaijerEtal92,ManningsBarlow98}. Interestingly,  debris discs detection frequency does not correlate with [M/H] of their host stars \citep{MaldonadoEtal12,MarshallEtal14,Moro-MartinEtal15}.
Observed debris discs also do not correlate with the presence of gas giants \citep[e.g.,][]{MMEtal07,BrydenEtal09,KospalEtal09}. It is not that debris discs do "not know" about planets: stars with an observed gas giant are half
{\it as likely} to host a detected debris discs than stars orbited by planets less massive than $30 M_\oplus$ \citep{Moro-MartinEtal15}. 


The suggestion that debris discs are destroyed by interactions with gas giants \citep{RaymondEtal11} could potentially explain why debris discs do not correlate with [M/H] or gas giant presence. However, the observed gas giants (for which the correlations were sought) are orbiting their hosts at separations $\lesssim 1$ AU, whereas the observed debris discs can be as large as tens and even hundreds of  AU, making their dynamical interaction (in the context of Core Accretion) unlikely. Further, radial velocity, microlensing and direct imaging results all show that there is of order $\sim  0.1$ gas giant planets per star \citep{SanterneEtal15,ShvartzvaldEtal15,BillerEtal13,BowlerEtal15,WittenmyerEtal16} {\it at both small and large separations from the host star}, whereas \cite{RaymondEtal11}'s scenario needs several giants in a debris disc-containing 
system to work.

In Tidal Downsizing, higher [M/H] discs provide higher pebble accretion rates, so that few gas fragments are destroyed.  Debris disc formation is hence infrequent at high metallicities compared to low [M/H] hosts. However, each disrupted fragment contains more metals in higher [M/H] than their analogs in low metallicity systems. 

\cite{FletcherNayakshin16a} found that the debris disc -- host metallicity correlation from Tidal Downsizing would dependent on the sensitivity of synthetic survey. A high sensitivity survey picks up low [M/H] hosts of debris discs most frequently because they are more frequent. So such surveys would find an anti-correlation between debris disc  of presence and host metallicity. A medium sensitivity surveys however would find no correlation, and a low sensitivity surveys shows preference for debris around high metallicity hosts. These results appear qualitatively consistent with observations of debris disc -- host star metallicity correlation.

\cite{FletcherNayakshin16a} also considered planet -- debris discs correlations in Tidal Downsizing. A detected gas giant planet implies that the parent fragment {\it did not} go through a tidal disruption -- hence not producing a debris disc at all. A detected sub-Saturn mass planet, on the other hand, means that there was an instance of debris disc formation. In a single migrating fragment scenario, that is when there is only one fragment produced by the parent disc, this would imply that gas giants and debris discs are mutually exclusive, but sub-Saturn planets and debris discs are uniquely linked. However, in a multi-fragment scenario, which is far more realistic based on numerical simulations of self-gravitating discs (\S \ref{sec:3D}), other fragments could undergo tidal disruptions and leave debris behind. Survival of a {\it detectable} debris disc to the present day also depends on where the disruption occurred, and the debris discs -- migrating gas fragment interactions, which are much more likely in Tidal Downsizing scenario than in Core Accretion because pre-collapse gas giants are widespread in Tidal Downsizing and traverse distances from $\sim 100$ AU to the host star surface. Therefore, we expect a significant wash-out of the single fragment picture, but some anti-correlation between debris discs and gas giants and the correlation between debris discs and sub-giants may remain. 

\subsection{Cores closest to their hosts}\label{sec:MM_valley}

\begin{figure}
\includegraphics[width=0.95\columnwidth]{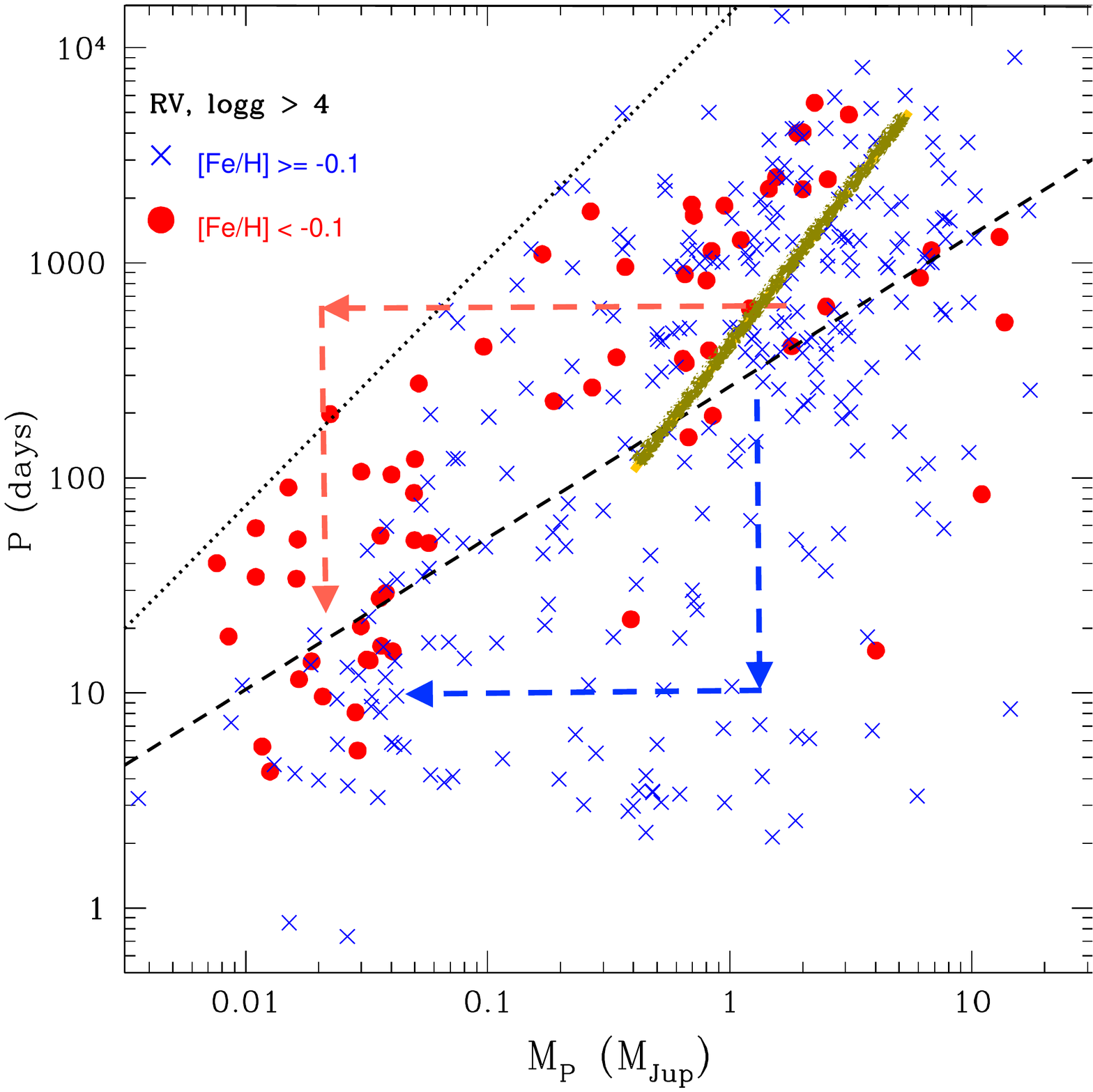}
\caption{The right panel of figure 1 from Adibekyan et al (2013), showing the planet period versus its mass. The sample is separated into the metal poor and metal rich sub-samples. The green, blue and red lines are added on the plot with permission from the authors. The green line is the exclusion zone boundary (eq. \ref{aex1}), which shows approximately how far a pre-collapse gas fragment of mass $M_{\rm p}$ can approach an $M_* = 1\msun$ star without being tidally disrupted. The blue and red lines contrast how gas fragments evolve in a metal-rich and a metal poor disc, respectively. See text in \S \ref{sec:MM_valley} for more detail.}
 \label{fig:Vardan}
 \end{figure}

\cite{AdibekyanEtal13}  shows that planets around low metallicity hosts tend to have larger orbits than their metal rich analogs. The trend is found for all planet masses where there is sufficient data, from $\sim 10 \mearth$ to $4\mj$. Their Fig. 1, right panel, reproduced here in Fig. \ref{fig:Vardan}, shows this very striking result. The dividing metallicity for the metal poor vs metal rich hosts was set at [M/H]$=-0.1$ in the figure. The figure was modified (see below) with permission. The blue crosses show metal rich systems whereas the red circles show metal-poor systems. \cite{AdibekyanEtal15} extended this result to lower mass/radius cores, showing that metal-rich {\it systems of cores} tend to be more compact than systems of planets around metal poor stars (see the bottom panels in their Fig. 1).

As noted by \cite{AdibekyanEtal13}, in the Core Accretion context, massive cores in metal poor discs are expected to appear later than they do in metal rich ones. At these later times, the protoplanetary discs may be less massive on average. Cores formed  in metal poor systems should therefore migrate slower (cf. eq. \ref{tmig1}). They also have less time before their parent gas discs are dissipated. Hence one may expect that cores made in metal-deficient environments migrate inward less than similar cores in metal-rich environment. 


However,  planet masses span a range of $\sim 1000$ in Fig. \ref{fig:Vardan}. This means that planet migration rates may vary by a similar factor -- from some being much longer than the disc lifetime, and for the others being as short as $\sim 10^4$ years. It is therefore not clear how a difference in timing of the birth of the core by a factor of a few would leave any significant imprint in the final distribution of planets {\it across such a broad planet mass range}.

In Tidal Downsizing, there is no significant offset in when the cores are born in metal rich or metal poor discs. All cores are born very early on. However, as described in \S \ref{sec:pebbles} and \ref{sec:giants_Z}, gas fragments in metal-poor discs tend to be disrupted by stellar tides when they migrate to separations of a few AU. This forms an exclusion zone barrier (cf. the red line in the right panel of fig \ref{fig:scatter} and the green line in Fig. \ref{fig:Vardan}), so that, as already explained (\S \ref{sec:cold_giants_Z}), moderately massive gas giants around metal poor hosts are to be found mainly above the green line in Fig. \ref{fig:Vardan}. Fragments in metal-rich systems, however, are more likely to contract and collapse due to pebble accretion {\it before} they reach the exclusion zone, so they can continue to migrate into the sub-AU regions. The exclusion zone hence forms a host metallicity dependent filter, letting gas giants pass in metal rich systems but destroying them in metal poor ones.

Further, as explained in \S \ref{sec:massive_giants_Z}, planets more massive than $\sim 5\mj$ cool rapidly radiatively, and thus they are able to collapse and pass the barrier without accreting pebbles. These high $M_{\rm p}$ planets are not expected to correlate wth [M/H] strongly at any separation (\S \ref{sec:massive_giants_Z}). This is consistent with Fig. \ref{fig:Vardan} -- note that a larger fraction of gas giants are metal-rich at high planet masses.

Let us now consider what happens with $M_{\rm p}\sim 1\mj$ fragments after they reach the exclusion zone in some more detail. The blue lines with arrows show what may happen to such a planet in the metal rich case. Since the planet is in the second, dense configuration, it may continue to migrate in as long as the gas disc is massive enough. The fragments will eventually enter the hot Jupiter regime (periods $P\lesssim 10$~days). Some just remain there when the gas disc dissipates; others are pushed all the way into the star. Yet others can be disrupted at about $a\sim 0.1$~AU by a combination of over-heating because of the very hot disc environment and disruption by stellar tides (this was called the "second disruption" in \S \ref{sec:2nd_dis}). The disrupted fragments then travel approximately horizontally in the diagram, as indicated by the blue horizontal arrow, becoming hot sub-Saturn or hot super Earth planets \citep{Nayakshin11b}.

In contrast, tidal disruption of gas fragments in metal-poor systems occurs at around the exclusion zone boundary. The planet also travels horizontally to lower planet mass regime, as shown with the horizontal red line in Fig. \ref{fig:Vardan}. After the disruption these low mass planets (usually dominated by massive cores), continue to migrate inward, now evolving vertically downward as shown in Fig. \ref{fig:Vardan} with the vertical red line. Planet migration rate in type I regime is relatively slow for core-dominated planets, thus one can then expect that the "red" cores will in general not migrate as far in as did the "blue" ones.


Focusing on the lowest mass cores, $M_{\rm core} \le 0.03 \mj \sim 10 \mearth$ in Fig. \ref{fig:Vardan}, we note quite clearly a dearth of metal rich (blue crosses) cores beyond the period of $\sim 10-20$ days, which corresponds to $a \approx 0.1 - 0.15$~AU. In principle, this could be a detectability threshold effect -- planets are progressively more difficult to detect at longer periods. However, the approximate (empirical) detection threshold is shown in the figure with the dotted line, which is a factor of several longer than the 10 day period; so these observational results are unlikely to be due to detection biases. 


Second disruptions have not yet been included in rigorous enough detail in the population synthesis.

%
%
%
%
%
%

\section{Planet compositions}\label{sec:comp}

\subsection{Metal over-abundance in gas giants}\label{sec:Zpl_giants}

Heavy element content of a giant planet can be found with some certainty by knowing just the planet  mass and radius \citep{Guillot05}, provided it is not too strongly illuminated \citep{MillerFortney11,ThorngrenEtal15}. Heavy elements contribute to the total mass of the planet, but provide much less pressure support per unit weight.


Metal over-abundance of gas giant planets is expected in Tidal Downsizing thanks to partial stripping of outer metal-poor layers \citep{Nayakshin10c} and pebble accretion. In \cite{Nayakshin15a}, it was estimated that accreted pebbles need to account for at least $\sim10$\% of planet mass for it to collapse via pebble accretion as opposed to the radiative channel. This number however depends on the mass of the fragment. As explained in \S \ref{sec:transition} and \S \ref{sec:massive_giants_Z}, more massive gas giants cool more rapidly at the same dust opacity. For this reason they are predicted to not correlate as strongly with the host star metallicity (see \S \ref{sec:massive_giants_Z}) {\it and} require less pebbles to accrete in order to collapse. 

\begin{figure}
 \includegraphics[width=1\columnwidth]{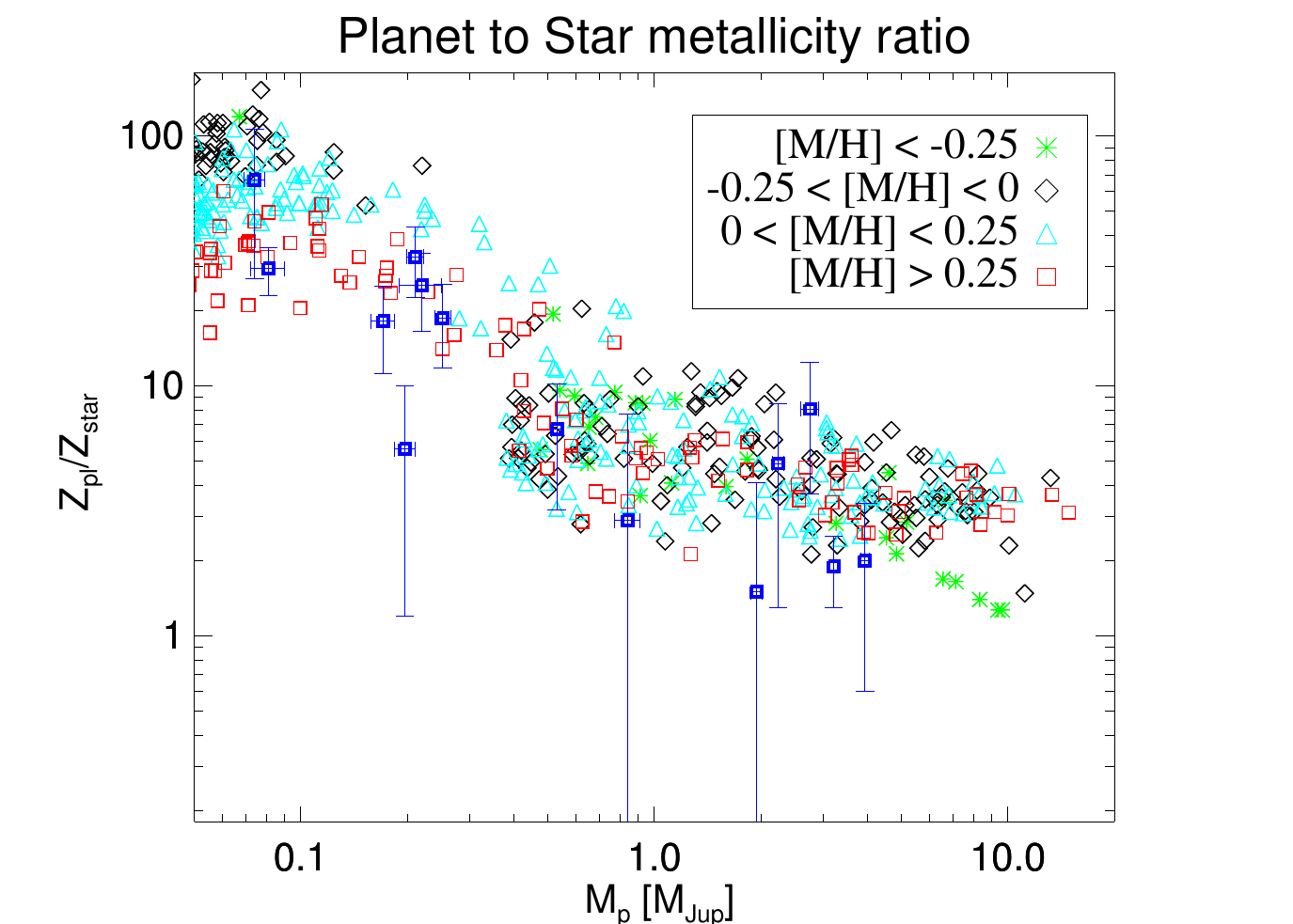}
 \caption{Metal over-abundance of gas giant planets versus their mass. Blue squares with error bars shows the results of Miller \& Fortney (2011). The other symbols are results from population synthesis, binned into four host star metallicity bins as detailed in the legend.}
 \label{fig:Zpl}
 \end{figure}

Fig. \ref{fig:Zpl} shows the relative over-abundance of gas giant planets, that is, the ratio $Z_{\rm pl}$ to star metal content, $Z_*$, as a function of the planet mass from population synthesis by \cite{NayakshinFletcher15} compared with the results of \cite{MillerFortney11}, who deduced metal content for a number of exoplanets using observations and their planet evolution code. No parameter of the population synthesis was adjusted to reproduce the \cite{MillerFortney11} results.

Fig. \ref{fig:Zpl} also shows that there is a continuous metallicity trend with $M_{\rm p}$ from $\sim 0.1 \mj$ all the way into the brown dwarf regime. The {\it continuous} transition in metal over-abundance from gas giants to brown dwarfs argues that brown dwarf formation may be  linked to formation of planets (\S \ref{sec:BD_vs_planets}).

\subsection{Core compositions}\label{sec:core_comp}


Tidal Downsizing predicts rock-dominated composition for cores \citep[][and \S \ref{sec:composition}]{NayakshinFletcher15}. Core Accretion scenario suggests that massive core formation is enhanced beyond the snow line since the fraction of protoplanetary disc mass in condensible solids 
increases there by a factor of up to $\sim 3$ \citep[e.g., see Table I in][]{PollackEtal96}. Most massive cores are hence likely to contain a lot of ice in the Core Accretion model. Reflecting this, Neptune and Uranus in the Solar 
System are often referred to as "icy giants" even though there is no direct observational support for their cores actually being composed of ice \citep[see \S 5.1.2 in][]{HelledEtal13a}. For example, for Uranus, the gravity and rotation data can be fit with models containing rock or ice as condensible material \citep{HelledEtal10}. When  SiO$_2$ is used to represent the rocks, Uranus interior is found to consist of 18\% hydrogen, 6\% helium, and 76\% rock. Alternatively, when H$_2$O is used, UranusÕ composition is found to be 8.5\% and 3\% of H and He, respectively, and 88.5\% of ice.

Composition of extrasolar cores is obviously even harder to determine. \cite{Rogers15} shows that most {\it Kepler} planets with periods shorter than 50 days are not rocky for planet radii greater than $1.6 R_\oplus$ as their density is lower than an Earth-like core would have at this size. Unfortunately, just like for the outer giants in the Solar System, the interpretation of this result is degenerate. It could be that these planets contain icy cores instead of rocky ones, but it is also possible that the data can be fit by rocky cores with small atmospheres of volatiles on top.

To avoid these uncertainties, we should focus on cores that are unlikely to have any atmospheres. Close-in ($a \lesssim 0.1$~AU) moderately massive cores \citep[$M_{\rm core}\lesssim 7\mearth$, see][]{OwenWu13} are expected to lose their atmosphere due to photo-evaporation. The observed close-in planets in this mass range all appear to be very dense, requiring Venus/Earth rock-dominated compositions \citep[e.g., Fig. 4 in][]{DressingEtal15}. \cite{EspinozaEtal16} present observations of a Neptune mass planet of radius $R_{\rm p}\approx 2.2 R_\oplus$, making it the most massive planet with composition that is most consistent with pure rock.   \cite{WeissEtal16} re-analyse the densities of planets in the Kepler-10 system and find that planet c has mass of $\approx 14\mearth$ and its composition is consistent with either rock/Fe plus $0.2$\% hydrogen envelope by mass or Fe/rock plus (only) 28\% water. There thus appears to be no evidence so far for ice-dominated massive cores in exoplanetary systems.

Another interesting way to probe the role of different elements in making planets is to look at the abundance difference between stars with and without planets. Observations show little difference in differential element abundances between "twin stars" {\it except} for refractory elements \citep{MaldonadoV16}, again suggesting that ices are not a major planet building material, whereas silicates could be. These results may be disputed, however, because the effects of Galactic stellar evolution \citep{GHEtal13} drive extra variations in abundance of metals. These effects are hard to deconvolve from the possible planet/debris disc formation signatures. 

Cleaner although very rare laboratories are the nearly identical "twin" binaries, which certainly suffer identical Galactic influences. \cite{SaffeEtal16} studies the $\zeta$~Ret binary which contains nearly identical stars separated by $\sim 4000$~AU in projection. One of the twins has a resolved debris disc of size $\sim 100$~AU \citep{EiroaEtal10}, whereas the other star has no planet or debris disc signatures. Refractory elements in the debris disc hosting star are deficient \citep{SaffeEtal16} compared to its twin by at least $3 \mearth$, which the authors suggest is comparable to the mass of solids expected to be present in a debris disc of this spatial size. Results of \cite{SaffeEtal16} are therefore consistent with that of \cite{MaldonadoV16} and could not be driven by the Galactic chemical evolution. This twin binary observation is especially significant since the debris disc size is $\sim 100$~AU, well beyond a snow line, so ices should be easily condensible into planets/debris. If ices were the dominant reservoir from which debris discs and planets are made then they should be missing in the star with the observed debris.

%
%
%
%
%

\section{Planet Mass Function}\label{sec:PMF}

\subsection{Mass function}\label{sec:CMF}

Small planets, with radius less than that of Neptune ($\sim 4 R_\oplus$) are ubiquitous \citep{HowardEtal12}. This planet size translates very roughly to mass of $\sim 20\mearth$ \citep{DressingEtal15}. Observations of close-in exoplanets show that planet mass function (PMF) plummets above this size/mass \citep{HowardEtal12,MayorEtal11}. These observations add to the long held belief, based on the Solar System planets'  observations, that theplanetary cores of mass $M_{\rm core}\sim 10-20 \mearth$ have a very special role to play in planet formation.

In the Core Accretion scenario, this special role is in building gas giant planets by accretion of protoplanetary disc gas onto the cores \citep{Mizuno80,PT99,Rafikov06}. In Tidal Downsizing, the role of massive cores in building gas giant planets is negative due to the feedback that the core releases (\S \ref{sec:feedback}). The observed dearth of gas giants and abundance of small planets means in the context of Tidal Downsizing that most of the gas fragments originally created in the outer disc must be disrupted or consumed by the star to be consistent with the data.

\begin{figure}
 \includegraphics[width=0.84\columnwidth]{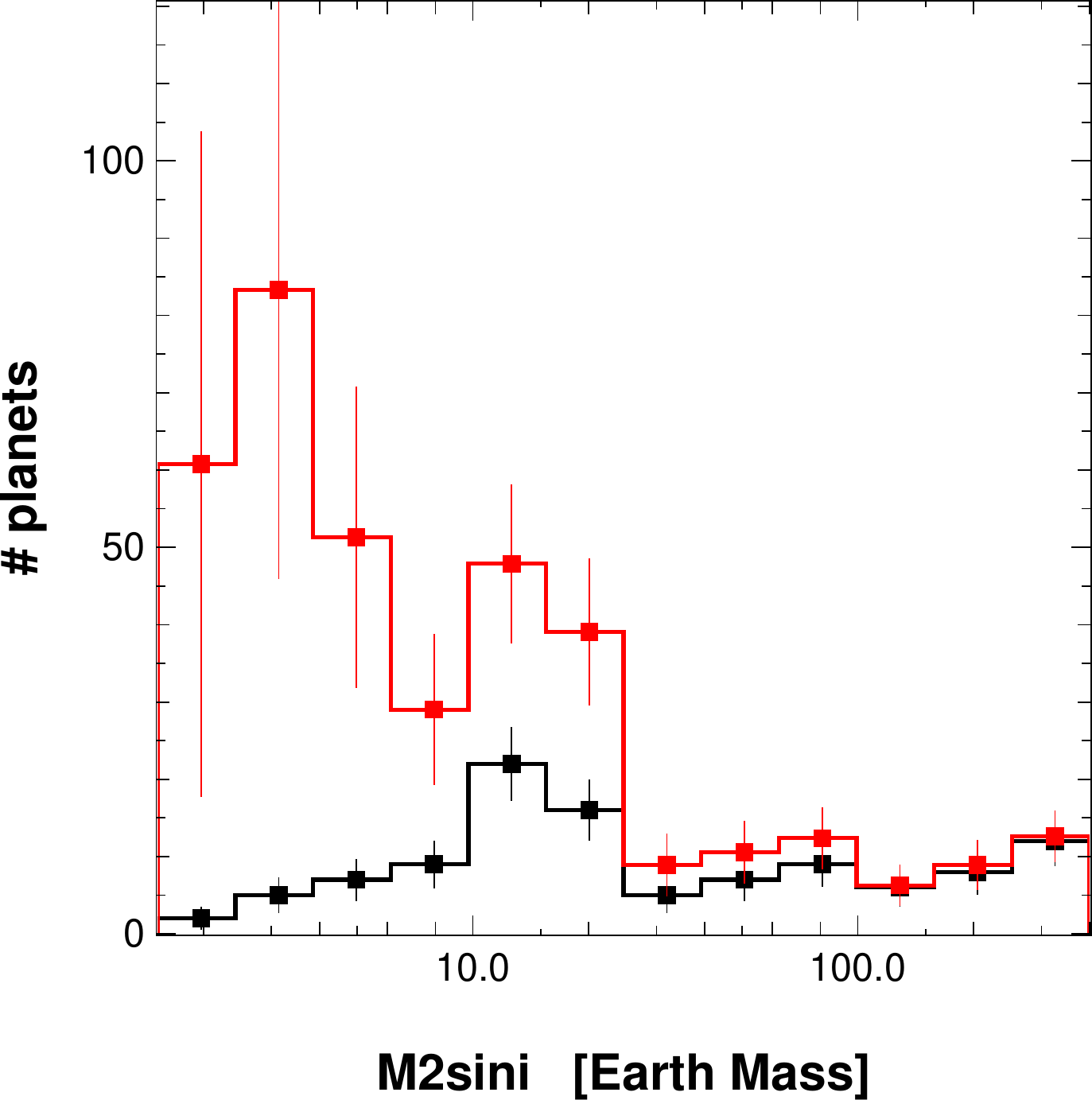}
 \medskip
 \includegraphics[width=1\columnwidth]{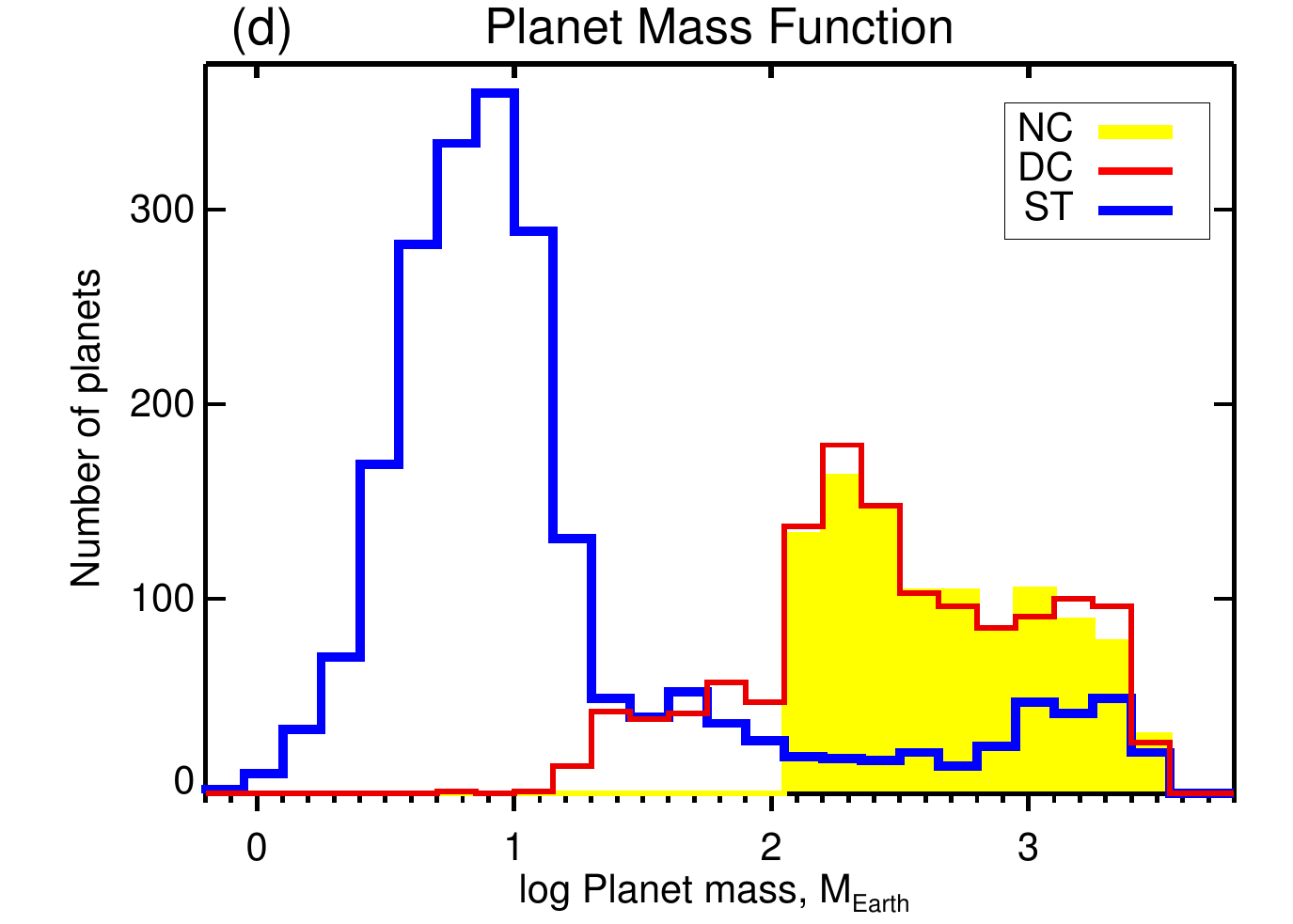}
 \caption{{\bf Top panel:} Planet mass function (PMF) from HARPS spectrograph observations from Mayor et al (2011). The black histogram gives  observed number of planets, whereas the red corrects for observational bias against less massive planets. {\bf Bottom panel:} PMF from the Tidal Downsizing population synthesis calculations, exploring the role of core feedback. The histograms are for runs without core formation (NC), with core formation but feedback off (DC) and standard (ST), which includes core feedback. Without feedback, the PMF of Tidal Downsizing scenario looks nothing like the observed mass function.}
 \label{fig:pmf_fb}
 \end{figure}
 
The top panel of Fig. \ref{fig:pmf_fb} shows the observed PMF from \cite{MayorEtal11}. The black shows the actual number of planets, whereas the red shows the PMF corrected for observational bias. The bottom panel of Fig. \ref{fig:pmf_fb} shows PMF from three population synthesis calculations performed with three contrasting assumptions about the physics of the cores \citep{Nayakshin16a} in the model, to emphasise the importance of core feedback in Tidal Downsizing. Simulation ST (standard) includes core feedback, and is shown with the blue histogram. This PMF is reasonably similar to the observed one in the top panel.

In simulation NC (no cores), shown with the yellow histogram, core formation is artificially turned off. In this case tidal disruptions of gas fragments leave behind no cores. Thus, only gas giant planets are formed in this simulation. In simulation DC (dim cores), shown with the red histogram in the bottom panel, core formation is allowed but the core luminosity is arbitrarily reduced by a factor of $10^5$ compared with simulation ST. 

By comparing simulations ST and DC we see that the core luminosity is absolutely crucial in controlling the kind of planets assembled by the Tidal Downsizing scenario. A strong core feedback leads to a much more frequent gas fragment disruption, reducing the number of survived gas fragments at all separations, small or large. This also establishes the maximum core mass ($10-20\mearth$, eq. \ref{Mcrit} and \S \ref{sec:feedback}), above which the cores do not grow because the parent clumps cannot survive so much feedback.

In simulation DC (dim cores),  cores grow unconstrained by their feedback and so they become much more massive \citep[see also Fig. 5 in ][on this]{Nayakshin16a} than in simulation ST, with {\it most} exceeding the mass of $10 \mearth$. Given that they are also dim, these cores are always covered by a massive gas atmosphere even when the gas fragment is disrupted (cf. the next section). This is why there are no "naked cores" in simulation DC.

One potentially testable prediction is this. As core mass approaches $\sim 10 \mearth$, feedback by the core puffs up the fragment and thus $dM_{\rm core}/dt$ actually drops. Therefore, growing cores spend more time in the vicinity of this mass. Since 
core growth is eventually terminated  by the fragment disruption or by the second collapse, whichever is sooner, the mass of cores should cluster around this characteristic mass. In other words, the core mass function should show a peak at around $\sim 10\mearth$ before it nose-dives at higher masses. 

There may be some tentative evidence for this from the data.  \cite{SilburtEtal15} looked at the entire {\it Kepler} sample of small planets over all 16 quarters of data, and built probably the most detailed to date planet radius function at $R_{\rm p} \le 4 R_\oplus$. They find that there is in fact a peak in the planet radius distribution function at $R_{\rm p} \approx 2.5 R_\oplus$, which corresponds to $M_{\rm core} \approx 15 \mearth$. 

\subsection{Atmospheres of cores: the bimodality of planets}\label{sec:atmo}

One of the most famous results of Core Accretion theory is the critical mass of the core,  $M_{\rm crit} \sim$ a few to $\sim 10-20 \mearth$, at which it starts accreting gas from the protoplanetary disc \citep{Mizuno80,Stevenson82,IkomaEtal00,Rafikov06,HoriIkoma11}.  For core masses less than $M_{\rm crit}$, the cores are surrounded by usually tiny atmospheres.

In \S \ref{sec:atm} it was shown that a massive core forming inside a self-gravitating gas fragment in the context of Tidal Downsizing also surrounds itself by a dense gas atmosphere for exactly same reasons, except that the origin of the gas is not the surrounding protoplanetary disc but the parent fragment. \cite{NayakshinEtal14a} calculated the atmosphere structure for a given central properties of the gas fragment (gas density, temperature, composition), core mass and luminosity. The population synthesis model of \cite{Nayakshin15c,NayakshinFletcher15} uses the same procedure with a small modification. To determine the mass of the atmosphere actually bound to the core, I consider the total energy of atmosphere shells. Only the innermost layers with a negative total energy are considered bound to the core. These layers are assumed to survive tidal disruption of the fragment.

\begin{figure}
 \includegraphics[width=1\columnwidth]{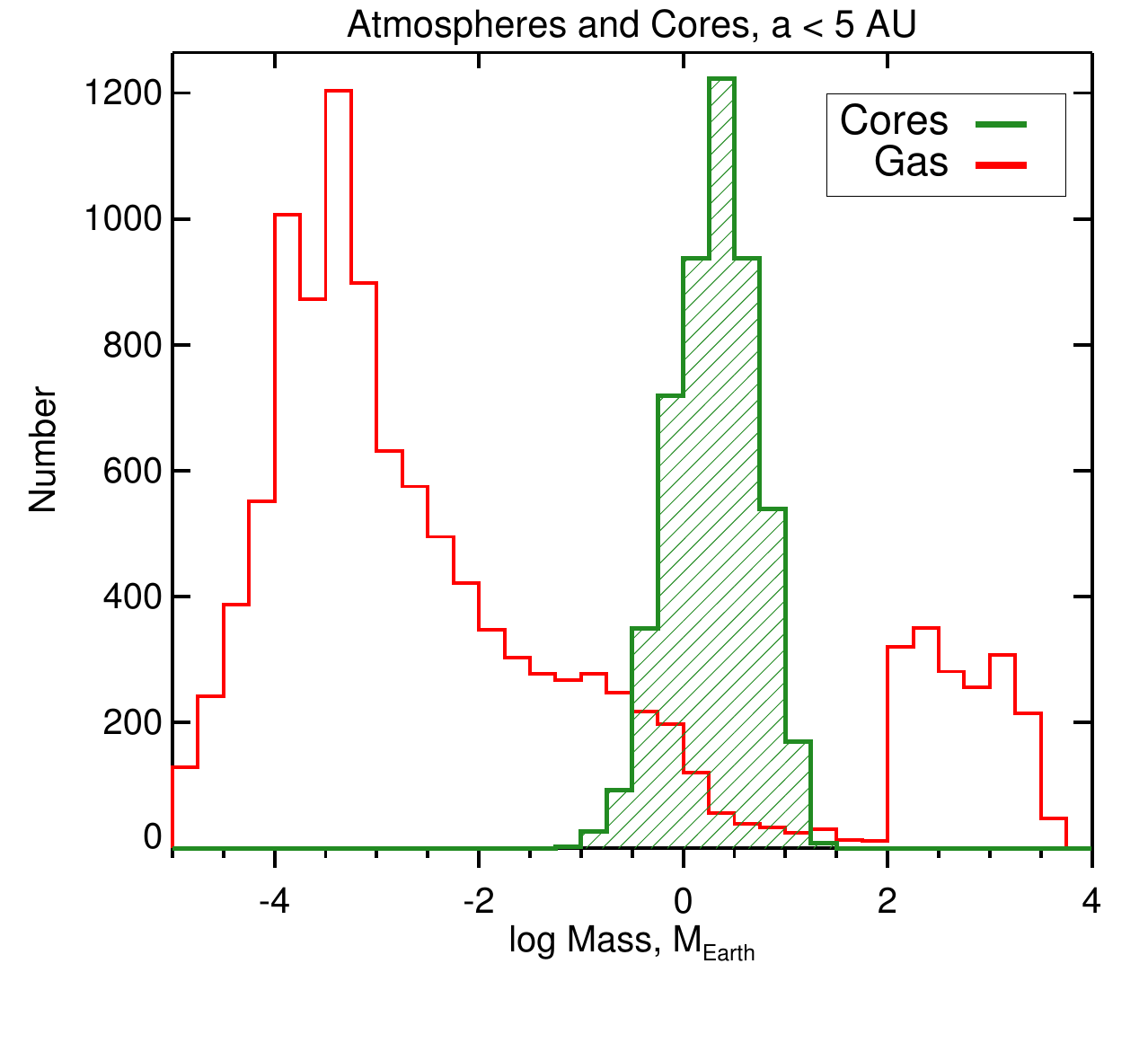}
 \caption{The distribution of core and gas masses for planets in the inner 5 AU from population synthesis calculations of Nayakshin \& Fletcher (2015). Note that the planets are either core-dominated with tiny atmospheres or gas giants. See \S \ref{sec:atmo} for more detail}
 \label{fig:atmo}
 \end{figure}

Figure \ref{fig:atmo} is reproduced from \cite{NayakshinFletcher15}, and shows the mass of all of the cores in the inner 5 AU from the host at the end of the simulations (green shaded), while the red histogram shows the mass distribution of {\it gas} in the same planets. Gas fragments that were not disrupted remain in the Jovian mass domain, within the bump at $\log (M_{\rm gas}/\mearth) > 2$. These planets are dominated by the gas but do have cores. The second, much more populous peak in the red histogram in Fig. \ref{fig:atmo} is at tiny, $\sim 10^{-3}\mearth$ masses. This peak corresponds to the gas fragments that were disrupted and became a few Earth mass cores with the small atmospheres.

Tidal Downsizing scenario thus also naturally reproduces the observed bi-modality of planets -- planets are either dominated by cores with low mass (up to $\sim 10$\% of core mass, generally) atmospheres, or are totally swamped by the gas. The conclusion following from this is that the special role of $\sim 10\mearth$ cores in planet formation  may dependent on how the planets are made only weakly. It is likely that the ability of massive ($M_{\rm core} \gtrsim 10\mearth$) cores to attract gas atmospheres of comparable mass is a fundamental property of matter (hydrogen equation of state, opacities) and {\it does not tell us much about the formation route of these planets}, at least not without more model-dependent analysis.

 %
 %
 %
 %
 %
 %
 %

 \section{Distribution of planets in the separation space}\label{sec:radial}

\subsection{Period Valley of gas giants}\label{sec:Pvalley}

The radial distribution of gas giant planets has a "period valley" at $0.1 < a < 1$ AU \citep{CummingEtal08}, which was interpreted as a signature of protoplanetary disc dispersal by \cite{Alexander2012}. In their model, photo-evaporation removes disc gas most effectively from radii of $\sim 1-2$~AU for a Solar type star, hence creating there a dip in the surface density profile. Therefore, planets migrating from the outer disc into the sub-AU region may stall at $a\sim 1-2$~AU and thus pile up there. 

The period valley issue has not yet been studied in Tidal Downsizing, but preliminary conclusions are possible. The photo-evaporation driven process of stalling gas giant planets behind $\sim 1$~AU should operate for both planet formation scenarios because it has to do with the disc physics. However, the timing of gas giant planet formation is different in the two models. Core Accretion planets are born late in the disc life, when the disc has lost most of its mass through accretion onto the star. Tidal Downsizing fragments are hatched much earlier, when the disc is more massive. Most of Tidal Downsizing fragments hence migrate through the disc early on, well before the photo-evaporative mass loss becomes important for the disc. During these early phases the disc surface density profile does not have a noticeable depression at $\sim 1-2$~AU \citep[see][]{Alexander2012}. Therefore, the photo-evaporative gap is probably not as efficient at imprinting itself onto the gas giant period or separation distribution in Tidal Downsizing as it is in the Core Accretion.

However, the exclusion zone boundary at $\sim 1$ to a few AU is a hot metallicity-dependent filter for the gas giant planets (\S \ref{sec:giants_Z} \& \ref{sec:MM_valley}). Current population synthesis calculations in the Tidal Downsizing scenario show that the surface density of planets decreases somewhat at $\sim 1$~AU for all masses $M_{\rm p}\gtrsim 1\mj$ (cf. Fig. \ref{fig:scatter}), and this effect is dominated by the tidal disruptions. The period valley should thus be stronger for metal poor hosts than for metal-rich ones in Tidal Downsizing scenario.

\subsection{On the rarity of wide separation gas giants}\label{sec:wide}

Although there are some very well known examples of giant planets orbiting Solar type stars at separations of tens to $\sim 100$~AU, statistically there is a strong lack of gas giant planets observed at wide separations \citep[e.g.,][]{ViganEtal12,ChauvinEtal15,BowlerEtal15}. For example, \cite{BillerEtal13}, finds that no more than a few \% of stars host $1-20\mj$ companions with separations in the range $10 - 150$~AU.  \SNc{\cite{GalicherEtal16} makes the most definitive statement, finding that the fraction of gas giants beyond 10 AU is $\approx 1$\%.} 

\SNc{Current population synthesis models \citep[e.g.,][]{Nayakshin16a} exceed these constraints by a factor of a few to 10. This may be due to (a) the models assuming migration rates slower than the 3D simulations find (\S \ref{sec:rapid}), so that more population synthesis planets remain at wide separations after the disc is removed; (b) neglect of gas accretion onto the planets which could take some of them into the brown dwarf regime (\S \ref{sec:AorM} and \S \ref{sec:desert}); (c) too rapid removal of the outer disc in the models. These issues must be investigated in the future with both 3D simulations and population synthesis.}

%
%
%
%
%
%

\section{The HL Tau challenge}\label{sec:HLT}


HL Tau is a young ($\sim 0.5-2$ Myr old) protostar that remains invisible in the optical due to obscuration on the line of sight, but is one of the brightest protoplanetary
discs in terms of its millimetre radio emission \citep{AndrewsWilliams05,KwonEtal11}. For this reason, Atacama Large Millimetre/Submillimetre Array (ALMA) observed HL Tau as one of the first targets, in the science verification phase, with baseline as long as 15 km \citep{BroganEtal15}. This yielded resolution as small as 3.5 AU at the distance for the source, and resulted in the first ever {\it image} of a planet forming disc. The image of HL Tau shows a number of circular dark and bright rings in the dust emissivity of the disc. Such rings  can be opened by embedded massive planets \citep[e.g.,][]{LinPap86,RiceEtal06,CridaEtal06}.

Note that it is the dust emission that observable in the radio continuum, the gas of the disc can only be traced by its CO and HCO$^+$ line emission. \cite{PinteEtal16} performed a detailed modelling of the dust component in HL Tau disc assuming circular orbits for the gas. The well-defined circular gaps observed  at all azimuthal  angles (HL Tau disc is inclined to the line of sight) imply that $\sim$ millimetre sized dust has settled in a geometrically thin, $H_{\rm dust}/R \sim 0.02$, disc. This is much thinner than the gas disc which has $H/R \sim 0.1$ at these radii. The strong degree of grain settling sets an upper limit on the viscosity coefficient of the disc, requiring $\alpha \sim 3\times 10^{-4}$. The observed CO and HCO$^+$ line profiles constrain the protostar mass, $M_* =1.7 \msun$. \cite{PinteEtal16} find hotter gas disc than \cite{ZBB15}, who argued that the observed rings are formed by grain condensation at ice lines of abundant molecular species, and therefore their condensation fronts do not coincide with the gaps' positions. The small but non zero eccentricity of the rings, the surprisingly small magnitude of disc viscosity, coupled with irregular spacings of the rings, probably rule out Rossby wave instabilities or zonal flows \citep{PinillaEtal12} as possible origins of the rings, leaving planets as the most likely origin of the gaps \citep{BroganEtal15}.

A number of authors performed detailed coupled gas-dust hydrodynamical simulations to try to determine the properties of planets that are able to open gaps similar to those observed in HL Tau \citep{DipierroEtal15,JinEtal16,PicognaK15,DipierroEtal16a,RosottiEtal16}. The main conclusion from this work is that the minimum planet mass to produce the observed signatures is $\sim 15 \mearth$, while the maximum appears to be around $0.5 \mj$. \cite{DipierroEtal16a} find that the best match to the data is provided by planets of mass $M_{\rm p} \approx 20 \mearth$, $30 \mearth$ and $0.5\mj$ orbiting the star at orbits with semi-major axes of $a \approx 13$, 32 and 69 AU, respectively.

These results challenge classical ideas of planet formation. It should take $\sim 100$ Myr to grow massive cores at tens of AU distances from the star via planetesimal accretion \citep[e.g.,][]{KobayashiEtal11,KB15}. The presence of massive cores in a $\sim 1$~Myr old disc at $\sim 70$~AU is unexpected and also contradicts the metallicity correlations scenario presented by \cite{IdaLin04a,IdaLin04b,MordasiniEtal09b,MordasiniEtal12}. In that scenario, core growth takes $\sim $ 3-10 Million years at separations $a\lesssim 10$~AU, which should be much faster than core growth at 70 AU. Therefore, in the Core Accretion framework, HL Tau observations strongly favour assembly of cores via pebble accretion \citep[e.g.,][]{OrmelKlahr10,LambrechtsJ12,JohansenEtal15b} rather than by the standard planetesimal accretion \citep{Safronov72}. 

Further, planets with masses greater than $10-15\mearth$ should be accreting gas rapidly \citep[e.g.,][]{PollackEtal96}. The largest problem here is for the outermost planet whose mass is estimated at $M_{\rm p} \sim 0.5\mj$. Such planets should be in the runaway accretion phase where gas accretion is limited by the supply of gas from the disc \citep[e.g.,][]{HubickyjEtal05}. Using equation (34) of \cite{GoodmanTan04} to estimate the planet accretion rate, $\dot M_{\rm p} \sim \Sigma \Omega_K R_{\rm H}^2$, we find that
\begin{equation}
\dot M_{\rm p} \sim 2\times 10^{-4} {\mj \over \text{yr}}\; {M_{\rm d} \over 0.03 \msun} \left({M_{\rm p}\over 0.5 \mj}\right)^{2/3}.
\label{mdotp}
\end{equation}
On the other hand, the accretion rate onto the planet should not be much larger than $\sim M_{\rm p}/(1 Myr) = 5 \times 10^{-6} \mj/$~yr, where 1 Myr is the planet likely age. Thus the accretion rate onto the $a\sim 70$~AU planet must be much smaller than the classical planet assembly picture predicts \citep{PollackEtal96}.

Classical Gravitational disc Instability model of planet formation also may not explain formation of the observed HL Tau planets because the innermost planets are too close in and their mass is much too low to form by direct gravitational collapse.

Tidal Downsizing predicts planets with properties needed to understand the observations of HL Tau \citep{Nayakshin16a}.  In \S \ref{sec:feedback} it was shown massive cores, $M_{\rm core}\sim 10\mearth$, release enough accretion energy to puff up the gas envelopes of $M_{\rm p}\sim 1 \mj$ pre-collapse gas fragments, and eventually destroy them.  Population synthesis calculations show that massive cores located at distances of tens of AU from the host star is a very frequent outcome (cf. the right panel of Fig. \ref{fig:scatter}), made even more frequent in realistic discs if dozens of fragments are born initially in its outskirts. The outermost planet in this picture has not yet (or will not) be disrupted because its core is not massive enough. It does not accrete gas as explained in \S \ref{sec:AorM}.

%
%
%
%
%
%
%
%

\section{Kepler-444 and other highly dynamic systems}\label{sec:kepler444}

\begin{figure*}
\includegraphics[width=0.95\columnwidth]{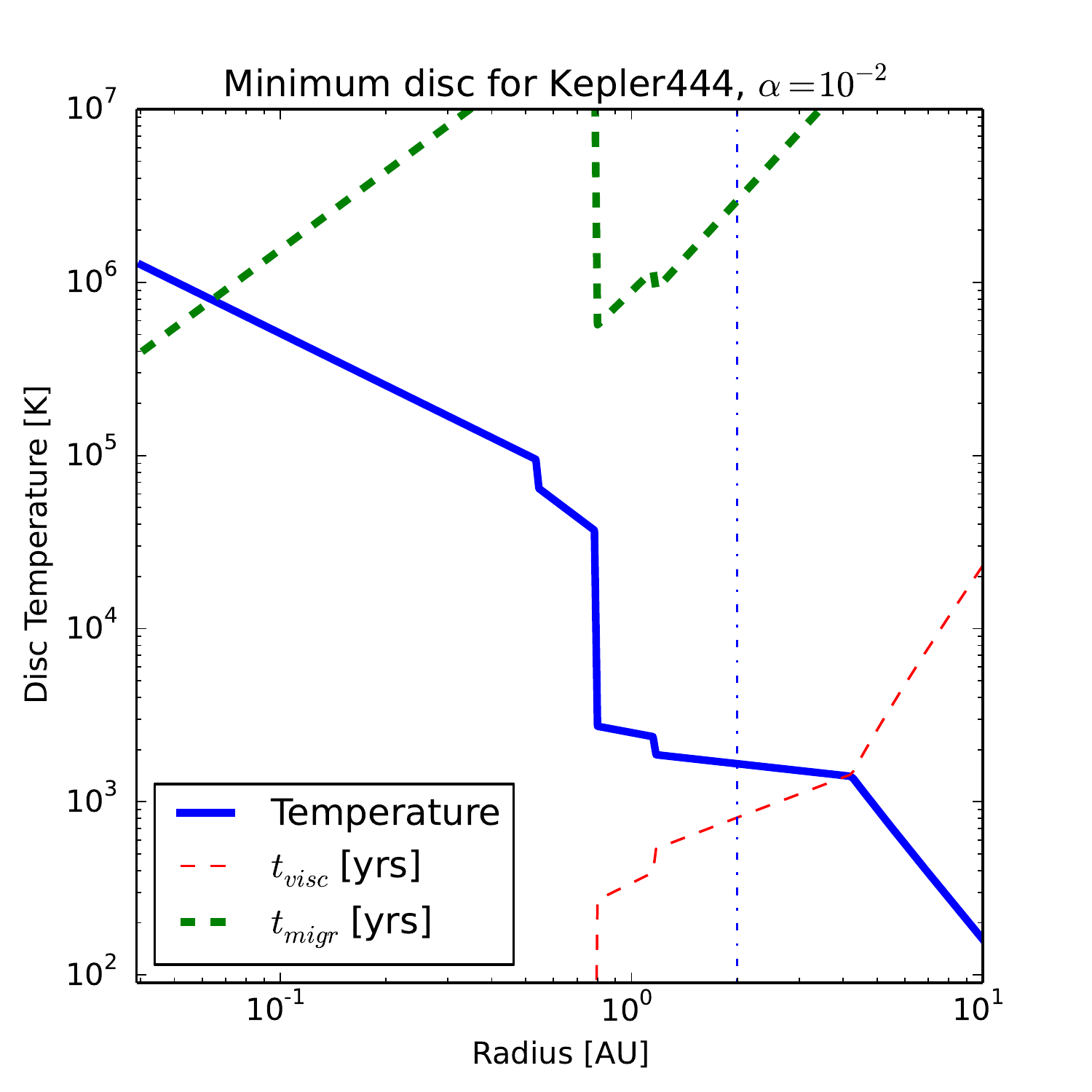}
 \includegraphics[width=0.95\columnwidth]{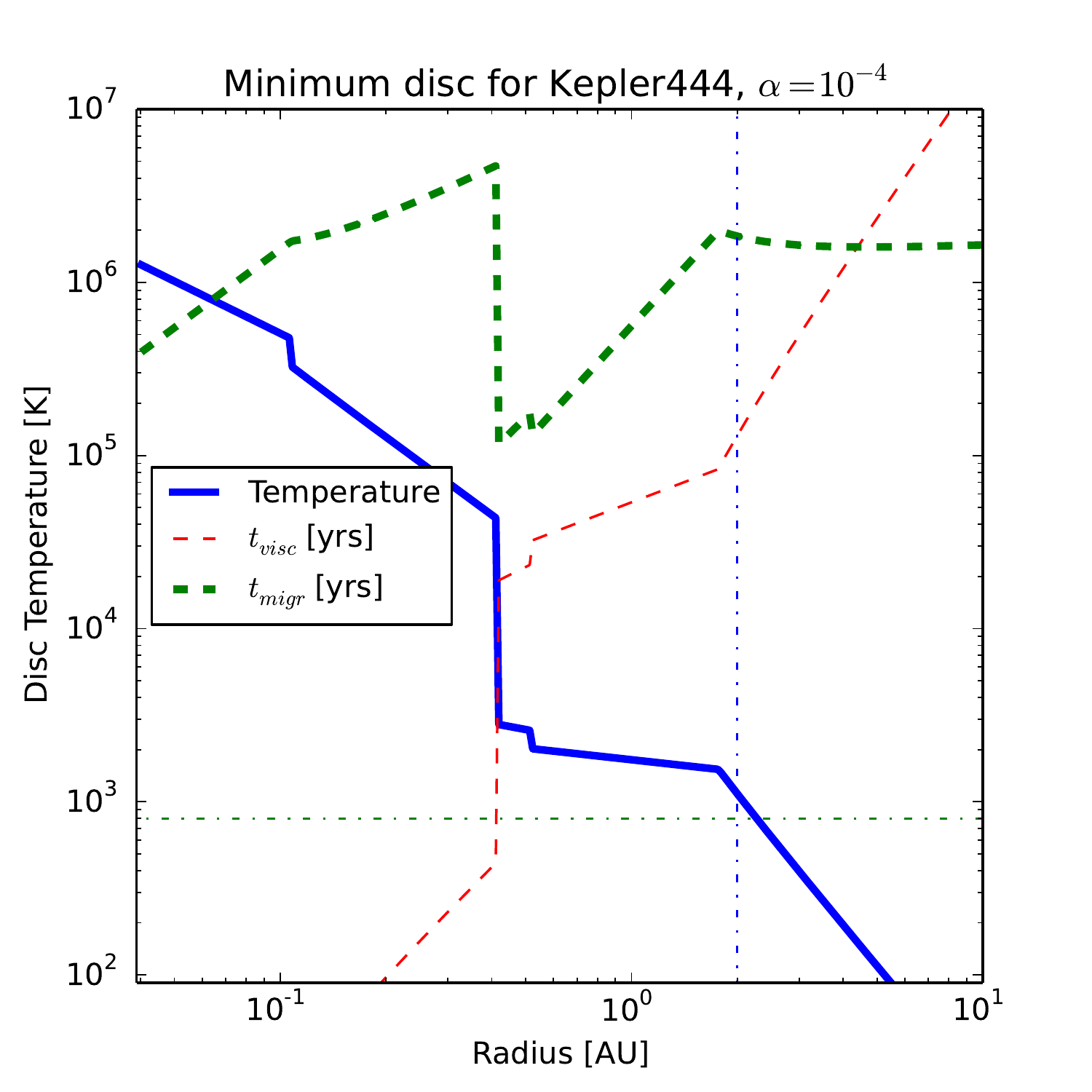}
 \caption{Minimum disc models for Kepler-444 system. {\bf Left:} Disc viscosity coefficient $\alpha=10^{-2}$. {\bf Right:} Same but for $\alpha =10^{-4}$.
 Solid curve shows disc midplane temperature,  while the dashed red and green show the disc viscous time and Kepler-444b migration time scales, respectively. Kepler-444 planetary system could not have formed anywhere inside $2$~AU disc.}
 \label{fig:Kep444}
 \end{figure*}

Kepler-444A is a solar type star with mass of $M_A =(0.76\pm 0.03)\msun$ widely separated from a tightly bound pair of M dwarf stars B and C with almost equal masses, $M_B+M_C \approx (0.54\pm 0.05)\msun$ \citep{CampanteEtal15}. The upper limit on separation of stars B \& C is 0.3 AU. The projected current separation of A and BC pair is $\approx 66$~AU. 
Star A has a very low metallicity, [Fe/H]~$\approx -0.69\pm 0.09$ which means that the metal content of the disc around A should have been $ 10^{0.7} \approx 5$ times lower than would be in a Solar composition disc \citep{CampanteEtal15}. Kepler-444A is orbited by 5 rather small planetary companions at separations ranging from 0.04 AU to 0.08 AU, with planet radii ranging from $0.4 R_\oplus$ to $0.74 R_\oplus$.

\cite{DupuyEtal16} were able to measure an unexpectedly small astrometric motion for the stellar system A-BC, suggesting that its orbit is very eccentric. They also measure a change in the radial velocity of the A-BC orbit, which allows the authors to constrain the orbit eccentricity as $e = 0.86\pm 0.02$. The pericentre separation of A-BC is only $a_{\rm peri} = 5\pm 1$~AU. The orbital planes of the planetary system and the stellar components coincide within a few degrees \citep{DupuyEtal16}. 

This high degree of the orbital alignment argues against the pair BC being captured in some kind of an N-body encounter after the planetary system formation \citep{DupuyEtal16} and is more likely to mean that the planets and the M dwarf pair were formed during a phase when a gas disc of some kind connected all the components of this puzzling system. 

The minimum mass of gas from which the $\approx 1.5 \mearth$ worth of planets in the system were made is approximately 5 Jupiter mass for Kepler-444A. In this estimate it is assumed that planets' composition is Earth-like, given that small exoplanets observed within $0.1$~AU appear to be very dense \citep[see][and discussion in \S \ref{sec:core_comp}]{RappaportEtal13,DressingEtal15}. Assuming that "only" half of refractories in the disc gets locked into the observed planets, we require a disc of initial mass $M_{\rm min}  = 10 \mj$ around Kepler-444A for the planets to be made.

We can now discuss at what separation from the star these planets could have formed. Suppose that the disc size was $R$ at the time of planet formation. This yields the disc surface density, $\Sigma \sim M_{\rm min}/(\pi R^2)$, at that radius. Assuming a value for the disc viscosity coefficient $\alpha < 1$, we can then calculate the disc midplane temperature and other interesting parameters from the \cite{Shakura73} disc theory. Of particular interest are the disc accretion rate, $\dot M$, and the scale-height $H$. Knowing these two we can calculate the disc viscous timescale, $t_{\rm visc} = M_{\rm min}/{\dot M}$, and the type I migration time for the planets (eq. \ref{tmig1}).

Figure \ref{fig:Kep444} presents two such calculations, for two different values of the viscosity parameter, $\alpha = 10^{-2}$ and $\alpha = 10^{-4}$ for the left and the right panels, respectively. The solid blue, the dashed red and green curves show the disc midplane temperature, the viscous and the (smallest) planet migration time scales, respectively, all as functions of distance $R$ from the star A. 

\subsection{In situ formation at $a\sim 0.04-0.1$~AU}\label{sec:Kep444_insitu}

The most obvious conclusion is that Kepler-444 planets could not have formed in situ as the gas would be simply too hot. $10\mj$ of gas at radii $R\lesssim 0.1$~AU yields a very large disc surface density $\gtrsim 5 \times 10^6$ g~cm$^{-2}$. This is larger than the disc surface density at which hydrogen in the disc must transition to the fully ionised state, that is, the upper branch of the well known "S-curve" for the disc \citep[][see point A in Fig. 1 of the latter paper]{Bell94,LodatoClarke04}, even for $\alpha$ as small as $10^{-4}$. In fact, with opacities from \cite{ZhuEtal09} that include more grain species than the \cite{Bell94} opacities did, the disc is even hotter and so I find the transition to the unstable branch at somewhat lower $\Sigma$ than given by eq. 6 in \cite{LodatoClarke04}.

As is well known from previous work, such values of $\Sigma$ would result in FU Ori like outbursts \citep[see \S \ref{sec:rapid} and \ref{sec:2nd_dis} and ][]{HK96,ArmitageEtal01}, during which even the surface layers of the disc are {\it observed} to be as hot as $\sim (2-5) \times 10^3$~K out to radii of $\sim 0.5-1$ AU \citep{EisnerH11}. In fact time-dependent model of discs push the disc onto the very hot branch for an order of magnitude lower values of the disc surface densities \citep[see figs. 13-16 in][]{NayakshinLodato12}.

At disc midplane temperature as high as $10^5$ or more Kelvin, not only grains but even km-sized or larger planetesimals will not survive for long\footnote{Interested reader may request detail of the calculation from the author}.

\cite{ChiangLaughlin13} propose that super-Earth mass planets orbiting their  host stars at separation $a$ as small as $0.1 AU$ formed in situ. However,\cite{ChiangLaughlin13} assume that the disc midplane temperature is 1000 K. Here, the accretion disc theory was used to evaluate the temperature for the requested $\Sigma$, and it is concluded that not only dust but planetesimals would be vaporised rapidly in the inner sub-AU region on Kepler-444. The nearly isothermal $T\sim 10^3$ K zone to which \cite{ChiangLaughlin13} appeal based on work of \cite{DAlessioEtal01} only exists for disc surface densities smaller than those needed for in-situ planet assembly inside $0.1$~AU by 2-3 {\it orders of magnitude} (see figs. 3-5 in the quoted paper). 

\subsection{Forming the planets in a few AU disc}\label{sec:Kep444_few_au}

We now assume that Kepler-444 planets must have migrated from further out. Let us try to estimate the minimum radius beyond which they could have formed. We have the usual constraint that the disc must be cooler than about 1500 K. In addition, the outer radius of the disc would have been truncated by the tidal torques from the Kepler-444BC pair, so that the outer radius of the disc, $R_{\rm out}$, is likely to be between 1 and 2 AU \citep{DupuyEtal16}. The vertical dot-dash line in Fig. \ref{fig:Kep444} shows $R_{\rm out} = 2$~AU constraint. This introduces two additional constraints: (1) the disc must be cold enough for dust coagulation {\it within} $R_{\rm out}$ and (2) the planet migration time to their final positions should be shorter than the disc viscous time.  Since the disc has a finite extent, there is a finite amount of mass, and once that gas accretes onto the Kepler-444A there is no more disc to keep pushing the planets in.

For the second constraint, it is the least massive planet Kepler-444b, the innermost one at $a=0.04$~AU with planet radius $R_{\rm p} =0.4 R_\oplus$ that places the tightest constraint since migration timescale in type I $\propto M_{\rm p}^{-1}$ (eq. \ref{tmig1}). The planet radius is just $\sim 5$\% larger than that of Mercury, whose mass is $M_{\rm p} = 0.055 \mearth$. I therefore estimate Kepler-444b mass as $M_{\rm p} = 0.07\mearth$.

Focusing first on the larger $\alpha$ case, the left panel of Fig. \ref{fig:Kep444}, we note that the disc is too hot in the inner few AU to allow grains of any composition to get locked into larger objects. Furthermore, even if it were possible to form Kepler-444b in such a disc, planet migration time is $\gtrsim 10^6$ years whereas the disc viscous time is just thousands of years or less \citep[again, recall that such high values of $\Sigma$ are above those needed to power FU Ori outbursts, which are known to wane rapidly by damping most of the disc mass onto the star; see ][]{LodatoClarke04}. Therefore, values of $\alpha$ as large as $10^{-2}$ are ruled out for Kepler-444 planetary system.

Shifting the focus to the right panel of Fig. \ref{fig:Kep444} now, the situation is somewhat better for $\alpha=10^{-4}$ but $t_{\rm visc}$ is still shorter than the migration time for Kepler-444b by more than an order of magnitude. Continuing the game of lowering $\alpha$, it is found that the value of $\alpha \lesssim 3\times 10^{-5}$ finally satisfies both constraints (1) and (2). 

Unfortunately, such a low viscosity parameter is not expected for discs hotter than about 800-1000 K because the ionisation degree of the gas becomes sufficiently high \citep{Gammie96,ArmitageEtal01} and the disc becomes MRI-active. Observations of Dwarf Novae systems show that $\alpha \gtrsim 0.1$ in the ionised state; even in quiescence, when H$_2$ molecules dominates the disc, the inferred values of $\alpha\gtrsim 0.01$ \citep[see][]{KingLLP13}. The corresponding region where the disc could be sufficiently cold for the disc to be "dead" is $R \gtrsim 2$~AU, clashing with condition (2). Therefore, there appears to be no corner in the parameter space $\alpha$ and $R < R_{\rm out}$ that would satisfy all the observational and physical constraints on formation of Kepler-444 planets.

\subsection{A TD model for Kepler-444 system}\label{sec:Kep444_TD}

Clearly, a detailed 3D simulation is desirable to study any formation scenario of this highly dynamic system. In the absence of such, any preliminary formation scenario that does not appear to contradict basic physics of star and planet formation is still a step in the right direction. 

Stars grow by gas accretion on first cores, first hydrostatic condensations of gas that form when the parent molecular cloud collapses \citep[][see also \S \ref{sec:term}]{Larson69}. First cores start off being as large as $\sim 10$~AU, and contract as they accrete more gas. This large initial size of the first cores suggests that the A -- BC system is unlikely to have formed on its present orbit because the peri-centre of the orbit is just 5 AU. 

More likely, the parent gas reservoir from which the triple star system formed had a strong $m=2$ perturbation \citep['bar type' in terminology of][]{MH03} which is best described as a filament. Filaments are observed in collapsing molecular clouds, see, e.g., \cite{HacarT11}.  For Kepler-444, the two main self-gravitating centres corresponding to A and BC could have formed on opposing sides of the filament/bar, roughly at the same time. They were probably separated initially by $R_{\rm bin,0}\sim 10^3$~AU or more. 

With time these two self-gravitating centres coalesce as the filament collapses along its length. Dissipation and accretion of gas onto the growing proto-stars shrinks the binary \citep[e.g.,][]{BateB97} on the timescale of a few free fall times from $R_{\rm bin,0}$, $t_{\rm ff} \sim R_{\rm bin,0}^{3/2}/(GM_{444})^{1/2} \sim 5\times 10^3 (R_{\rm bin,0}/1000)^{3/2}$ years, where $M_{444} = 1.3\msun$. This means that during some $10^4$ years the systems A and BC evolve independently, accreting gas mainly from their immediate environment rather than exchanging it. 

If star A possessed a disc larger than $\sim 30$~AU, the disc may fragment on multiple fragments. Migration of gas fragments from those distances would take only $\sim 1000$ years in a strongly self-gravitating disc (\S \ref{sec:rapid}). The fragments are presumably disrupted in the inner disc and leave behind their low mass cores -- ready made planets Kepler-444b though Kepler-444f.

When the filament collapses, and the configuration of A-BC system becomes comparable to the current one, the planets are already in the inner $\sim 1$~AU region from star A. Their eccentricities are pumped up every time BC passes its pericentre, but the gas disc acts to dump their eccentricities and in doing so forces the planets to migrate in faster than the type I rate. The eccentricity dumping time scale for type I migrating planets is known to be shorter by as much as factor $(H/R)^2$ than the canonical migration time scale for circular orbits \citep[e.g.,][]{BitschKley10}. This mechanism may perhaps bring the planets to their current location faster than the disc would dissipate. 

Note that eccentricity pumping migration scenario proposed here would not work for the classical Core Accretion scenario cores because  core growth by planetesimal accretion would be too slow for the eccentric orbits.

%
%
%
%
%
%
%
%
%
%
%
%

\section{The Solar System}\label{sec:SS}


In \S \ref{sec:SS_basic}, a schematic model for formation of the Solar System (SS) was presented. The main difference of the Solar System from many of the exoplanetary systems observed to date, many of which have very close-in planets, is that the Solar System protoplanetary disc should have been removed before the planets had time to migrate closer to the Sun. 

\subsection{Rotation of planets}\label{sec:SS_rotation}\label{sec:CA_spin}

Five out of eight Solar System planets rotate rapidly in the prograde fashion,
that is, in the direction of their revolution around the Sun (the Sun spins in
the same direction too). The spins of the two inner planets,
Mercury and Venus, are thought to have been strongly affected by
the tidal interactions with the Sun. Another exception to the prograde rotation is Uranus,  with its spin inclined at more than $90^\circ$ to the Sun's
rotational axis. Therefore, out of the major six planets not strongly
affected by the Solar tides, the only exception to the prograde rotation is
Uranus. The planets spin with a period of between about half a day and a day.

The origin of these large and coherent planetary spins is difficult to
understand \citep[e.g.,][]{LissauerKary91,DonesTremaine93} in the context of the
classical Earth assembly model \citep[e.g.,][]{Wetherill90}.  A planet accreting planetesimals should
receive similar amounts of positive and negative angular momentum \citep{Giuli68,Harris77}. For this reason, the large spins of
the Earth and the Mars are most naturally explained by one or a few ``giant''
planetesimal impacts \citep[][]{DonesTremaine93}. The impacts would have to be very specially oriented to give the Earth and the Mars
  similar spin directions, also consistent with that of the Sun. \cite{JohansenLacerda10} show that accretion of pebbles onto bodies larger than $\sim 100$ of km from the disc tends to spin them up in the prograde direction. Provided that planets accreted $\sim 10-50$\% of their mass via pebble accretion their spin rates and directions are then as observed. In the case of the Earth, a giant impact with the right direction is still needed to explain the Earth-Moon system angular momentum.
  
  In Tidal Downsizing,  gas clumps formed in 3D simulations of fragmenting discs rotate in the prograde direction \citep[][]{BoleyEtal10,Nayakshin11b}. Massive cores formed inside the clumps would inherit the rotational direction of the parent. An exceptional direction of planetary spin, such as that of Uranus, may arise if the host fragment interacted with another fragment and was spun up in that non-prograde direction during the interaction. Such interactions do occur in 3D simulations \citep[e.g., there were a number of such interactions in simulations presented in][]{ChaNayakshin11a}.

\subsection{The Moon}\label{sec:Moon}

The Moon is thought to have formed due to a giant impact of a large solid body on the Earth \citep{HartmannDavis75,CA01}. However, Earth-Moon compositional constraints present a very tough challenge.  In Core Accretion, composition of planetesimals change as a function of distance from the Sun, so Theia (the impactor) is expected to have a similar yet somewhat different composition from the proto-Earth. However, the Moon and the Earth have not just similar, they have undistinguishable isotopic compositions for oxygen  \citep{WiechertEtal01}, and very close isotopic ratios for  chromium \citep{LugShu98}, silicon \citep{GeorgEtal07} and tungsten \citep{TouboulEtal07}.
 This motivated suggestions of complicated and highly efficient mixing processes during the Earth-Theia collision \citep{PahlevanStevenson07}. Numerical simulations of giant impacts indicate that the Moon would have been mainly made of the impactor \citep[$\sim$ 80\%, see][]{Canup08}. The situation has not been improved by the use of much more sophisticted numerical simulation methods \citep[see][]{HosonoEtal16}.
  
In the framework of Tidal Downsizing, (a) assembly of the Earth and the Moon in the centre of the same parent gas clump may also account for the nearly identical isotope compositions, and (b) the prograde orientation of the Earth-Moon angular momentum is the record of the prograde rotation of its parent gas clump \citep{Nayakshin11a}.

 \subsection{Satellites of giant planets}\label{sec:SS_sat_giants}
 
 In the Solar System, giant planets have many satellites, while terrestrial planets, with the exception of the Earth-Moon system, have no significant satellites to speak of. This is usually interpreted as evidence of satellite assembly in a circum-planetary disc that surrounded giant planets during their formation. 
 
 Circum-planetary discs also form in Tidal Downsizing after second collapse of the rotating parent gas fragment \citep{GalvagniEtal12}. 3D numerical simulations of these authors show that the central hydrostatic core (accounting for only $\sim 50$\% of the total fragment mass) is initially surrounded by a thick gas disc. These circum-planetary disc may form the satellites via collapse of the grains rather than H/He phase. The satellites made in this way would be "regular", i.e., those rotating around the planet in the same way as the planet spin axis. Irregular satellites may be those solid bodies that orbited the solid core before the gas envelope of the parent gas fragment was destroyed. When the envelope is removed, the bodies that are weakly bound to the core obtain much more irregular orbits \citep{NayakshinCha12}.

\subsection{Bulk composition of planets}\label{sec:SS_comp1}

As explained in \S \ref{sec:composition}, due to the high temperature  ($T\gtrsim 500$~K or so) in the centres of the host gas fragments, water ice and organic grains are not likely to sediment all the way into the centre of gas fragments and get locked into the core \citep{HelledEtal08,HS08}. This means that cores made by Tidal Downsizing are dominated by rocks and Fe \citep{ForganRice13b,NayakshinFletcher15}. This prediction is consistent with the rock-dominated composition of the inner four planets in the SS.

 In \cite{Nayakshin14b} it has been additionally shown that mechanical strength of grains may also regulate which grains get locked into the core first. In this model, proposed to explain the observed Fe-dominant composition of Mercury \citep{PeplowskiEtal11,SmithEtal12a}, Fe grains sediment before the silicates because their mechanical strength is higher, so that their settling velocity is larger. Most of the silicates remain suspended in gas in the form of small grains, and are removed with the envelope when the parent gas fragment of Mercury is disrupted.


The cores of the Solar System giants Neptune and Uranus are often considered to be icy. However, as shown by \cite{HelledEtal10}, current observations and theoretical calculations of the structure of these two planets do not constrain the core composition (and even its mass) uniquely. Models in which the cores contain only rock or only ice both produce reasonable fits to the data with slightly different fractions of mass in hydrogen and helium (cf. \S \ref{sec:core_comp}).

The fact that gas giant planets Saturn and Jupiter are over-abundant in metals, containing $\sim 30-40 \mearth$ of solids, compared to the Sun is well known \citep{Guillot05}. Tidal Downsizing scenario is consistent with this result (see \S \ref{sec:Zpl_giants}), predicting a similar amounts of solids inside gas giant planets of Saturn and Jupiter masses (see Fig. \ref{fig:Zpl}).

\subsection{The Asteroid and the Kuiper belts}\label{sec:SS_belts}

In the context of Tidal Downsizing, planetesimals are born inside pre-collapse gas fragments \citep[\S \ref{sec:hier} and \ref{sec:planetesimals}, and][]{NayakshinCha12}, and are released into the disc when these fragments are disrupted. \cite{NayakshinCha12} suggested that this model may explain (a) the eccentricity versus semi-major axis correlation for the classical Kuiper Belt objects; (b) the presence of two distinct populations in the belt; (c) the sharp outer edge of the Kuiper belt. In addition, as is well known,  $\sim 99.9$\% of the initial planetesimals are required to have been removed from the Kuiper belt \citep{PfalznerEtal15}  in order to reconcile its current small mass with the existence of bodies as large as Pluto. In Tidal Downsizing, however, massive bodies are assembled inside the environment of a gas fragment, not a disc, so that this "mass deficit" problem of the Kuiper belt does not apply.

For the astroid belt, Tidal Downsizing correctly predicts its location (see eq. \ref{aex1}). Additionally, asteroids are observed to have orbital eccentricities $e\sim 0.1$ and inclinations of 10-20$^\circ$. Tidal disruption of a Jupiter mass gas fragment naturally creates orbits with such properties simply because the size of the Hill radius is $\sim 0.1$ of the orbital separation at the point of the fragment disruption \citep{NayakshinCha12}.

Since the asteroids result from disruptions in the inner few AU of the Solar System, their host fragments must have been rather dense and therefore hot, with gas temperatures likely exceeding $\sim 1000$~K. This predicts refractory composition for both planetary cores and the asteroids. On the other hand, asteroids on orbits beyond the snow line could have accreted water and other volatiles on their surfaces by sweeping the latter up inside the disc, although efficiency of this process needs to be clarified.

Kuiper belt objects (KBO) would result from tidal disruption of more extended and therefore cooler parent fragments. Volatiles (CHON) may now be available for contributing material to building large solid bodies, so Kuiper belt objects made by Tidal Downsizing may contain a larger fraction of ices and volatiles than the asteroids. 

The NICE model for the Solar System architecture \citep[e.g.,][]{GomesEtal05,TsiganisEtal05} has been very successful, especially in its outer reaches \citep{Morbidelli10}. The model is based on the Core Accretion ideas, in particular on the presence of  a massive Kuiper belt that drives migration of Neptune and Uranus.  Without detailed calculations it is difficult to assess whether a similarly successful theory of the Solar System structure could be build starting from the end product of a Tidal Downsizing phase. This is a widely open issue.

\subsection{Timing of planet and planetesimal formation}\label{sec:pl_ages}

The inner terrestrial planets are usually believed to have grown in gas-free environment because their formation ages are found to be in tens of Million years after the formation of the Sun. For example, the age of the Earth is estimated between $\sim 30$ and $\sim 100$ Million years from Hf-W and U-Pb chronometry \citep[e.g.,][]{Patterson56,KoenigEtal11,RudgeEtal10}. If this is true then a Tidal Downsizing origin for the Earth is ruled out since the Earth is nearly coeval with the Sun in this scenario.

However, terrestrial samples provide us with information about only the upper hundreds of km of the Earth. It may well be that the bulk of the planet, that is, $\sim 99\%$ of the mass, is significantly older than the Earth's surface. In confirmation of this, recent research \citep[e.g.,][]{BallhausEtal13} indicates that the Earth accreted lots of volatiles tens of million years after the core formation, suggesting that the U-Pb system of the Earth's silicate mantle has little chronological significance \citep[e.g., \S 2.5 in][]{PfalznerEtal15}. Measured "formation ages" for the other planets and the Moon suffer from similar uncertainties in their interpretation.

%
%
%
%
%
%

\section{Discussion}\label{sec:discussion}

\subsection{Tidal Downsizing, summary of outcomes}\label{sec:exo_basic}

\begin{figure*}
\includegraphics[width=0.95\textwidth]{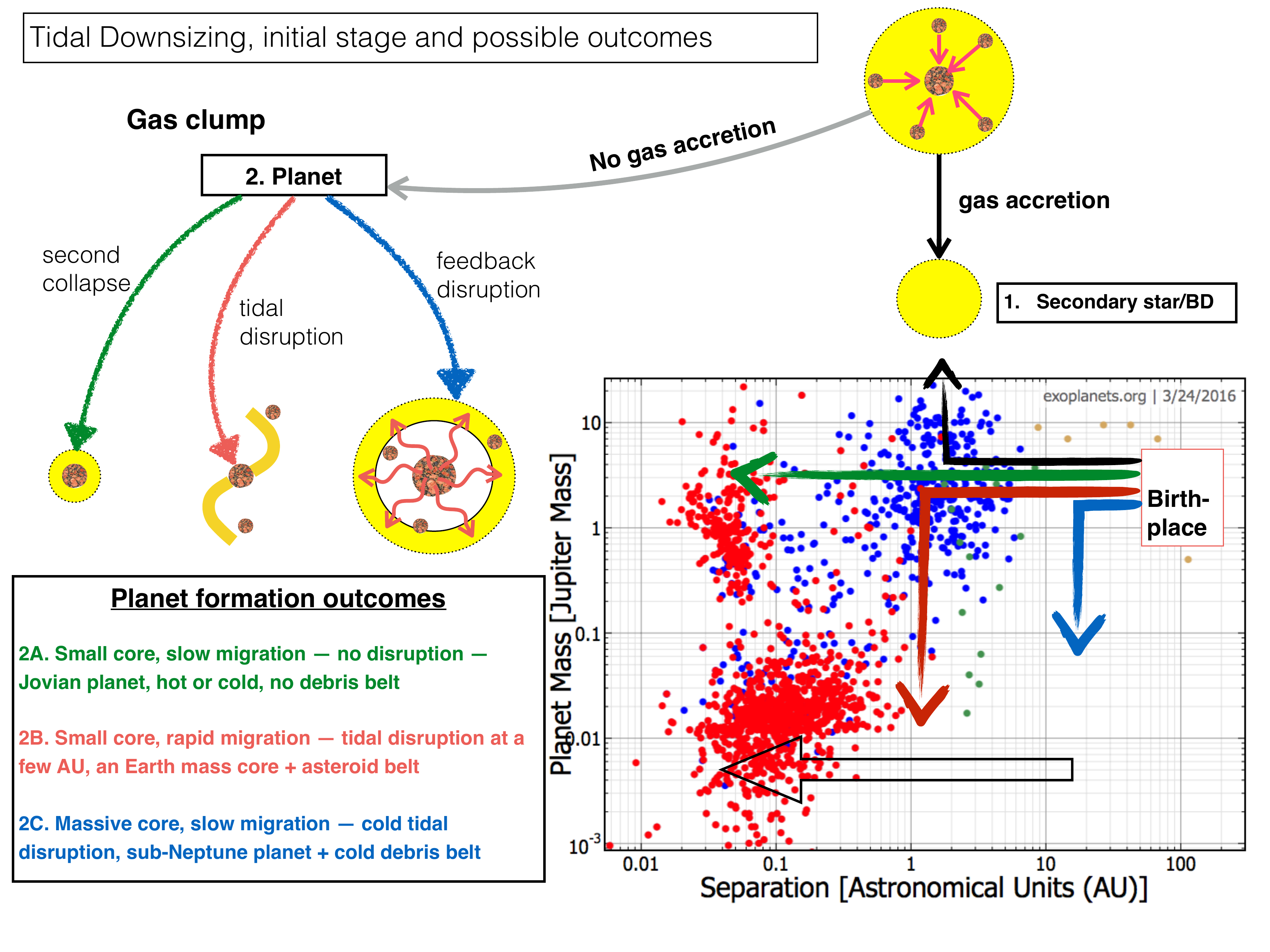}
 \caption{A schematic illustration of how Tidal Downsizing scenario may relate to the observed companions to stars, from planets to low mass stars, as described in \S \ref{sec:exo_basic}.}
 \label{fig:sketch2}
 \end{figure*}

 \SNc{Fig. \ref{fig:sketch2} illustrates as gas clumps born at separations of $\sim 100$~AU from the host star by gravitational disc instability could evolve to produce sub-stellar objects with masses from asteroids and comets to brown dwarfs an host separations from a few stellar radii to tens and even hundreds of AU. The evolutionary paths taken by the objects are shown with arrows on top of the planet mass versus separation diagram from "exoplanets.org"  \citep{HanEtal14}.}
 
 In the top right corner of the figure, the main object of Tidal Downsizing, a pre-collapse gas clump with an ongoing grain sedimentation and core formation is shown. The two arrows  pointing away from the clump show the first important bifurcation in the fate of the clump. If the \SNc{clump accretes gas rapidly (see \S \ref{sec:AorM}), it becomes a brown dwarf or a low stellar mass companion to the host star (path 1, black, pointing down from the clump in the Figure). This evolutionary path is quite analogous to the first--second core evolution of protostars \citep{Larson69}, except it takes place inside a massive protoplanetary disc.}
 
\SNc{ If the clump does not accrete gas, it evolves towards becoming a planet or planetary remnant(s) (grey, to the left from the clump in the figure). Three main outcomes could be distinguished here:}

(2A) {\it A gas giant planet} (green arrows in the sketch). If the inward radial migration of the fragment is slower than planet contraction, and if the core feedback is sufficiently weak, the fragment contracts and survives as a gas giant planet. Usually, this requires the core mass to be below a Super Earth mass ($\lesssim 5 \mearth$, \S \ref{sec:feedback}). Planet migration may bring the planet arbitrarily close to the host star, including plunging it into the star. No debris ring of planetesimals is created from this clump since it is not disrupted.

(2B){\it A low mass solid core planet}, $M_{\rm p}\lesssim$ a few $\mearth$ (red arrows). Similar to the above, but the fragment is migrating in more rapidly than it can collapse. In this case it fills its Roche lobe somewhat outside the exclusion zone boundary and gets tidally disrupted. This results, simultaneously, in the production of a small rocky planet and an Asteroid belt like debris ring at a few AU distance from the host star.

(2C) {\it A high mass solid core planet}. If the fragment is able to make a massive solid core, $M_{\rm core}\gtrsim 5-10\mearth$, its feedback on the fragment may unbind the fragment at separations as large as tens of AU. This process is shown with the blue arrow and leaves behind the massive core, plus a Kuiper-belt like debris ring. 

All of the planets and even stars so created may continue to migrate in, as shown by the black open arrow on the bottom right of the sketch, until the disc is finally removed. Note that a much more massive disc is needed to move a brown dwarf or a star into the inner disc region as opposed to moving a planet. Because very massive gas discs cannot be very common, this predicts that brown dwarfs and stellar mass companions are more likely to be found at large (tens of AU or more) separations; gas giant planets  are more likely to migrate closer in to the host star.

\subsection{Observations to test this scenario}

Dozens of independent numerical simulations (\S \ref{sec:rapid}) show that Jupiter mass planets migrate from $\sim 100$~AU into the inner $\sim 10$~AU or less in about 10,000 years or even less. Therefore, the popular idea \citep[e.g.,][]{Boley09} of dividing the observed population of planets onto "made by Core Accretion" (inside the inner tens of AU) and "made by Gravitational Instability" (outside this region) is not physically viable. Based on the rapid migration speeds found in the simulations, a giant planet observed at $\sim 0.1$~AU is as likely to have migrated there from a few AU as it is to have migrated there from $100$~AU. Likewise, due to tidal disruptions, Tidal Downsizing produces a numerous supply of core-dominated planets, many of which may end up at same distances as normally reserved for the Core Accretion planets.

We thus need to be crystal clear on which observables can be used to differentiate between the two scenarios and which are actually less discriminating  than previously thought. 

\subsubsection{Similarities between the two scenarios}

The observed planets naturally divide into two main groups -- those dominated by solid cores, usually below mass of $\sim 20\mearth$, and those dominated by gas, usually more massive than Saturn ($\sim 100 \mearth$). This has been interpreted as evidence for gas accretion runaway \citep[e.g.,][]{MordasiniEtal09b,MayorEtal11} above the critical mass for the core-nucleated instability \citep{Mizuno80,Stevenson82,Rafikov06}. However, a similar bi-modality of planets is found in Tidal Downsizing (Fig. \ref{fig:atmo}). When the parent gas fragment is disrupted, the mass of the gas remaining bound to the core is usually a small fraction of the core mass for reasons quite analogous to those of Core Accretion (\S \ref{sec:atm}). This implies that the observed dichotomy of planets may be driven by the fundamental properties of matter (equation of state and opacities) rather than by how the planets are made. 

The bulk composition of planets is another example where the predictions of the two theories are not so different. In Core Accretion, the more massive the planet is, the smaller the fraction of the total planet mass made up by the core. This may account for the observed over-abundance of metals decreasing with the planet mass \citep{MillerFortney11}. In Tidal Downsizing, the more massive the gas giant is, the smaller is the "pebble accretion boost" needed for it to collapse, and this may also account for the observations (see Fig. \ref{fig:Zpl} \& \S \ref{sec:Zpl_giants}).

The strong preference amongst gas giants to orbit metal rich rather than metal poor hosts is well known \citep[e.g.,][]{Gonzalez99,FischerValenti05,SanterneEtal15}, and is normally attributed to the more rapid assembly of massive cores in metal rich discs \citep{IdaLin04b,MordasiniEtal09b}. However, if gas giants collapse due to "metal loading" \citep{Nayakshin15a} rather than due to the classical radiative collapse \citep{Bodenheimer74}, then the frequency of their survival is also a strong function of the host disc metallicity \citep{Nayakshin15b,NayakshinFletcher15}. These observations cannot be claimed to support one of the two planet formation scenarios.

\subsubsection{Observable differences between the theories}


Tidal Downsizing however predicts that beyond the exclusion zone at $a\sim$  a few AU, there should be no correlation between the gas giant presence and the host star metallicity because the tidal disruption "filter" does not apply or at least applies not as strongly there (\S \ref{sec:cold_giants_Z}). Observations \citep[][]{AdibekyanEtal13} started to probe the few-AU region of the parameter space, and there is a hint that this prediction is supported by the data \citep[][see also Fig. \ref{fig:Vardan}]{AdibekyanEtal15}, but more observations are needed.

Similarly, planets more massive than $\sim 5-10\mj$ and brown dwarfs should not correlate with the metallicity of the host in the Tidal Downsizing model (\S \ref{sec:transition}), whatever the separation from the star. Currently, this prediction is clearly supported by observations of brown dwarfs and low mass stellar companions to stars \citep{RaghavanEtal10,TroupEtal16} but the transition region between planets and brown dwarfs is not well studied. Massive gas giant planets do appear to become less sensitive to the host metallicity above the mass of $5\mj$ (\S \ref{sec:massive_giants_Z} and Fig. \ref{fig:Z_massive}), but more data are desirable to improve the statistics.

At the lower mass end, there are differences between the models too. In the framework of Tidal Downsizing, planetary debris is only made when the gas clumps -- the future gas giant planets -- are disrupted (see \S \ref{sec:planetesimals} \& \ref{sec:hier}). Since tidal disruption of the clumps anti-correlates with the host metallicity as explained above, no simple correlation between the debris disc presence and host [M/H] is predicted \citep{FletcherNayakshin16a}. Secondary predictions of this picture (see \S \ref{sec:Z_debris}) include a possible correlation of the debris disc presence with that of a sub-Saturn planet (that is, any downsized planet), and an anti-correlation with the presence of  gas giant planets. 

\SNc{Further, post-collapse planets are too hot to permit existence of asteroid or comet like debris inside of them. Pre-collapse planets are disrupted not closer than the exclusion zone, as mentioned above, so that debris belts made by Tidal Downsizing must be never closer than $\sim 1$~AU to the host solar type star. This is different from Core Accretion where planetesimals are postulated to exist as close as $\sim 0.1$~AU from the host star \citep[e.g.,][]{ChiangLaughlin13}. \cite{KenyonEtal16} identifies the very low frequency of observed {\em warm} debris discs ($\sim 2-3$\%) in young debris discs as a significant puzzle for Core Accretion, and offers a solution. Another difference is the likely much smaller mass of the debris rings made by Tidal Downsizing, and their significant birth eccentricities \citep[up to $e\sim 0.1$;][]{NayakshinCha12}.}

For cores, the host star metallicity correlation is predicted to depend on the core mass in Tidal Downsizing. Low mass cores, $M_{\rm core} \lesssim$ a few $\mearth$, are most abundant around low metallicity hosts because of the already mentioned tendency of the parent gas clumps to be disrupted more frequently at low metalicites. High mass cores, on the other hand, are mainly made in disruptions of gas clumps made by metal-rich discs \citep[e.g., see the black curve in Fig. 3 in][]{FletcherNayakshin16a}. Therefore cores more massive than $\sim 10-15\mearth$ are likely to correlate with the metallicity of the host. For a broad range of core masses, one gets no strong correlation with [M/H], somewhat as observed \citep{NayakshinFletcher15}. Future observations and modelling of core correlations with metallicity of the host are a sensitive probe of the two planet formation scenarios.

While some of the Core Accretion population synthesis models also predict no strong correlation between core-dominated planets and the host star metallicity \citep[e.g.,][]{MordasiniEtal09b}, the degeneracy between the two models may be broken in two areas. Tidal Downsizing predicts that massive core formation is a very rapid process, even at $\sim 100$ AU, requiring less than $\sim 10^5$ years \citep{Nayakshin16a}, whereas Core Accretion takes $\sim 1-3$~ Million years even at distances $a\lesssim 10$~AU. ALMA observations of protoplanetary discs such as HL Tau (\S \ref{sec:HLT}), showing signs of very early planet formation, is key to constrain the timing of massive core growth and is a challenge to the classical version of Core Accretion.\footnote{As an aside, the recently discovered rapid core growth via pebble accretion \citep[e.g.,][]{JohansenEtal14a,JohansenEtal15a,LevisonEtal15} may solve the HL Tau mystery in the context of Core Accretion, but then the classical framework for the metallicity correlations suggested by \cite{IdaLin04b,MordasiniEtal09b} is in doubt because it is based on a long core growth time scale. Therefore, at the present it appears that Core Accretion may account for either the well known gas giant planet -- host star metallicity correlations (\S \ref{sec:giants_Z}) or the HL Tau young cores, but not both.}

Another area where the two models differ is the expected core composition. Core Accretion predicts that ices may be the dominant contributor to the mass budget of massive cores \citep{PollackEtal96}. While these cores would form beyond the snow line, many would migrate all the way into the inner tenths of an AU region that is accessible to modern observations \citep[e.g., see Fig. A1 in][]{ColemanNelson16}. Tidal Downsizing predicts that ices and organics are less likely to contribute to making planetary cores than silicates because the  ices and organics are too volatile to sediment into the centres of hot pre-collapse fragments \citep[][also \S \ref{sec:composition}]{HS08,HelledEtal08}. 

Cores that are further away than $\sim 0.1$~AU from their hosts, including the Solar System giants, do not present us with a clean composition test because their mass-radius relation is degenerate due to the unknown H/He mass fraction \citep[e.g., see \S 5.1.2 in][]{HelledEtal13a}. However, moderately massive cores \citep[$M_{\rm core}\lesssim 7\mearth$, see][]{OwenWu13} lose their H/He envelopes due to photo-evaporation at separations less than $\sim 0.1$~AU. It is thus sensible to concentrate on these close-in cores when pitting Tidal Downsizing against Core Accretion. The close-in cores are (so far) observed to have a rocky Earth-like composition (\S \ref{sec:core_comp}), but the current data are still scarce.

Observations show a strong roll-over in frequency of planets more massive than $\sim 20 \mearth$ \citep{MayorEtal11} or larger than $\sim 4 R_\oplus$ \citep{HowardEtal12}. Building solid cores via accretion of planetesimals or via giant impacts has no obvious limit at this mass range except for the run away by gas accretion \citep{PollackEtal96,MordasiniEtal09b}. This scenario should however not apply to metal-poor systems: if these are made in gas-free discs \citep{IdaLin04b}, then their cores should be free to grow more massive than $M_{\rm crit}$. Very massive solid cores are however not observed around metal-poor stars. In Tidal Downsizing, the drop above the mass of $\sim 20 \mearth$ may be due to the strong feedback unleashed by the massive cores onto their host gas fragments (\S \ref{sec:feedback} and Fig. \ref{fig:pmf_fb}). This mechanism should affect both metal rich and metal poor systems. Observations of stars more massive than the Sun may be helpful here, as these are expected to have more massive discs \citep{MordasiniEtal12}, and thus their cores should be more massive if made by Core Accretion and not if made by Tidal Downsizing.

Finally, planet formation in extreme systems such as binaries is a very tough test for any planet formation scenario. Kepler-444 may be an example of a system where the observed planets could not have been made by Core Accretion, as argued in \S \ref{sec:kepler444}, due to the inner disc being both too hot to make the planets in situ, and yet not long lived enough to move them in place if made further out. However, it remains to be seen if detailed simulations in the framework of Tidal Downsizing could produce such an extreme planetary system.

\subsection{Open issues}\label{sec:dis_ass}


The population synthesis model of \cite{NayakshinFletcher15} assumes, for simplicity, that gas fragments evolve at a constant {\it gas} mass until they are disrupted or they collapse. The disruption is assumed to remove all of the gas envelope except for the dense layers of gas strongly bound to the core, the core atmosphere (\S \ref{sec:atm}). This is based on the fact that a polytropic gas clump with index $n=5/2$ is strongly unstable to the removal of mass as it expands as $R_{\rm p}\propto M_{\rm p}^{-3}$ when the mass is lost. Within these assumptions, the model requires gas clumps with the minimum initial mass min[$M_{\rm in}$]$\sim (0.5-1)\mj$ to account for the observed gas giant planets, many of which have mass around that of Jupiter or less. This is somewhat uncomfortable since most authors \citep[e.g.,][and \S \ref{sec:AorM}]{ForganRice13} find that the minimum initial mass of a gas clump born by gravitational instability of a protoplanetary disc is $M_{\rm in} \sim 3-10\mj$, and that gas clumps may accrete more gas \citep[e.g.,][]{KratterEtal10}.

This important disagreement needs to be investigated with 3D numerical simulations of both fragmenting discs and individual gas clumps. Similarly, 3D numerical simulations of gas fragment collapse are needed to ascertain angular momentum evolution of gas clumps, which is of course not resolved in the current 1D population synthesis. This evolution may dictate how much of the clump collapses into the planet proper and how much into the circum-planetary disc \citep{BoleyEtal10,GalvagniEtal12}, and what the spins of the planets and the core are \citep{Nayakshin11a}. Formation of the circum-planetary disc is key to formation of planet satellites. Further, grain sedimentation, core formation and especially planetesimal/debris formation within the fragment are certainly not spherically symmetric (e.g., see Fig. \ref{fig:num3D}), so 3D coupled gas-grain simulations of gas clumps are urgently needed.


Another unsolved issue is gas accretion onto gas clumps, which is likely to control the frequency with which planets are made as opposed to brown dwarfs   \citep[][see also \S \ref{sec:AorM}]{ZhuEtal12a,NayakshinCha13,Stamatellos15}. \SNc{Preliminary work (\S \ref{sec:desert} shows that efficiency of gas accretion strongly depends on the cooling rate of gas in the Hill sphere of the planet. This suggests that this issue will remain uncertain for some time since dust opacity of the gas is uncertain.}

3D simulations are also needed to address how the presence of multiple gas clumps changes the predictions of population synthesis \citep[][allowed multiple gas fragments in their protoplanetary discs, but it was not possible to track stochastic clump-clump interactions or orbit interchanges]{ForganRice13}. So far, 3D numerical simulations of fragmenting discs did not resolve the internal processes within the fragments, and have also been performed for a relatively small number of test cases \citep[e.g.,][]{BoleyEtal10,ChaNayakshin11a}. Ideally, the strengths of the 1D isolated clump models (grain physics, long term evolution of the clumps and the disc) should be imported into the 3D simulations of global discs with self-consistent fragment formation in order to overcome the shortcomings.

Another assumption made in the population synthesis presented here is that dust opacity has not been modified much by grain growth inside the clumps. This is an approximation only.  Grain growth clearly occurs in protoplanetary discs and should be included into the models. Numerical experiments of \cite{Nayakshin15c} suggest that grain opacity reduction by a factor of $\sim 3$ can be tolerated, but factors of tens would be too large. Self-consistent models of fragment evolution with grain growth \citep[in the style of][]{HB11} and metal loading are needed to explore these issues better.

Tidal Downsizing hypothesis is very young and is so far untested on dozens of specific planet formation issues in the Solar System and beyond, such as formation of short period tightly packed systems \citep[e.g.,][]{HandsEtal14}, the role of ice lines in the model, etc. and etc. \SNc{One may clearly critique the model for failing to address these systems. However,  } these issues have not been covered here not because of the author's desire to hide away from the data but rather due to a lack of detailed work on these specific issues. Commenting on these without performing a thorough calculation first would amount to speculation one way or another. The author plans, and invites the community, to examine these additional constraints in the future.

\begin{acknowledgements}
This research was funded by STFC and made use of ALICE and  DiRAC High Performance Computing Facilities at the University of Leicester. I thank Ed Vorobyov, Ravit Helled, Richard Alexander, Vardan Adibekyan, Duncan Forgan, Dimitris Stamatellos, Lucio Mayer, Eugene Chiang, Alexandre Santerne, and Mark Fletcher for valuable discussions and comments on the draft. Christoph Mordasini is thanked for providing his data for one of the figures. The Chief Editor of PASA, Daniel Price, is thanked for the invitation to collect the rather broad material into one, hopefully coherent, story, and for his encouragement and patience while this review was completed. A special thank you is to an anonymous referee whose detailed report improved presentation of this review significantly.
\end{acknowledgements}


\end{document}